\documentclass[fleqn,usenatbib]{mnras}
\usepackage{amssymb}
\usepackage{pifont} 
\usepackage{newtxtext,newtxmath}
\usepackage{subcaption}
\usepackage[version=4]{mhchem}
\usepackage{tabularray}
\usepackage{graphicx} 
\usepackage{xspace}
\usepackage[T1]{fontenc}
\usepackage{siunitx}
\usepackage{gensymb}
\usepackage{fix-cm}
\usepackage{tikz}
\usetikzlibrary{arrows.meta, shapes.geometric, positioning, fit, calc}

\DeclareRobustCommand{\VAN}[3]{#2}
\let\VANthebibliography\thebibliography
\def\thebibliography{\DeclareRobustCommand{\VAN}[3]{##3}\VANthebibliography}

\usepackage{amsmath}
\usepackage{lipsum}

\newcommand{\one}{3DOneExpl\xspace}

\newcommand{\two}{3DTwoExpl\xspace}

\newcommand{\mass}{\mbox{M$_{\odot}$}}

\newcommand{\submch}{sub-$M_\mathrm{Ch}\xspace$}

\newcommand{\microns}{\ensuremath{\,\text{\textmu}\mathrm{m}}\xspace}
\newcommand{\dsix}{D$^6$\xspace}
\newcommand{\kms}{$\,\mathrm{km}\,\mathrm{s}^{-1}$\xspace}
\newcommand{\threedvm}{\mbox{3DViolentMerger}\xspace}
\newcommand{\onedvm}{\mbox{1DViolentMerger}\xspace}
\newcommand{\vm}{\mbox{Violent Merger}\xspace}

\title[3D Nebular Signatures of SNe~Ia] {
  Nebular Fingerprints of a Violent White Dwarf Merger: 3D NLTE Modelling of Type Ia Supernovae}
\author[J. M. Pollin et al.]{J. M. Pollin,$^{1}$
R. Pakmor,$^{2}$
L. J. Shingles,$^{3}$
F. P. Callan,$^{3}$
S. A. Sim,$^{3}$
L. A. Kwok,$^{4}$
C. E. Collins,$^{5}$
\newauthor
F. K. Röpke$^{6,7,8}$ and
A. L. McGarrity$^{3}$\\
$^{1}$Department of Physics, Oregon State University, 301 Weniger Hall, Corvallis, OR 97331-6507, USA\\
$^{2}$Max-Planck-Institut für Astrophysik, Karl-Schwarzschild-Str. 1, D-85748, Garching, Germany\\
$^{3}$Astrophysics Research Center, School of Mathematics and Physics, Queen’s University Belfast, Belfast BT7 1NN, Northern Ireland, UK\\
$^{4}$Center for Interdisciplinary Exploration and Research in Astrophysics (CIERA), 1800 Sherman Ave., Evanston, IL 60201, USA\\
$^{5}$School of Physics, Trinity College Dublin, The University of Dublin, Dublin 2, Ireland\\
$^{6}$Heidelberger Institut für Theoretische Studien, Schloss-Wolfsbrunnenweg 35, 69118 Heidelberg, Germany\\
$^{7}$Zentrum für Astronomie der Universität Heidelberg, Institut für Theoretische Astrophysik, Philosophenweg 12, 69120 Heidelberg, Germany\\
$^{8}$Zentrum für Astronomie der Universität Heidelberg, Astronomisches Rechen-Institut, M{\"o}nchhofstr.\ 12--14, 69120 Heidelberg, Germany
}

\date{Accepted XXX. Received YYY; in original form ZZZ}

\pubyear{2026}

\begin{document}
\label{firstpage}
\pagerange{\pageref{firstpage}--\pageref{lastpage}}
\sloppy
\maketitle

\begin{abstract}
Binary systems composed of two carbon-oxygen white dwarfs (WDs) are a leading progenitor candidate for Type Ia supernovae. 
One widely discussed scenario is the dynamically driven double-degenerate double-detonation (\dsix) of a sub-Chandrasekhar-mass WD binary, where detonations are triggered by dynamical interaction.
However, some systems are expected to undergo violent mergers, in which the primary ignites through direct carbon ignition as the secondary strikes its surface.
We present the first 3D nebular-phase radiative-transfer calculations of a violent merger, using a $1.1\mass$ and $0.7\mass$ sub-Chandrasekhar binary.
Our simulations employ a full NLTE (non local thermodynamic equilibrium) treatment of excitation and ionisation, including non-thermal electron contributions.
By comparing 1D and 3D realisations, we show that multidimensional modelling improves the ionisation state and reveals features absent from 1D calculations, most notably [\ion{O}{I}] from unburned material associated with the secondary.
The model reproduces much of the panchromatic spectrum of the normal SN~2021aefx, but underpredicts [\ion{Ni}{II}] while producing strong high-ionisation stable-Ni features, illustrating that stable-Ni signatures depend not only on abundance, but also on ionisation state.
Although the model does not reproduce the strong [\ion{Ar}{II}] and [\ion{Ne}{II}] emission observed in the 03fg-like SN~2022pul, our calculations suggest that this event may require a similar merger configuration, involving full disruption of the secondary or a more massive companion with more extensive burning.
Finally, viewing-angle variation is substantial, with signatures distinct from \dsix-like scenarios, suggesting that JWST nebular samples, combined with multidimensional modelling, can discriminate between channels.
\end{abstract}

\begin{keywords}
Radiative transfer – Transients: supernovae – Methods: numerical - Stars: binaries - White Dwarfs
\end{keywords}
\section{Introduction}
\label{sec:Introduction}

There is broad agreement that Type Ia supernovae (SNe~Ia) arise from the thermonuclear explosion of a carbon–oxygen (CO) WD in a close binary system \citep{Hoyle1960}. However, it remains a subject of debate whether the majority of SNe~Ia originate from progenitors near the Chandrasekhar mass ($M_\mathrm{Ch}$) or from sub-Chandrasekhar mass (sub-$M_\mathrm{Ch}$) WDs (see e.g.\,\citealt{Liu2023,Ruiter2025} for a review). The nature of the companion star also remains uncertain, and there is no consensus on whether a non-degenerate companion in the single-degenerate scenario \citep{Whelan1973} or another WD in the double-degenerate scenario \citep{Iben1984} is favoured. 

An increasing body of evidence supports the double-degenerate scenario. This includes the absence of hydrogen \citep{Leonard2007}, the relatively weak contribution of helium in observed spectra \citep{Jiang2017,Noebauer2017,De2019,Collins2023,Callan2025He}, the consistency of predicted rates of sub-$M_\mathrm{Ch}$ WDs with observed SNe~Ia rates \citep{Ruiter2009}, and the ability of these systems to reproduce the observed delay-time distribution \citep{Maoz2012}. Further support also comes from soft X-ray limits on steadily accreting $M_\mathrm{Ch}$ WDs in nearby galaxies \citep{Gilfanov2010}.

Idealised sub-$M_\mathrm{Ch}$ CO WD detonation models have been explored extensively and have demonstrated good agreement with observations of normal SNe~Ia \citep[e.g.,][]{sim2010,Blondin2017,Shen2018,Shen2021a}. However, the uncertain nature of the companion introduces ambiguity in the detonation mechanism. Among the proposed scenarios, double detonations of the underlying CO core have received particular attention \citep{Taam1980,Livne1990,Nomoto1980,Hoeflich1996,Nugent1997}. Modern double detonation models \citep{Bildsten2007,Shen2009,kromer2010,fink2010,Shen2010,Townsley2019,Polin2019,Gronow2020,Gronow2021,Boos2021,Shen2024} have achieved detonations with increasingly lower-mass helium shells compared to earlier work \citep{Livne1990,Livne_Glasner1990,Livne1995}, resulting in synthetic observables that are more consistent with observations. As a result, this mechanism remains an active area of investigation.

Recent work by \cite{Shen2024} demonstrated that nearly all CO WDs in close binaries may explode if they ignite a helium detonation on the primary WD during the dynamical inspiral phase, leading to both double and quadruple detonations \citep[i.e., disruption of both the primary and secondary WD;][]{Tanikawa2018,Tanikawa2019,Roy2022,pakmor_2021,Boos2024}. However, achieving the required numerical resolution to model a shell detonation at low helium shell masses remains challenging, and current 3D simulations still require unphysically large initial helium masses.
Using high-resolution 1D calculations, \cite{Shen2024} demonstrated that only the most massive WDs ($\gtrsim1.0\mass$) are unable to dynamically ignite the helium shell they form at birth, and must instead accrete helium prior to binary interaction. While the photospheric, nebular, and hydrodynamic signatures of these dynamically driven double-degenerate double-detonation scenarios have been extensively explored in recent years \citep[e.g.,][]{Boos2024,Pollin2024a,Pollin2025,Kumar2025,Simotas2025,Ferrand2025,Fujimaru2026}, the exact dynamics of merging WDs remains uncertain, particularly those at high masses (i.e., $\gtrsim1.0\mass$). Therefore, several explosion scenarios continue to be viable.

One such alternative is the violent merger scenario \citep[e.g.,][]{Pakmor2010,Pakmor2011,Pakmor2012b,Pakmor2013,Pakmor2026}, in which the secondary WD is tidally disrupted during the final stages of the inspiral and accelerates towards the primary. The disrupted secondary then strikes the surface of the primary, directly triggering a carbon detonation that propagates through the primary and unbinds the system. Violent mergers were initially proposed as a pathway to produce normal SNe~Ia, as they can reproduce several key photospheric features \citep{Pakmor2012b}. However, subsequent work has suggested that they are more likely associated with peculiar subclasses \citep[see][for an in-depth review]{stefan_sn_review}, due to their large polarisation signatures \citep{Bulla2016} and their ability to reproduce events such as the 02es-like SN~2010lp \citep{Kromer2013}. In particular, these models were expected to account for late-time optical [\ion{O}{I}] emission in peculiar events, arising from unburned material located near the centre of the ejecta. However, this feature was not reproduced in recent nebular calculations by \cite{Blondin2023}, which instead yielded spectra more similar to those of normal SNe~Ia.

Violent mergers have been proposed as promising progenitors for several peculiar subclasses of SNe~Ia, in particular the 03fg-like events. These objects exhibit light-curve morphologies broadly similar to normal SNe~Ia, but possess higher peak luminosities and slower declines \citep{stefan_sn_review}. As violent mergers are expected to occur for relatively massive WDs, their intrinsically high luminosities, combined with interaction with ejected material from the secondary \citep{Dimitriadis2023,Siebert2023,Siebert2024,Kwok2024}, make them strong candidates for this subclass. They have also been suggested as progenitors of subluminous 02es-like SNe~Ia \citep{Maguire2011,Ganeshalingam2012,Kromer2013,Srivastav2023_ilv}. Notably, both subclasses show early-time excess emission (i.e., `bumps') in their light curves, which is absent in most normal SNe~Ia and may point towards a unified progenitor scenario \citep{Hoogendam2024}. There may also exist a continuum between these subclasses and normal SNe~Ia, with transitional objects such as SN~2006bt \citep{Foley2010} and SN~2022ywc \citep{Srivastav2023_ywc}. While such events can be reproduced by double- and quadruple-detonation models with large helium shells \citep{Polin2021,pakmor_2021}, these scenarios may not account for the circumstellar material required to explain the early excess emission.

Recently, \cite{Pakmor2026} revisited the violent merger scenario, by simulating a $1.1\mass$ and $0.7$\mass WD binary in \textsc{arepo} \citep{Springel2010,Pakmor2016,Weinberger2020}. When compared to an earlier investigation of a violent merger of massive WD \citep{Pakmor2011}, the secondary WD was found to not be fully disrupted, but instead survives and is ejected as a bound object. At the same time, circumstellar material is produced during the inspiral, and unburned material remains in the central ejecta. The overall sequence and structure of this explosion model is described in detail in Section~\ref{sec:Method}. It is worth noting that $\sim1\%$ of SNe~Ia are estimated to be 03fg- or 02es-like \citep{Dimitriadis2025}, which is comparable to the observed fraction of hypervelocity ($\gtrsim2,000$\kms) WDs \citep{Shen2018,El_Badry2023,Shen2025}. \cite{Pakmor2026} and \cite{Bhat2026} found that the surviving WD is consistent with the hypervelocity objects D6-1 and D6-3, and that the low rate of violent mergers is compatible with the lack of detected surviving companions in nearby supernova remnants \citep{Schaefer2012,GonzalezHernandez2012,Kerzendorf2018}.

Understanding how explosion models appear at late times is essential for assessing their viability as SNe~Ia progenitors. As the ejecta expand and become optically thin, emission from the inner regions becomes visible during the nebular phase. This allows for a direct probe of both the composition and geometry of the central ejecta, which are considerably different from the outer ejecta in many explosion models \citep{Pakmor2024}. At these epochs (hundreds of days post-explosion), excitation via radiative and collisional processes becomes inefficient, such that only low-lying energy levels are significantly populated and forbidden emission lines dominate the spectra \citep{Spyromilio1992}. Observed nebular features in SNe~Ia are often shifted relative to their rest wavelengths \citep{Motohara2006,Gerardy2007,Maeda2010,Maguire2018,Flores2020}, reflecting an underlying asymmetric ejecta \citep{Maeda2010}. These asymmetries also produce variations in line widths and profile morphologies due to geometric projection effects \citep[see][for a review]{Jerkstrand2017}. While simplified geometries are often used for interpretation, modern 3D explosion models predict highly complex ejecta structures that do not always correspond to these simplified structures \citep[e.g.,][]{Baal2023,Pollin2025}.

Since the one-zone models of \cite{Axelrod1980}, numerous studies have investigated the nebular-phase properties of SNe~Ia \citep[e.g.][]{Ruiz1992,Liu1997,Mazzali2001,Hoflich2004,Kozma2005,Maeda2010,Maurer2011,Li2012,Diamond2015,Fransson2015,Botyanszki2017,Mazzali2018,Shingles2020,Mazzali2020,Polin2021,Shingles2022a,Blondin2023}. However, many of these models assume spherical symmetry, fully optically thin emission, or simplified ejecta geometries, and are therefore limited in reproducing the diversity shown by observations. Recently, we performed the first multidimensional nebular-phase radiative transfer calculations for a modern 3D explosion model \citep{Pollin2025}. These simulations demonstrated that 3D effects have significant implications for line profile morphologies, velocities, full width at half maximum (FWHM), and ionisation state. Crucially, they can also reveal spectral features that are entirely absent in 1D calculations and introduce strong viewing-angle dependence that can bring explosion models into better -- and worse -- agreement with observed SNe~Ia.

Interpreting late time spectral features is particularly challenging as the optical and near-infrared (NIR) suffer from heavy line blending from multiple species and ionisation stages, as well as telluric absorption. However, with the arrival of the \textit{James Webb Space Telescope} (JWST) it is now possible to explore the mid-infrared (MIR), where line blending is significantly reduced. Recent simultaneous panchromatic (optical + NIR + MIR) JWST observations span both normal and peculiar SNe~Ia, including SN~2021aefx \citep{Kwok2023,Ashall2024,DerKacy2023}, SN~2022xkq \citep{DerKacy2024}, SN~2024pxl and SN~2024vjm \citep{Kwok2025Iax}, SN~2022aaiq and SN~2024gy \citep{Kwok2025}, SN~2023qov \citep{Macrie2026}, and SN~2022pul \citep{Kwok2024} and provide a powerful dataset against which theoretical models can be tested.
However, all aforementioned JWST observations exhibit significant asymmetry which cannot be robustly understood by employing a 1D approach. It is therefore only by accounting for the multidimensional nature of these explosions that theoretical models can be meaningfully compared to observations and used to constrain progenitor scenarios \citep{Pakmor2024}.
Of the SNe~Ia observed with JWST, SN~2022pul remains the only 03fg-like event with panchromatic coverage. It exhibits several distinctive features, including optical [\ion{O}{I}], strong [\ion{Ca}{II}], and prominent MIR [\ion{Ar}{II}] and [\ion{Ne}{II}] emission, alongside comparatively weak stable Ni features. The presence of [\ion{O}{I}] is particularly notable, as it is exceedingly rare and has previously only been identified in a small number of events, including SN~2012dn \citep{Taubenberger2019} and SN~2021zny \citep{Dimitriadis2023}. Recent work by \cite{Blondin2023} showed that among existing models, only the violent merger scenario of \cite{Pakmor2012b} is able to produce strong MIR [\ion{Ne}{II}] emission. Combined with the expected emission of optical [\ion{O}{I}], this has led to the suggestion that 03fg-like SNe~Ia may originate from violent mergers.

In this work, we investigate the nebular-phase predictions of a modern violent merger explosion model to assess this connection to 03fg-like SNe~Ia. Our primary aim is to determine the generic spectral signatures of this scenario and how these predictions depend on observer orientation. In particular, we aim to examine how a full 3D treatment alters the ionisation state, line profiles, and emergent features when compared to a spherically averaged 1D calculation. We also aim to determine how the predicted orientation-dependent signatures compare to those from double and quadruple detonation models \citep{pakmor_2021,Pollin2025}, and assess whether specific intermediate-mass element (IME) and iron-group element (IGE) features can be related to the initial progenitor system. Finally, we aim to determine if this explosion scenario can reproduce key characteristics of both normal and peculiar SNe~Ia.
More broadly, this work aims to expand the current set of 3D predictions available for modern explosion models, providing a resource that can be used to interpret both existing ground-based observations and the growing JWST sample. By establishing clear and testable MIR predictions, we seek to motivate further JWST observational campaigns.  
The explosion model and radiative transfer setup are described in Section~\ref{sec:Method}. In Section~\ref{sec:angleaveraged}, we present the direction-averaged and spherically averaged synthetic spectra from both 3D and 1D realisations of the \cite{Pakmor2026} model. Section~\ref{sec:3DViolentMerger_Orientation_Effects} explores the viewing-angle dependence of both IME and IGE features. In Section~\ref{sec:comparison}, we compare our predictions with the panchromatic spectra of SN~2021aefx, SN~2022aaiq, SN~2024gy, and SN~2022pul at their respective epochs. Finally, we summarise our findings and present our conclusions in Section~\ref{sec:Discussion_and_Conclusions}.  

\section{Methods}
\label{sec:Method}

\subsection{Hydrodynamical Ejecta Models}
\label{sec:Models}

\begin{figure*}
\centering

\begin{subfigure}[b]{1\textwidth}
\hfill
   \includegraphics[width=0.97\linewidth]{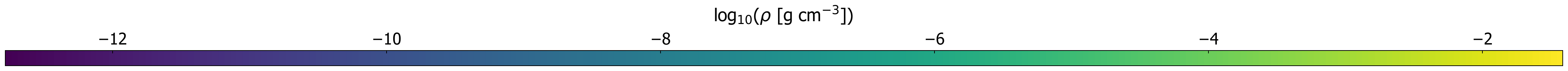}  
\end{subfigure}

\begin{subfigure}[b]{1\textwidth}
   \includegraphics[width=1\linewidth]{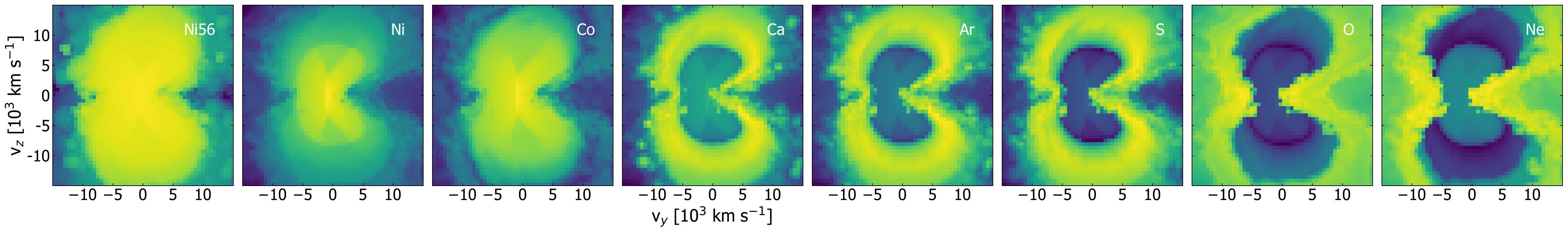}
\end{subfigure}
\begin{subfigure}[b]{1\textwidth}    
   \includegraphics[width=1\linewidth]{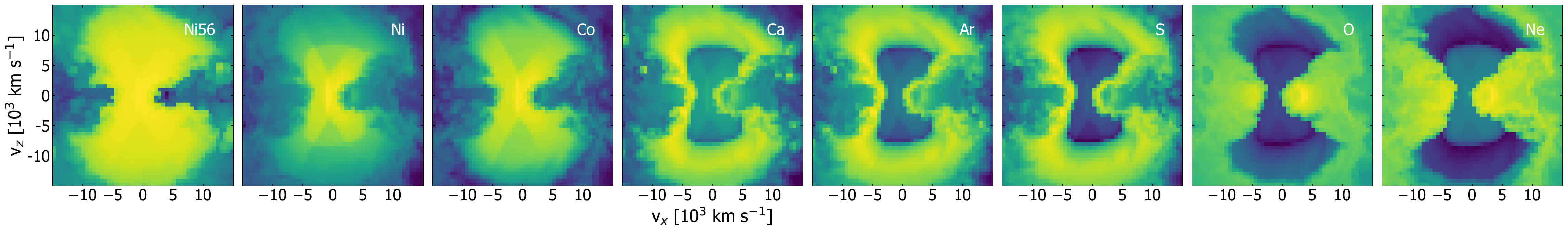}
\end{subfigure}
\begin{subfigure}[b]{1\textwidth}    
   \includegraphics[width=1\linewidth]{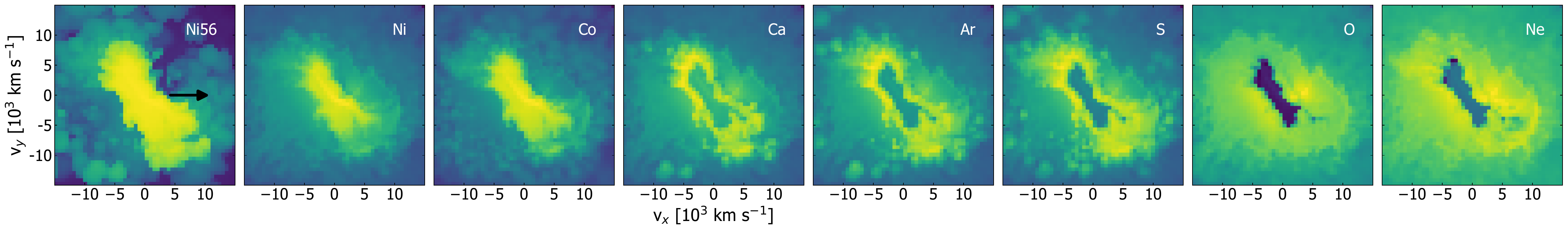}
\end{subfigure}

\caption{Density of key species in three planes: $Y$-$Z$ (top), $X$-$Z$ (middle), and $X$-$Y$ (bottom; $\cos(\theta)=0$, i.e., the merger plane). The arrow in the bottom left panel indicates the direction of $\phi=0\degree$. The colour scale indicates the logarithmic density of each species. Note this snapshot is at 270 days after explosion, apart from\xspace $^{56}$Ni, which is shown at 0.002 days.}
\label{fig:model_abundances}
\end{figure*}

The hydrodynamical explosion model investigated in this work is the 3D violent merger model of \cite{Pakmor2026}. We briefly summarise the model here; a more detailed description is provided by \cite{Pakmor2026}.
This scenario consists of an initial binary WD system with a primary mass of $1.1\,\mathrm{M_\odot}$ and a secondary mass of $0.7\,\mathrm{M_\odot}$. The primary consists of approximately equal amounts of carbon ($0.54\,\mathrm{M_\odot}$) and oxygen ($0.54\,\mathrm{M_\odot}$), mixed with $0.016\,\mathrm{M_\odot}$ of $^{22}\mathrm{Ne}$, and a helium shell of $3\times10^{-4}\,\mathrm{M_\odot}$. The secondary WD is composed of $0.32\,\mathrm{M_\odot}$ of carbon, $0.36\,\mathrm{M_\odot}$ of oxygen, $0.011\,\mathrm{M_\odot}$ of $^{22}\mathrm{Ne}$, and a helium shell of $0.009\,\mathrm{M_\odot}$.

The merger begins with an initial orbital period of approximately 74 seconds and decays due to an explicit angular momentum loss term, mimicking a sped up version of an inspiral driven by the emission of gravitational waves. During the inspiral phase the more extended secondary WD fills its Roche lobe and undergoes phases of helium mass transfer. Over the course of the inspiral, all helium initially present on the surface of the secondary is transferred to the primary. Some of this helium burns in bursts on the surface of the primary WD, and as a result a fraction of the ashes and unburned helium is ejected ($5\times10^{-3}\,\mathrm{M_\odot}$). This material would naturally appear as circumstellar material interaction at very early times. As we focus on the late phase in this work we do not model this interaction. Unlike that of other recent binary WD merger models with helium shells, the total helium shell masses in these calculations is insufficient to sustain a helium detonation that could wrap around the core. Consequently, the primary WD is not disrupted through the double detonation mechanism.

After several orbits the secondary WD is eventually tidally disrupted and consequently accelerated towards the primary. During this disruption, part of the secondary's material collides with the primary. The resulting compression of the carbon–oxygen material produces extremely high densities and temperatures, creating conditions suitable for direct carbon ignition. Unlike helium shell detonations, which can in principle be resolved in similar simulations, carbon ignition occurs on centimetre scales and therefore cannot currently be resolved.

When carbon ignition occurs in the compressed region, a detonation wave forms and propagates through the primary, rapidly burning material and completely unbinding the primary WD. Note that at the time of carbon ignition the secondary is already in the process of being disrupted and therefore has a considerably lower density than it began with. As the detonation propagates through the primary, a strong shock forms and collides with the material from the secondary. However, when this shock crosses the secondary WD, the already expanded and low density material leads to weak or incomplete burning, producing some IMEs in the outer layers of the secondary.

Unlike older violent merger simulations \citep{Pakmor2012b}, not all of the secondary WD is burned. Instead, a small ($0.16\,\mathrm{M_\odot}$) remnant core survives\footnote{We do not model this remnant in our investigation, but note that future theoretical studies should explore the effect of its presence on nebular spectra.}. It remains gravitationally bound and accretes a small amount of ash from the explosion. This results in a final ejecta mass of around $1.65\,\mathrm{M_\odot}$. This explosion model is highly asymmetric as the secondary WD shields the expansion in the orbital plane. As a result, the ejecta expand more easily perpendicular to the orbital plane, producing an elongated structure. This leads to a strong global asymmetry that is expected to produce observer-orientation-dependent synthetic observables at both early and late times. The structure of the inner ejecta and its composition during homologous expansion are shown in Figure~\ref{fig:model_abundances}.

\subsection{Radiative Transfer}
\label{sec:Radiative Transfer}

We perform our radiative transfer calculations using the 3D Monte Carlo radiative transfer code \textsc{artis} \citep{sim_2007,Kromer2009}. The methods used by \textsc{artis} are based on \cite{Lucy2002,lucy2003,Lucy2005}, which divides the radiation field into indivisible Monte Carlo packet quanta. This work utilises the full NLTE approach developed by \cite{Shingles2020}, which includes an NLTE population and ionisation solver and treatment of non-thermal leptons. To follow the population of leptons with non-thermally distributed energies, \textsc{artis} solves the Spencer-Fano equation (as formulated by \citealt{Kozma1992a}). Treatment of all levels in NLTE for every ion has a significant memory and computational cost. As such, we restrict the number of NLTE levels. For most ions, we treat the first 80 levels in NLTE but increase this to 197 NLTE levels for \ion{Fe}{II} to ensure all its associated metastable levels are treated in NLTE. We place other levels into an additional NLTE level that can vary in population, known as a `superlevel' \citep{Anderson1989a}; the absolute population of the superlevel is determined by the NLTE solver, while the relative populations of the stages within the superlevel are set by a Boltzmann distribution at the electron temperature. We also include the heating, ionisation and excitation due to Auger electrons from ionisations of inner shells. 

The atomic data sets we use are based on the compilation of \textsc{cmfgen} \citep{hillier1990a, hillier1998a} and are similar to that described by \cite{Shingles2020} but with some additional species and ions from the most recent compilation of \textsc{cmfgen} (see \citealt{Blondin2023}).  In our simulation we include \ion{C}{I--IV}, \ion{O}{I--IV}, \ion{Ne}{I--IV}, \ion{Mg}{I--IV}, \ion{Si}{I--V}, \ion{S}{I--V}, \ion{Ar}{I--V}, \ion{Ca}{I--V},
\ion{Ti}{II--IV}, \ion{Fe}{I--V}, \ion{Co}{I--V}, and \ion{Ni}{I--V}.
We note that there is no \ion{Ti}{I} data in our atomic data set. We do not expect \ion{Ti}{I} to be a dominant ion in our simulations, nor do we expect that it would significantly impact our conclusions.
In this work we adopt the hybrid photoionisation scheme \citep{Pollin2025} for calculating photoionisation rate estimators, where the detailed treatment is used for bound–free transitions whose lower levels are included in the NLTE solution and the integral over the binned radiation–field model \citep{Shingles2020} for all others.

The models are assumed to be in homologous expansion and each simulation utilises 75 logarithmically spaced time steps from 190 to 410 days post-explosion. We simulate photons produced within the ejecta during this epoch. Given the light-crossing time for the line-forming region of the ejecta, our calculations accurately represent the observable emission between $\sim210$ and $\sim360$ days. These calculations use $3.2 \times 10^{10}$ Monte Carlo packets, and transport is performed on a 3D Cartesian grid that co-expands with the ejecta. The calculations carried out here maintain the same grid resolution as our typical early time calculation \citep[e.g.,][]{Pollin2024a}. In order to accommodate the considerably larger memory requirements associated with storing the full set of 3D NLTE Monte Carlo estimators for the updated atomic data set in our previous 3D calculations, we exclude cells with $\max(v_x, v_y, v_z)  >\sim11{,}000\,\mathrm{km}\,\mathrm{s}^{-1}$. This adjustment preserves the inner ejecta while removing the outermost cells, which have minimal impact on the nebular spectra as the fast, diffuse outer ejecta have low densities and receive little to no energy deposition at these epochs \citep[see][]{Pollin2025}. This yields a maximum radial velocity of $\sim19{,}000\,\mathrm{km}\,\mathrm{s}^{-1}$ in the corners of the ejecta, which sets the range of observer arrival times for which we have predictions after accounting for light travel time. The computational cost of the simulations is discussed in Appendix~\ref{apen:comp_cost}.

To investigate direction-averaged synthetic observables we use all emergent packets to make the direction-averaged spectra (see Section~\ref{sec:angleaveraged}). 
To examine specific lines-of-sight (see Section~\ref{sec:3DViolentMerger_Orientation_Effects} and~\ref{sec:comparison}) we use the virtual packet scheme developed by \cite{Bulla2015}. The virtual packets are enabled for the entire calculation and are active between 0.35--30\,\microns. We have enabled virtual packets for 60 orientations, corresponding to $\cos{\theta} = -0.9, -0.6, -0.3, 0.0, 0.3, 0.6, 0.9$, with 10 equally spaced $\phi$ angles ranging from 0--360\degree. Where $\theta$ is the angle from the positive $z$-axis, and $\phi$ is the degree of rotation of the projection in the $xy$-plane. Hence, rotation of $\phi$ occurs anti-clockwise from $\phi=0\degree$ and is indicated by the directional arrow in Figure~\ref{fig:model_abundances}. Finally, when referring to the \onedvm and \threedvm models collectively we describe them as the \vm scenario.

\section{Results}
\label{sec:Results}

In this section, we investigate the synthetic observables of the explosion models across the optical, NIR and MIR regions.
In Section~\ref{sec:angleaveraged}, we examine the spectra from the spherically averaged ejecta and direction-averaged spectra of the 3D ejecta. In Section~\ref{sec:3DViolentMerger_Orientation_Effects} we explore the orientation effects of both IGEs and IMEs for individual lines of sight, and compare these to our previous investigation of dynamical mergers. Finally in Section~\ref{sec:comparison} we compare specific orientations from the explosion models to optical, NIR and MIR observations of both normal and peculiar SNe~Ia. We have divided the spectra into three wavelength regions to be consistent with \cite{Blondin2023} and \cite{Pollin2025}: the optical (0.35--1\microns), the NIR (1--5\microns), and the MIR (5--30\microns).

\subsection{Direction-averaged spectra} 
\label{sec:angleaveraged}

\begin{figure*}
\centering
\begin{subfigure}{\textwidth}
    \centering
    \includegraphics[width=0.99\textwidth,height=6.5cm]{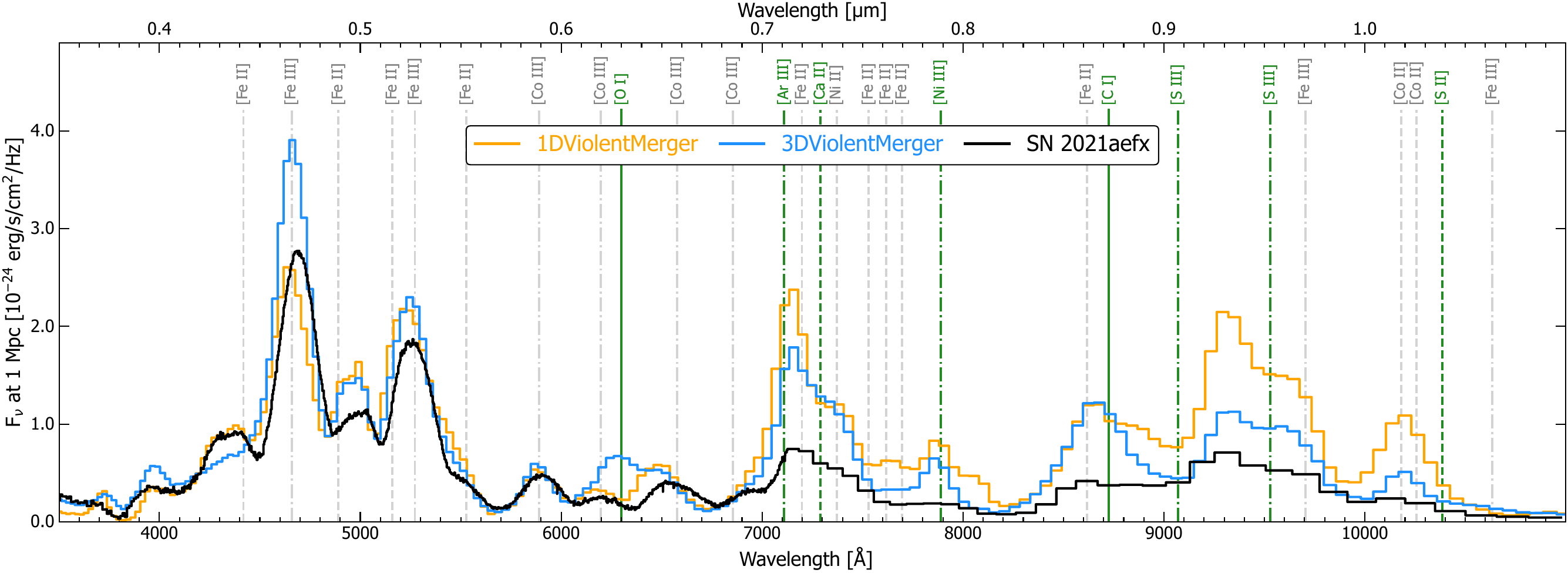}
\end{subfigure}

\vspace{1em}

\begin{subfigure}{\textwidth}
    \centering
    \includegraphics[width=0.99\textwidth,height=6.5cm]{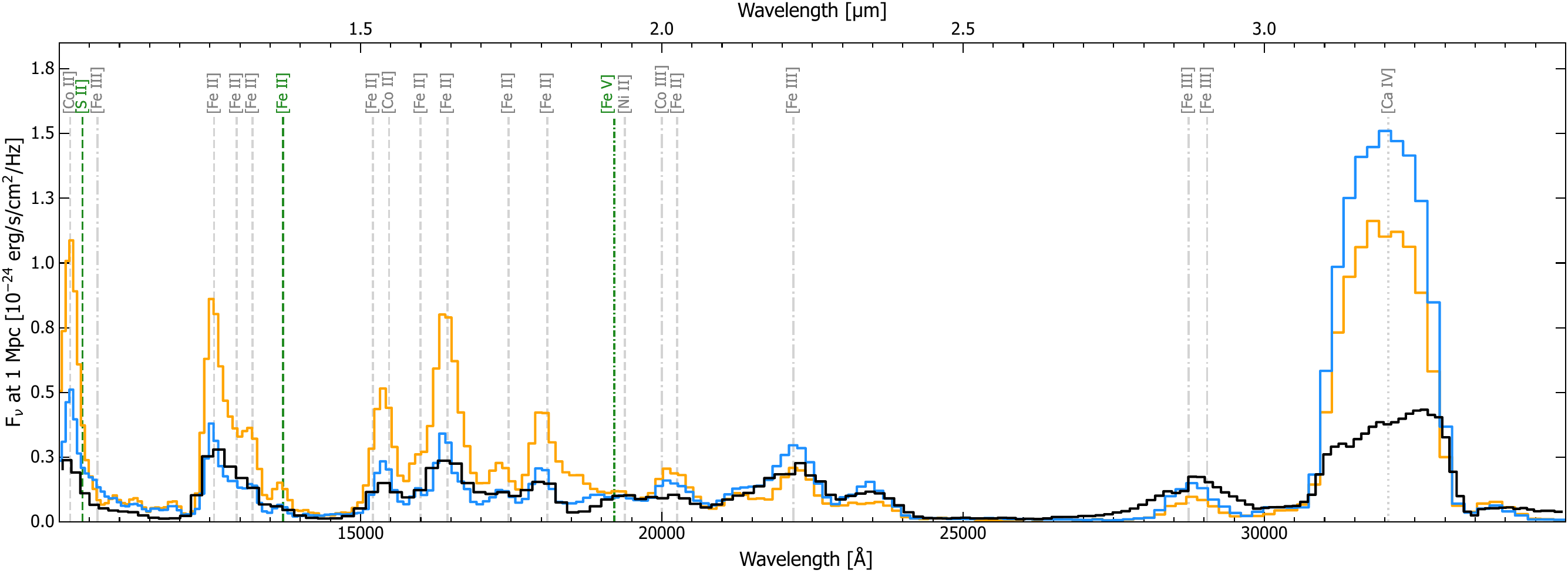}     
\end{subfigure}

\vspace{1em}

\begin{subfigure}{\textwidth}
    \centering
    \includegraphics[width=0.99\textwidth,height=6.5cm]{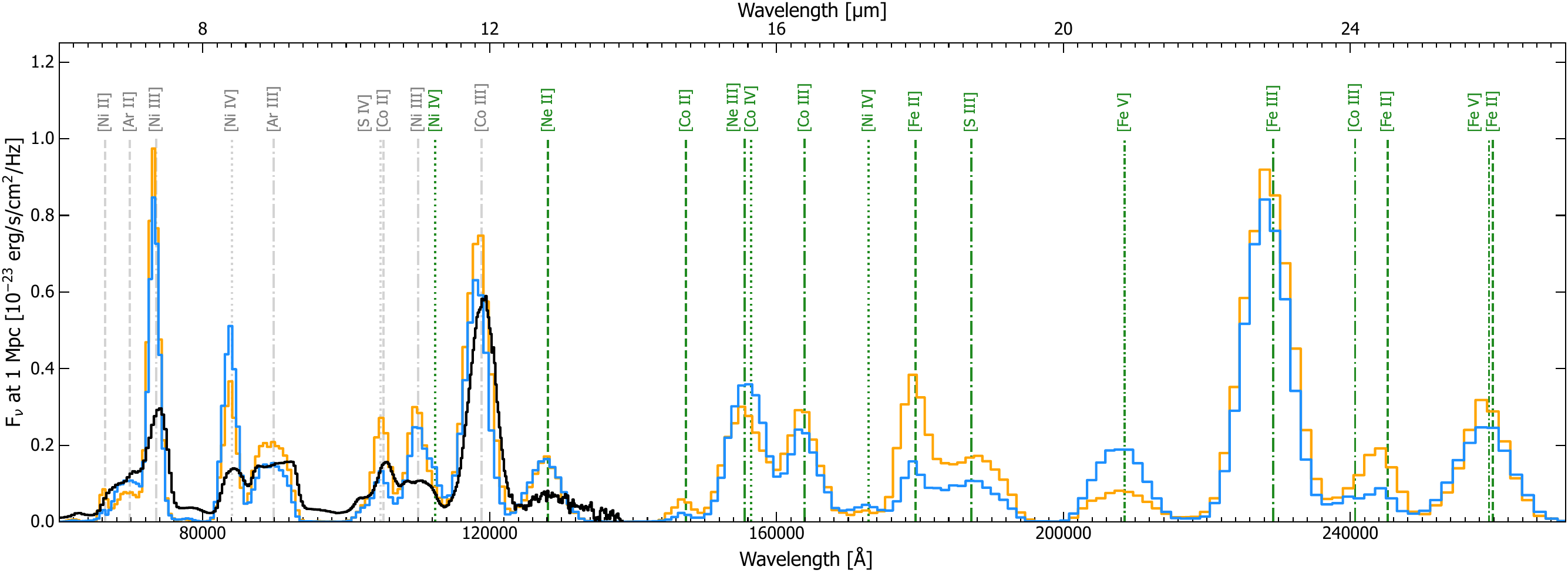}    
\end{subfigure}
                \caption{1D and direction-averaged 3D optical (top; $\sim$0.35--1\microns), NIR (middle; $\sim$1--3.5\microns), and MIR (bottom; $\sim$6--27\microns) spectra for the \threedvm and \onedvm models at 270 days post-explosion, compared to SN~2021aefx \protect\citep{Kwok2023}. The observational data have been corrected for a redshift of $z = 0.005017$ and for reddening due to host-galaxy extinction, $E(B-V)_{\text{host}} = 0.097$ mag \protect\citep{Hosseinzadeh2022}, as well as Milky Way extinction, $E(B-V)_{\text{MW}} = 0.008$ mag, and all spectra are scaled to a distance of 1 Mpc. Vertical grey lines indicate the rest wavelengths of prominent features identified by \protect\cite{Flores2020} and \protect\cite{Kwok2023}. In contrast, green lines highlight significant model features that diverge from observations and those which lie outside the spectral range of SN~2021aefx. The linestyles of the vertical lines indicate ionisation stages: solid for neutral species, dashed for singly ionised, dash--dotted for doubly ionised, dotted for triply ionised species, and a short dash--dot pattern for quadruply ionised features.}
    \label{fig:combined_spectra}
\end{figure*}

\begin{figure*} 
    \centering
    \includegraphics[width=0.99\textwidth,height=6.5cm]{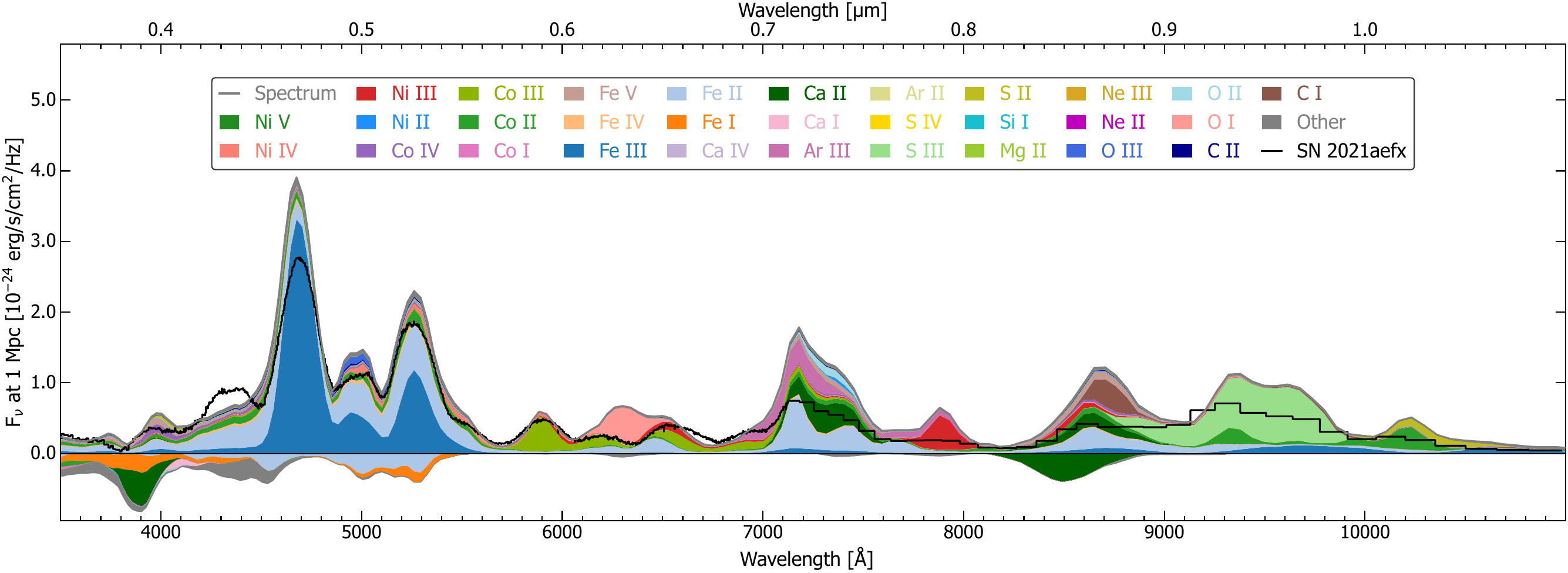}
    
    \includegraphics[width=0.99\textwidth,height=6.5cm]{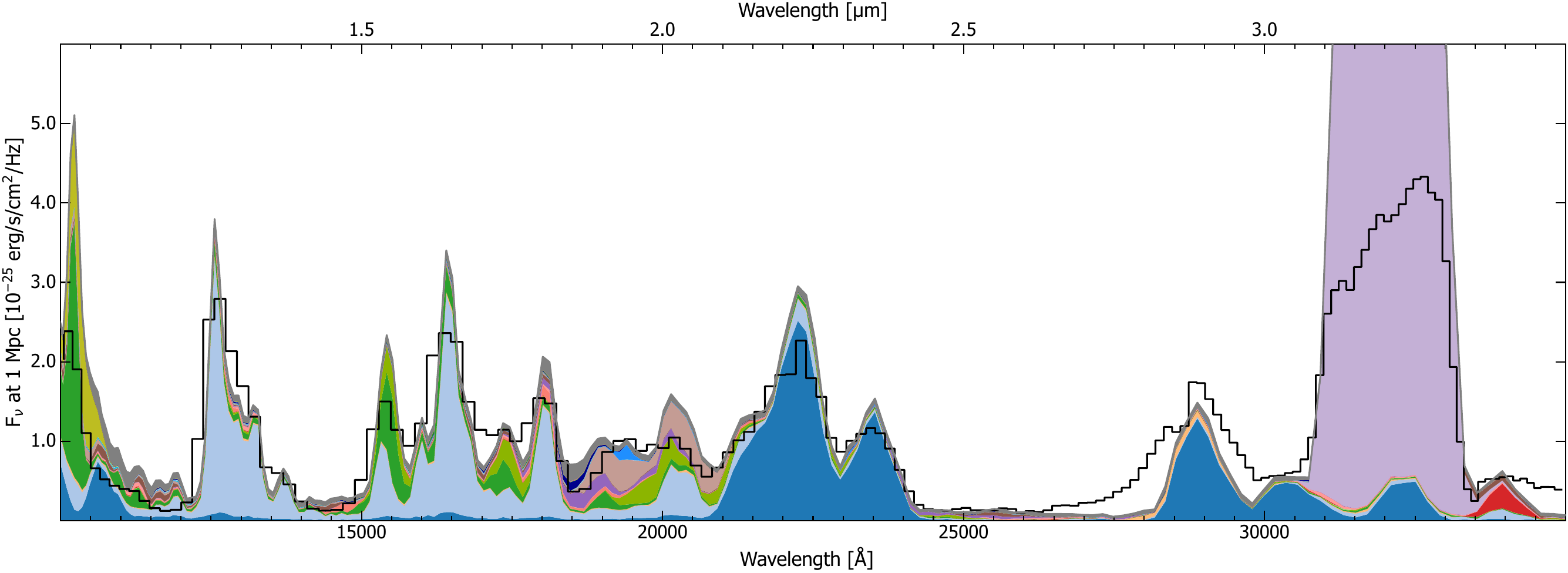}   
    
    \includegraphics[width=0.99\textwidth,height=6.5cm]{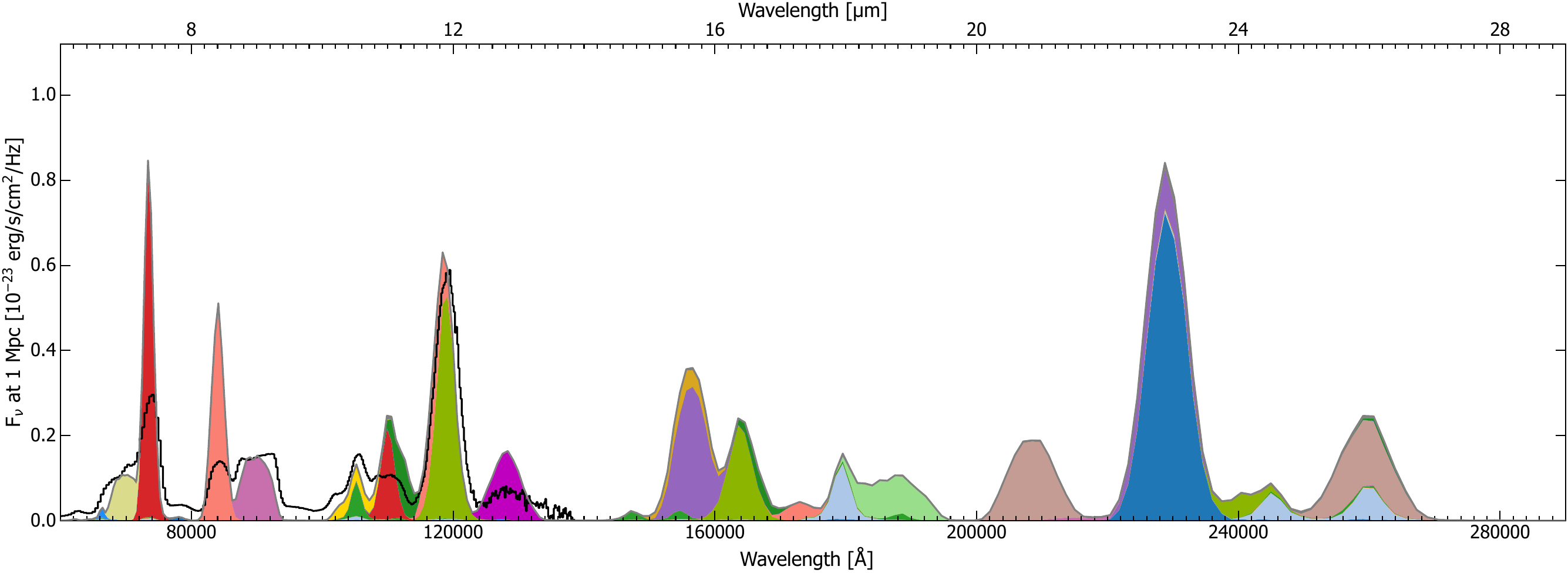}   
    
    \caption{Nebular emission and absorption spectra for the \threedvm model at 270 days across all wavelength ranges, from top to bottom: optical, NIR, and MIR.
    The positive axis is colour-coded to indicate the emitting ions, based on each Monte Carlo packet’s thermal emission type. The negative axis shows the corresponding absorption contributions from each ion, which is only present in the optical region. The total spectrum is overlaid as a thick grey curve, with the shaded regions indicating the contribution of individual ions. Observations of SN~2021aefx \protect\citep{Kwok2023} are included for comparison and have been corrected for redshift and reddening.}
    \label{fig:Kromer_3DViolentMerger}
\end{figure*}

\begin{figure*} 
    \centering
    \includegraphics[width=0.99\textwidth,height=6.5cm]{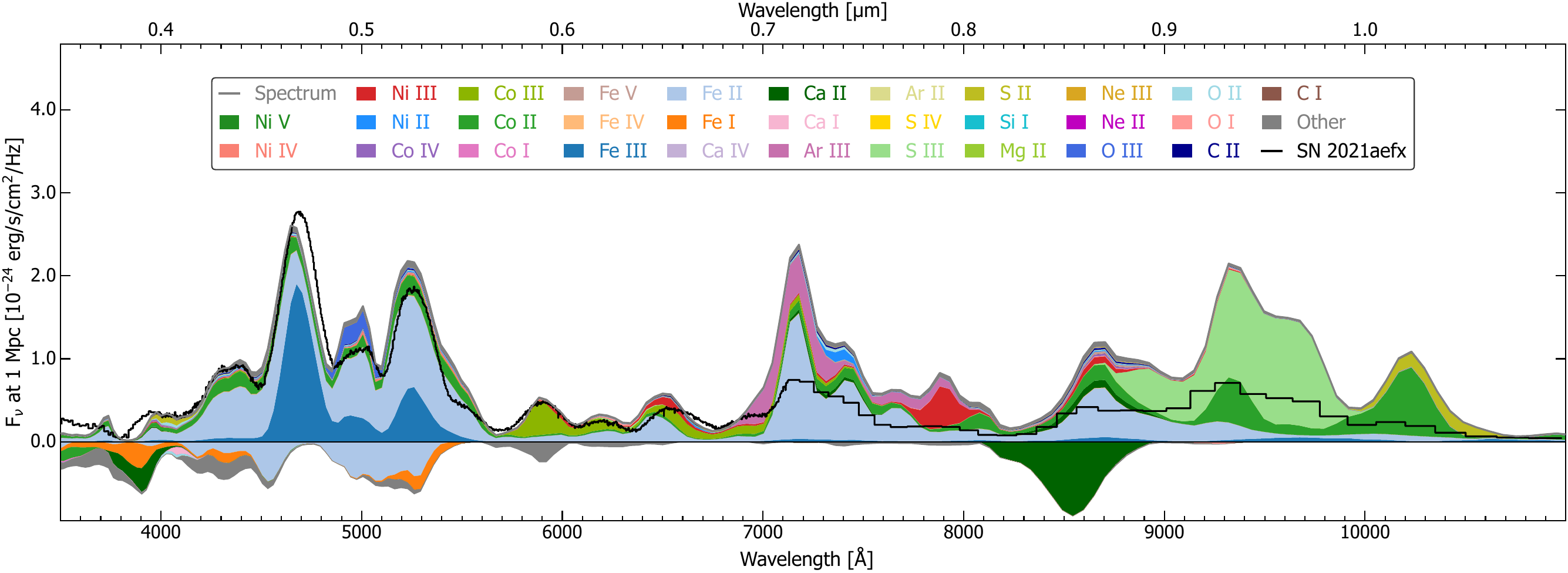}
    
    \includegraphics[width=0.99\textwidth,height=6.5cm]{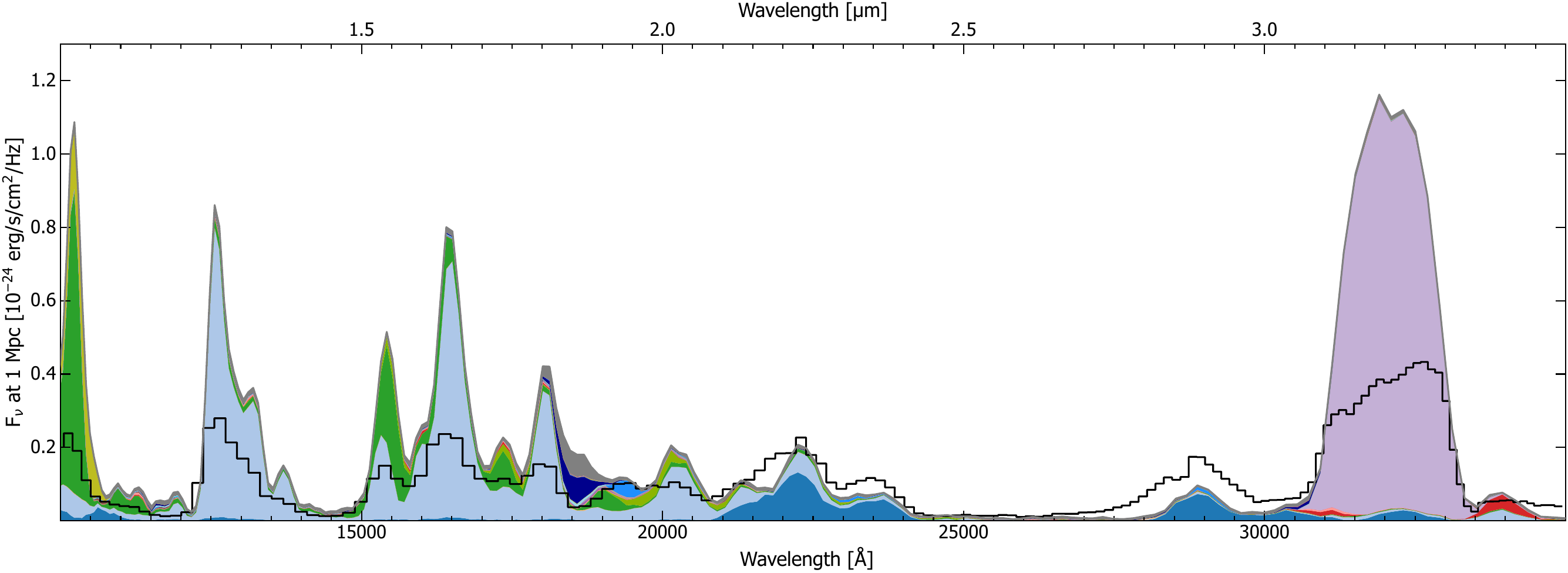}   
    
    \includegraphics[width=0.99\textwidth,height=6.5cm]{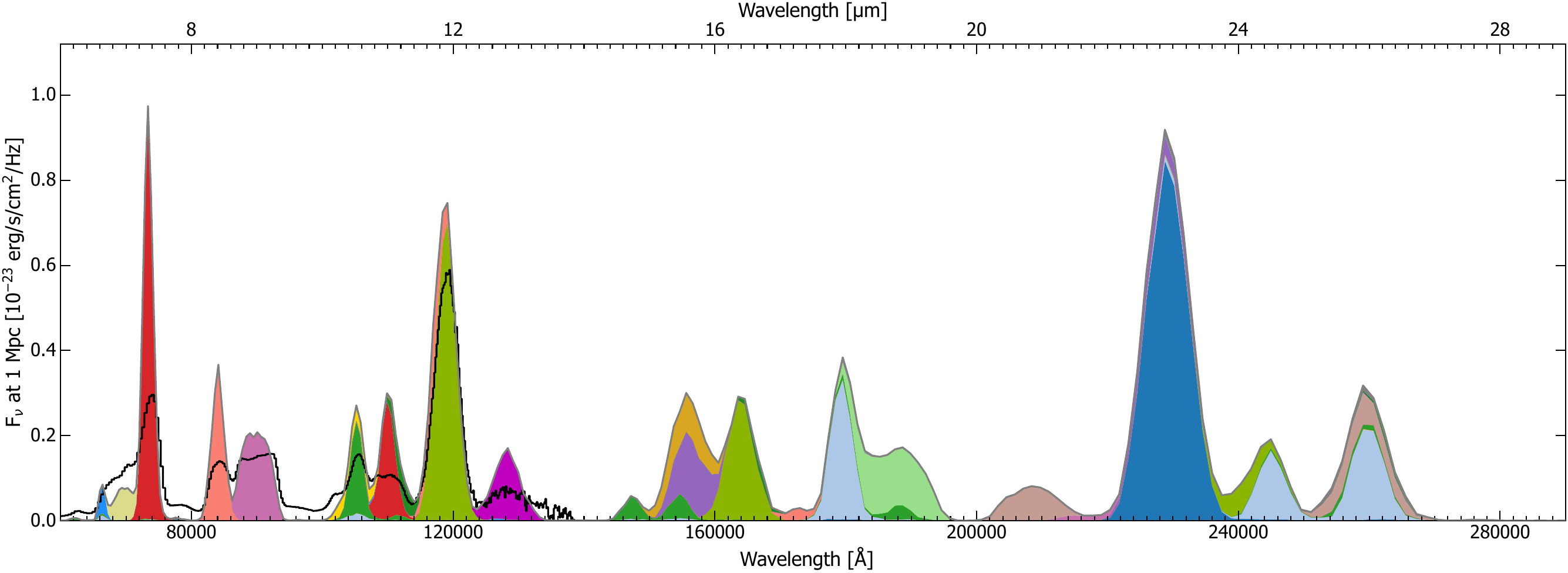}   
    
    \caption{Same as Figure~\ref{fig:Kromer_3DViolentMerger} but for the \onedvm model.}
    \label{fig:Kromer_1DViolentMerger}
\end{figure*}

We first introduce the figures that provide context for our analysis. Figure~\ref{fig:combined_spectra} compares the panchromatic spectra of the \onedvm and \threedvm models at 270 days post explosion with SN~2021aefx (see \citealt{Kwok2023} for observational details). We present spectra in flux per unit frequency, following \cite{Kwok2023,Blondin2023,Pollin2025}, to improve visual clarity. The figure also marks prominent lines identified in nebular SNe~Ia \citep{Flores2020,Kwok2023}, together with model features that differ from the observations. To identify the species responsible for these differences, we also use each Monte Carlo packet's thermal emission type to construct emission and absorption decomposition figures for the \threedvm and \onedvm models, shown in Figures~\ref{fig:Kromer_3DViolentMerger} and~\ref{fig:Kromer_1DViolentMerger}. Notably, even at 270 days post explosion, absorption remains present in the optical for both models. 

\begin{figure*}
\begin{subfigure}{\textwidth}
    \centering
    \includegraphics[width=0.98\textwidth,height=21.7cm]{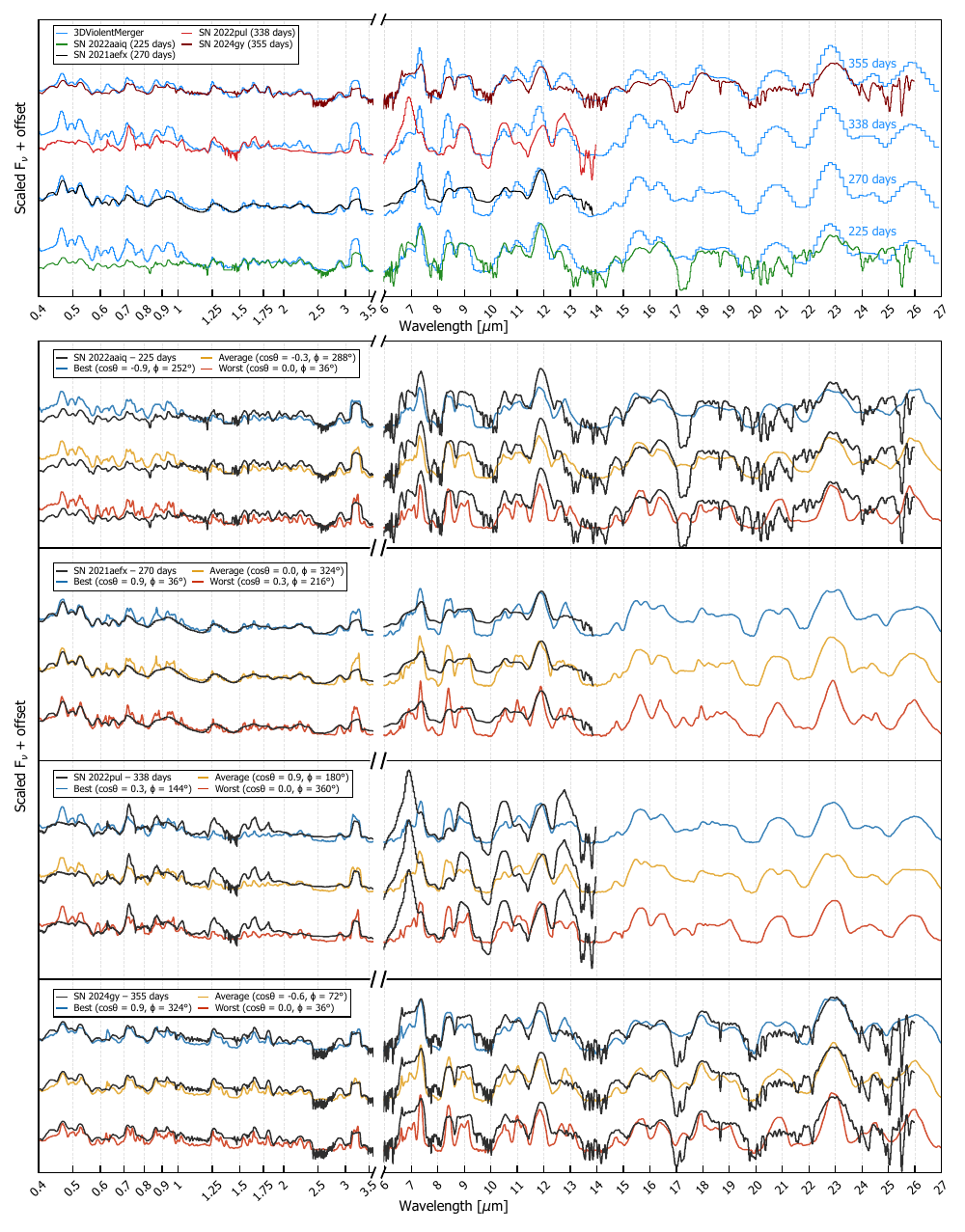}    
\end{subfigure}
\caption{Panchromatic optical, NIR, and MIR spectral time series of the direction-averaged spectrum (top panel) and viewing-angle orientations ranked by median absolute deviation (bottom panel) from the \threedvm model. Observational spectra of the normal SN~2022aaiq, SN~2021aefx, and SN~2024gy, and the 03fg-like SN~2022pul are shown; all have been corrected for redshift and reddening. The flux is normalised and scaled, with each observation–model pair offset for clarity. A modified wavelength scaling is applied to give different spectral regions comparable visual prominence. The wavelength range $\sim3.5$--$6\,\microns$ is omitted as it contains no prominent spectral features.}

    \label{fig:Panchromatic_avg}
\end{figure*}

To contextualise the direction-averaged spectrum within the broader JWST sample, Figure~\ref{fig:Panchromatic_avg} compares the \threedvm model with the normal SNe~2022aaiq, 2021aefx, and 2024gy, and the 03fg-like SN~2022pul at their respective observational epochs \citep{Kwok2023,Kwok2024,Kwok2025}. The spectra were normalised and scaled at each observational epoch. For each comparison, we calculated the median absolute deviation and selected representative viewing angles corresponding to the best, average, and worst rankings. These orientations are intended to illustrate the range of diversity produced by the model, rather than to definitively identify the viewing angles with the best agreement, since alternative metrics or scaling choices may favour different orientations \citep[see][for an in-depth discussion of metric and scaling choices]{Blondin2023}. As the model was not designed to reproduce any individual event, the most informative comparisons are line-profile shapes and the presence or absence of key features, rather than exact agreement in line strengths or overall flux. The direction-averaged synthetic spectra are discussed in Section~\ref{sec:angleaveraged}, while the viewing-angle spectra are examined primarily in Section~\ref{sec:comparison}.

\begin{figure*}
\begin{subfigure}{\textwidth}
    \centering
    \includegraphics[width=0.99\textwidth]{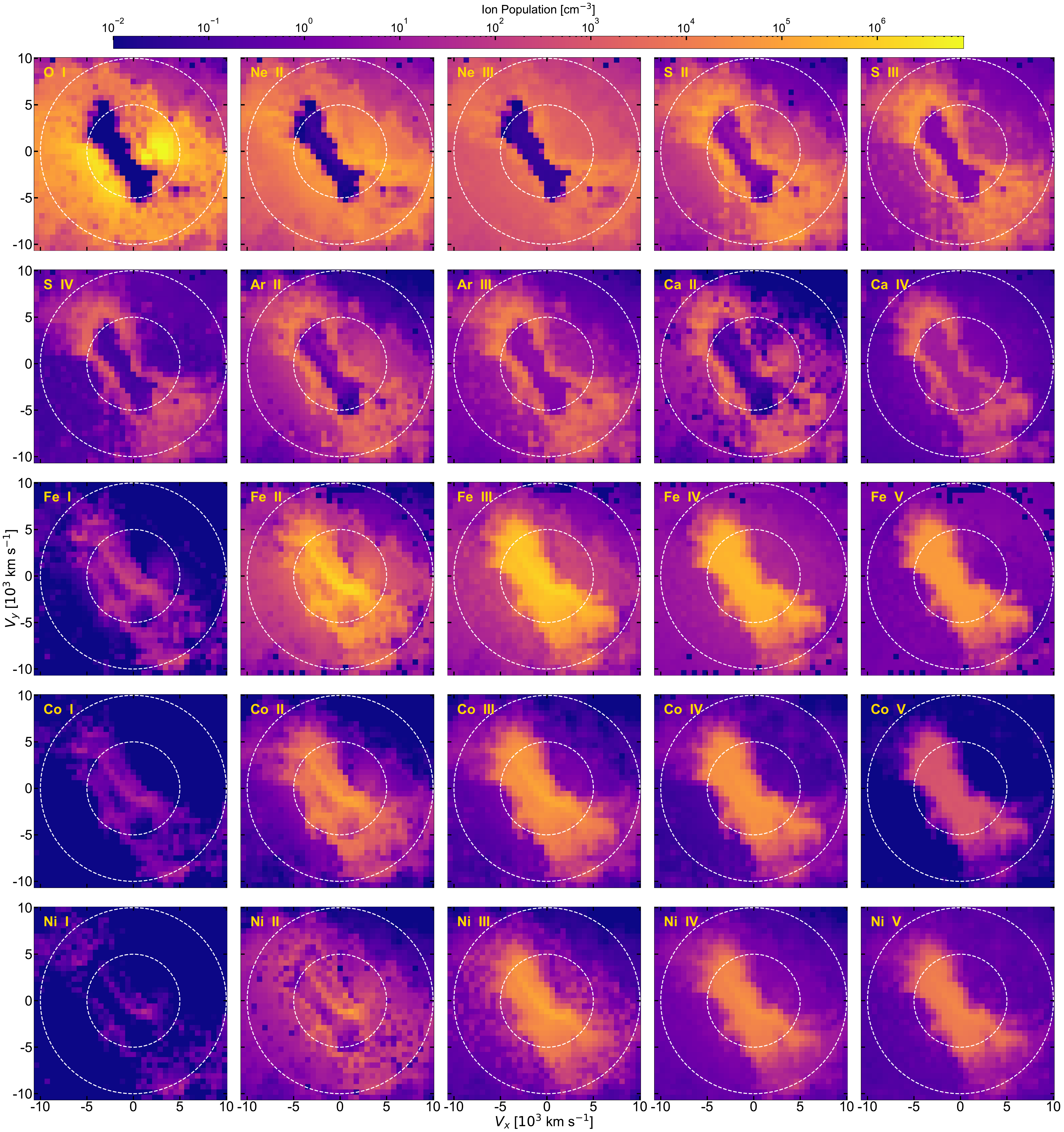}    
\end{subfigure}
        \caption{Ion populations for a slice through the \threedvm model ($\cos(\theta)=0$, i.e. the merger plane) at 270 days post explosion. Each panel shows a 2D slice of the 3D ejecta mapped into velocity space, illustrating the ion populations of key species: \protect\ion{O}{I}, \protect\ion{Ne}{II--III}, \protect\ion{S}{II--IV}, \protect\ion{Ar}{II--III}, \protect\ion{Ca}{II}, \protect\ion{Ca}{IV}, \protect\ion{Fe}{I--V}, \protect\ion{Co}{I--V}, and \protect\ion{Ni}{I--V}. Dashed circles indicate radial velocities of 5,000$\,\mathrm{km}\,\mathrm{s}^{-1}$ and 10,000$\,\mathrm{km}\,\mathrm{s}^{-1}$. The colour bar matches that of Figure~\ref{fig:1d_ion_plot} to allow direct comparison between the 1D and 3D models. A lower limit of $10^{-2}$ has been applied to the ion populations, as values below this threshold have a negligible effect on the spectra. Populations below this value have been clipped to improve clarity and to facilitate clearer comparisons across the panels.
        }
\label{fig:3d_ion_plot}
\end{figure*}

\begin{figure*}
\begin{subfigure}{\textwidth}
    \centering
    \includegraphics[width=0.95\textwidth]{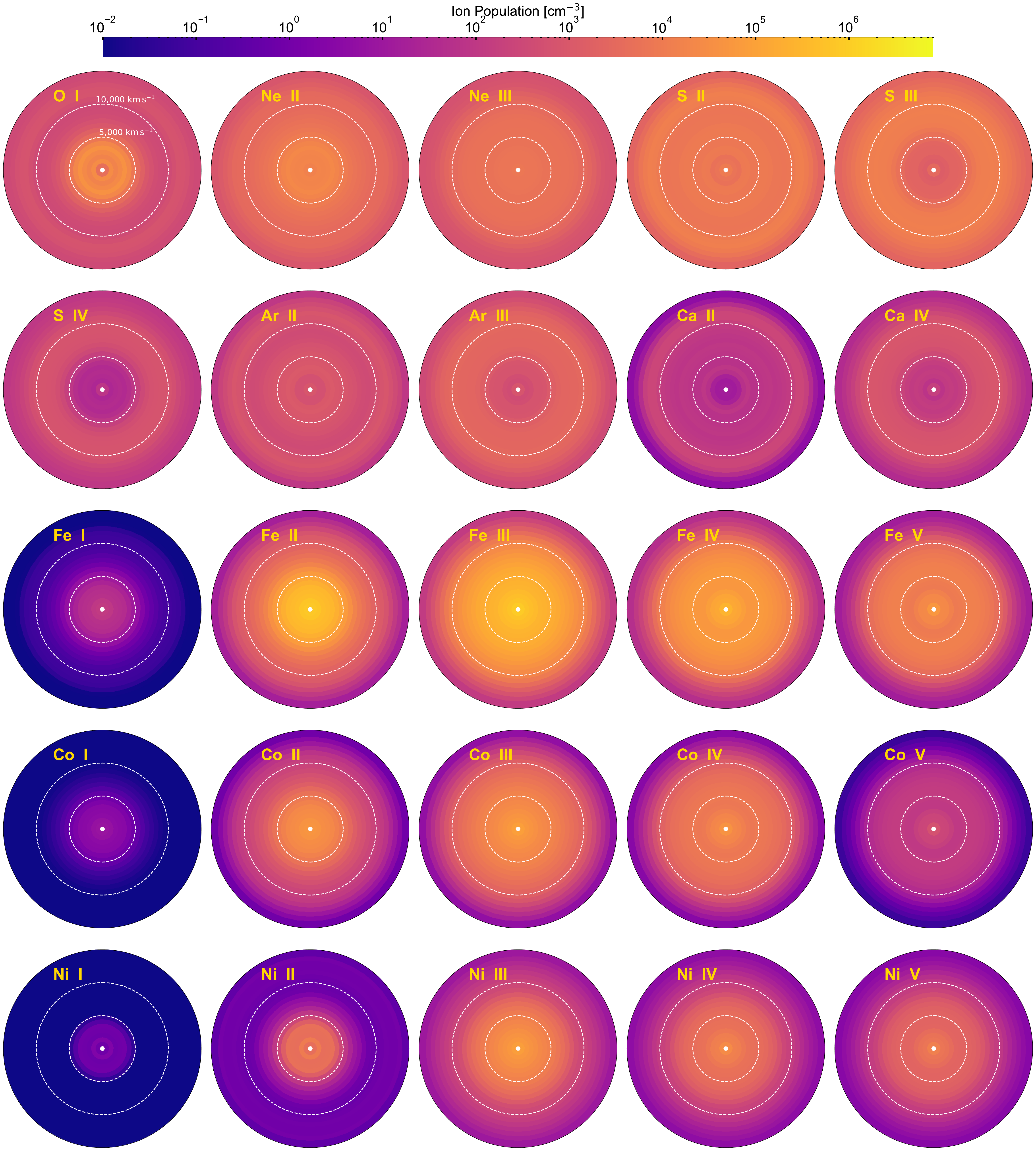}    
\end{subfigure}
    \caption{Ion populations for the \onedvm model at 270 days post explosion. Each panel shows a 2D projection of the 1D ion population, assuming spherical symmetry, as a function of radial velocity for key species: \protect\ion{O}{I}, \protect\ion{Ne}{II--III}, \protect\ion{S}{II--IV}, \protect\ion{Ar}{II--III}, \protect\ion{Ca}{II}, \protect\ion{Ca}{IV}, \protect\ion{Fe}{I--V}, \protect\ion{Co}{I--V}, and \protect\ion{Ni}{I--V}.      
    All panels share a common radial velocity scale, with inner and outer dashed circles marking velocities of 5,000$\,\mathrm{km}\,\mathrm{s}^{-1}$ and 10,000$\,\mathrm{km}\,\mathrm{s}^{-1}$, respectively. The colour bar matches that of Figure~\ref{fig:3d_ion_plot}, allowing direct comparison between the 1D and 3D models.     
    A lower limit of $10^{-2}$ has been applied to the ion populations, as values below this threshold have a negligible effect on the spectra. Populations below this value have been clipped to improve clarity and facilitate comparison across the panels.
}
\label{fig:1d_ion_plot}
\end{figure*}

To investigate the origin of the differences between the 3D and 1D spectra, we examine the ion populations and ejecta structure shown in Figures~\ref{fig:3d_ion_plot} and~\ref{fig:1d_ion_plot}. For the 1D model, the ion populations are projected into 2D assuming spherical symmetry, while for the \threedvm calculation we show a slice through the merger plane. 
We stress that neither a 1D average or line of sight, nor any single 2D slice through the ejecta, can fully capture the diversity of the 3D structure. We therefore provide additional 2D slices along different axes in Appendix~\ref{apen:Additional Viewing Angles}. Figure~\ref{fig:3d_ion_plot} demonstrates that the multidimensional calculation produces substantial departures from spherical symmetry in both the IME and IGE distributions, providing a physical basis for many of the spectral differences discussed below.

\subsubsection{Optical Comparison}
\label{sec:average_spectra_optical}

We begin by identifying the features generically predicted by the explosion models and the spectral regions that differ between the observations and the corresponding 1D and 3D calculations. We stress, however, that the direction-averaged spectrum is a superposition of multiple viewing angles and therefore does not correspond to a single line of sight in the 3D calculation. We therefore avoid describing these features as centrally peaked or shell-like, as such morphologies can arise from emission structures that vary with observer orientation. We assess line-profile morphologies and velocities in Section~\ref{sec:3DViolentMerger_Orientation_Effects}.

Both the 1D and 3D spectra of the \vm scenario show a considerably improved ionisation state compared to the \dsix scenario, with previous work also showing that overionisation seen in 1D can also persist in 3D \citep{Pollin2025}. This improvement is most evident in the main Fe complex between 0.40--0.55\microns, where the \vm models exhibit significantly stronger [\ion{Fe}{II}] emission than the \dsix scenario (see Figure~\ref{fig:combined_spectra}; see also figure~2 of \citealt{Pollin2025}). The \threedvm model also shows much weaker [\ion{Fe}{I}] emission and absorption than the \two model \citep[see figure~11 of][]{Pollin2025}. In both the \threedvm and \onedvm models, several other species make small individual contributions in this wavelength range, but together contribute appreciably to the flux.

Despite the improved ionisation state, both models still predict several features that are not seen in the observations. 
Notably, the 1D and 3D calculations produce strong [\ion{Ni}{III}] 0.789\microns emission, which is absent from nebular SNe~Ia observations (e.g., Figure~\ref{fig:Panchromatic_avg}). They also produce an [\ion{O}{III}] contribution at 0.496\microns, heavily blended with nearby Fe lines. This has not been observed in SN~2021aefx \citep{Kwok2023} or SN~2022pul \citep{Kwok2024}, although it has been reported in previous modelling efforts \citep{Blondin2023,Pollin2025}. In contrast, the [\ion{Co}{III}] feature at 0.589\microns is particularly well reproduced in both scenarios (see Figure~\ref{fig:combined_spectra}).

In the 3D calculation, a strong [\ion{O}{I}] feature emerges at 0.630\microns and 0.636\microns. In Figure~\ref{fig:3d_ion_plot}, the [\ion{O}{I}] emission originates primarily in the north-east and south-west quadrants at velocities below 5,000\kms. This \ion{O}{I}-rich material borders the oxygen-free IGE-rich core. However, in the 1D radially averaged model, this region is mixed with the elongated IGE-dominated region that extends along the north-west and south-east quadrants. This averaging produces an artificial \ion{O}{I} distribution that does not reflect the true 3D structure. Although the central regions of the 1D ejecta shown in Figure~\ref{fig:1d_ion_plot} contain higher \ion{O}{I} populations than the outer ejecta, these remain more than an order of magnitude lower than those in the 3D model. In comparison, \cite{Kwok2024} required an additional central mass and a micro-clumping factor in their 1D calculations, with the latter increasing local recombination rates and shifting the ionisation balance towards lower ionisation states, to reproduce the [\ion{O}{I}] emission; in our 3D calculation, this feature instead arises naturally from the ejecta structure.

A relatively strong [\ion{C}{I}] contribution is also present in the 3D calculation at around 0.873\microns, enhancing the emission in this region compared to the 1D calculation. This has not been identified in SN~2021aefx or SN~2022pul, nor has it been reported in previous theoretical investigations of violent mergers \citep[e.g.,][]{Blondin2023}. In our scenario, this unburned [\ion{C}{I}] emission originates from the tidally disrupted secondary WD. Notably, it is absent in the 1D calculation and only appears in the 3D model, again due to spherical averaging in 1D, as the [\ion{C}{I}] emission is largely co-spatial with the [\ion{O}{I}]. Other [\ion{C}{I}] lines have been identified in SN~2022pul at 0.982\microns and 0.985\microns \citep{Liu2025}. While these features are not prominent at the epoch shown here, they do emerge at later epochs in our model spectra.

There is also a small contribution from [\ion{O}{II}] in the $\sim$0.70--0.75\microns region. Together with the [\ion{O}{I}] and [\ion{O}{III}] discussed above, this suggests that multiple oxygen ionisation stages may be present in the optical spectra of such mergers, although only [\ion{O}{I}] contributes significantly, with the others being difficult to detect due to heavy blending. Oxygen emission in the optical is not entirely unique to violent mergers, as quadruple detonation models also predict oxygen features, although at much lower strengths \citep{Pollin2025}.
[\ion{O}{I}] has been detected in only three 03fg-like events to date \citep{Taubenberger2019,Dimitriadis2023,Kwok2024}, with velocity distributions that vary significantly between events \citep{Kwok2024}. We therefore defer further discussion of this feature to Section~\ref{sec:3DViolentMerger_Orientation_Effects}, where we explore the viewing-angle dependence of this explosion model.

The region around $\sim$0.70--0.75\microns is one of the most studied in nebular SNe~Ia investigations, as observationally identified [\ion{Fe}{II}] and [\ion{Ni}{II}] features are commonly used as diagnostics of the stable Ni synthesised in the explosion \citep[e.g.,][]{Maguire2018,Flores2020}. Many models that aim to reproduce normal SNe~Ia do not predict strong [\ion{Fe}{II}] or [\ion{Ni}{II}] emission in this region, instead showing only minor contributions from these species \citep[e.g.,][]{Shingles2020,Shingles2022a,Blondin2023,Pollin2025}. They often instead produce a strong [\ion{Ar}{III}] feature at 0.714\microns, which is too blue and narrow to account for all of the observed emission. In contrast to our previous calculations \citep{Shingles2020,Shingles2022a,Pollin2025}, we find a significant contribution from [\ion{Fe}{II}], primarily from the 0.715\microns and 0.745\microns transitions, with only a relatively small contribution from [\ion{Ni}{II}]. While [\ion{Fe}{II}] dominates the emission, [\ion{Ca}{II}] is the second strongest contributor, followed by [\ion{Ar}{III}].

Although some double-detonation models predict [\ion{Ca}{II}] emission in this region \citep[e.g.,][]{Mazzali2015,Blondin2018,Polin2021,Blondin2023}, observational studies of normal SNe~Ia generally do not require a strong [\ion{Ca}{II}] contribution to reproduce this feature \citep[e.g.,][]{Maguire2018,Flores2020,Kwok2023}. In contrast, the violent merger candidate SN~2022pul displays a particularly strong emission feature in this region (see Figure~\ref{fig:Panchromatic_avg}), for which \cite{Siebert2024} infer a significant [\ion{Ca}{II}] contribution. Both our 1D and 3D explosion models produce strong [\ion{Ca}{II}] emission at 0.729\microns and 0.732\microns. This feature exhibits considerable viewing-angle dependence and is explored in Section~\ref{sec:3DViolentMerger_Orientation_Effects}. The \threedvm model does, however, still show some degree of overionisation, as suggested by both its inability to reproduce the full strength of the [\ion{Ca}{II}] feature observed in SN~2022pul and the presence of optical [\ion{Ni}{III}] emission at 0.789\microns.

\subsubsection{NIR Comparison}
\label{sec:average_spectra_NIR}

The middle panel of Figure~\ref{fig:combined_spectra} shows that both the \onedvm and \threedvm models reproduce the main observed NIR features, including [\ion{Co}{II}] 1.019\microns, [\ion{Fe}{II}] 1.2\microns and 1.6\microns, [\ion{Fe}{III}] 2.2\microns and 2.9\microns, [\ion{Ca}{IV}] 3.2\microns, and a feature around 1.9\microns commonly associated with [\ion{Ni}{II}]. The main difference between the 1D and 3D calculations is the ionisation state. The 1D calculation overproduces singly ionised species, with [\ion{Fe}{II}] features around twice as luminous as observed in SN~2021aefx, while the doubly ionised species are slightly too faint. The 3D calculation is more consistent with observations, reducing the [\ion{Fe}{II}] flux while producing only a marginal increase in the NIR [\ion{Fe}{III}] features.
The main remaining discrepancy is the strength of the NIR [\ion{Ca}{IV}] emission. The 3D calculation produces an integrated [\ion{Ca}{IV}] flux that is $\sim25$\% larger than in the 1D case, which is already larger than observed. Several nearby transitions will also influence the detailed line-profile morphology for different orientations, including [\ion{Fe}{II}] at 3.043 and 3.230\microns, and [\ion{Ni}{III}] at 3.394\microns, with the [\ion{Fe}{II}] 3.230\microns transition likely to have the greatest impact. As shown in Figure~\ref{fig:3d_ion_plot}, [\ion{Fe}{III}] and [\ion{Ca}{IV}] occupy distinct regions of the ejecta, so their spatial separation will produce complex line-profile morphologies. Apart from the strong [\ion{Ca}{IV}], the 3D model provides an excellent match to the NIR spectra of SN~2021aefx.

Figures~\ref{fig:Kromer_3DViolentMerger} and~\ref{fig:Kromer_1DViolentMerger} show that most ions commonly associated with NIR SNe~Ia \citep[see e.g.,][]{Maguire2018,Flores2020,Kwok2024} are well reproduced, with one notable exception. Around $\sim1.9$\microns, the \threedvm model predicts only a small [\ion{Ni}{II}] contribution and instead shows strong [\ion{Fe}{V}] emission, dominated by transitions at 1.921\microns and 2.040\microns. This [\ion{Fe}{V}] contribution has not been reported in previous nebular violent merger calculations \citep[see figure~6 of][]{Blondin2023}, and is essentially absent in our 1D calculation. The resulting morphology appears incompatible with SN~2021aefx, where the [\ion{Fe}{V}] 2.040\microns feature is clearly absent, but may be present in SN~2022pul\footnote{See Appendix~\ref{apen:pul_em} for emission and absorption decomposition information at the same epoch of SN~2022pul.}.

Overall, the 3D calculation appears too highly ionised to match SN~2022pul, producing [\ion{Fe}{II}] and [\ion{Ni}{II}] features that are too weak, while [\ion{Fe}{III}] and [\ion{Ca}{IV}] are too strong. Reducing the ionisation state, for example through micro-clumping (see e.g., \citealt{Wilk2018,Wilk2020,Blondin2023}; see also \citealt{Shingles2022a}), may therefore improve the NIR agreement. A similar shift in the ionisation state would also suppress the optical [\ion{Fe}{III}] peak and produce a [\ion{Ni}{II}]-[\ion{Fe}{II}]-[\ion{Ca}{II}] region more consistent with observations. However, a lower ionisation state may also produce unobserved features, such as strong [\ion{S}{I}] and [\ion{S}{II}] emission around 1\microns. Figure~\ref{fig:Kromer_3DViolentMerger_pul} also shows a mismatch slightly blueward of the 1.2\microns [\ion{Fe}{II}] feature, which was present in the \two model of \cite{Pollin2025} and may be recovered with a slightly lower ionisation state. 

\subsubsection{Lower MIR Comparison}
\label{sec:average_spectra_MIR}

In this section, we discuss the 6--14\microns region, which corresponds to the upper wavelength limit of the SN~2021aefx and SN~2022pul observations at 270 and 338 days past explosion, respectively, and shown in the third panel of Figure~\ref{fig:combined_spectra}. This region contains several key IME and IGE features. The prominent IGE features are [\ion{Ni}{II}] 6.636\microns, [\ion{Ni}{III}] 7.349\microns and 11.002\microns, [\ion{Ni}{IV}] 8.405\microns and 11.726\microns, [\ion{Ni}{V}] 11.242\microns, [\ion{Co}{II}] 10.523\microns, and [\ion{Co}{III}] 11.888\microns. The prominent IME features are [\ion{Ar}{II}] 6.985\microns, [\ion{Ar}{III}] 8.991\microns, [\ion{Ne}{II}] 12.815\microns, and [\ion{S}{IV}] 10.511\microns.

When compared to SN~2021aefx (see Figure~\ref{fig:combined_spectra}), both the \onedvm and \threedvm models perform well, reproducing all observed features with broadly proportional strengths, particularly for the IME features. The 1D and 3D calculations nevertheless differ notably for both IME and IGE species. In the 3D treatment, the integrated flux of [\ion{Ni}{III}] 7.349\microns decreases by 17\%, while [\ion{Ni}{IV}] 8.405\microns increases by 40\%. The [\ion{Co}{III}] feature, which primarily traces the decay of the initial radioactive $^{56}$Ni rather than the stable Ni synthesised in the explosion, decreases by $\sim25\%$.
Both the \onedvm and \threedvm models are overionised in comparison to SN~2021aefx, generally producing [\ion{Ni}{III}] and [\ion{Ni}{IV}] features that are too strong. The models also predict a [\ion{Ni}{V}] 11.242\microns contribution on the red shoulder of the [\ion{Ni}{III}] 11.002\microns feature, which is not seen in SN~2021aefx. Together with the NIR ionisation state discussed above, these discrepancies indicate that the models are too highly ionised. Improving the agreement with SN~2021aefx would require a stronger [\ion{Ni}{II}] feature around $\sim$1.9\microns, together with weaker [\ion{Ni}{III}], [\ion{Ni}{IV}], and [\ion{Ni}{V}] features, implying that a lower ionisation state is likely required.
The different 1D and 3D behaviour and strengths of [\ion{Co}{III}], [\ion{Ni}{III}], and [\ion{Ni}{IV}] of these IGE features between the 1D and 3D calculations can be understood from Figures~\ref{fig:3d_ion_plot} and~\ref{fig:1d_ion_plot}. Although the stable and radioactive IGEs are largely co-spatial, and are predominantly distributed along the south-east and north-west quadrants, some regions show stable Ni ion populations that exceed those of the radioactive material. These regions occur between 5,000\kms and 10,000\kms in the north-east and south-west quadrants of the merger plane.

For the IME features, the integrated luminosity of [\ion{Ar}{III}] 8.991\microns is reduced by 30\% in the 3D calculation, bringing it closer to the level observed in SN~2021aefx. The [\ion{Ar}{III}] emission originates from a shell-like region on the outskirts of the IGE-rich core. As shown in Figure~\ref{fig:3d_ion_plot}, its spatial distribution differs from that in other explosion models, such as double- or quadruple-detonation scenarios \citep[see figures~4 and~6 of][]{Pollin2025}. The \ion{Ar}{III}-rich material spans velocities of approximately 5,000--10,000\kms around the IGE-rich core, so radial averaging blends regions of strong and weak emission, producing the artificial \ion{Ar}{III} distribution seen in the 1D model in Figure~\ref{fig:1d_ion_plot}. Although the merger plane illustrates the fundamentally aspherical nature of the ejecta, variations also exist along other axes of the explosion\footnote{See Appendix~\ref{apen:Additional Viewing Angles} for additional ion population axis.}. Caution is therefore required when relating features in this slice to the direction-averaged spectra, as this mapping is not exact and more closely reflects observer orientations in the merger plane. The radially averaged 1D ejecta also exhibit a different density structure due to the off-centred explosion, which affects the distribution of radioactive material and, consequently, the temperature and ion populations.

Compared to SN~2022pul (see Figure~\ref{fig:Panchromatic_avg}; see also Appendix~\ref{apen:pul_em}), the model shows substantial differences in both ionisation state and feature strengths. The strongest discrepancy is the [\ion{Ar}{II}] emission, which dominates the MIR spectrum of SN~2022pul but is not reproduced at comparable strength by our model. The observed [\ion{Ar}{II}] morphology also suggests a large central region of IMEs, more similar to that expected in quadruple detonations.
The prominent [\ion{Ne}{II}] 12.815\microns feature, which is a key piece of evidence supporting the violent merger interpretation of SN~2022pul, is also significantly weaker in our model than observed. In the calculations of \cite{Blondin2023}, where the secondary WD is fully disrupted during the merger, this feature is strong and consistent with the observations. In contrast, it is not reproduced at comparable strength in our model or in our previous work \citep{Pollin2025}. This difference reflects the reduced Ne mass in the ejecta rather than the ionisation state, as [\ion{Ne}{II}] is already the dominant ionisation stage of neon in our simulation. In our model, the neon mass is approximately one third lower, explaining the diminished strength of this feature. 
This suggests SN~2022pul likely arose from a scenario in which the secondary WD was fully unbound, or from a system with a more massive secondary WD that undergoes more extensive burning and increases the neon mass in the ejecta.
Our model also predicts [\ion{Ni}{III}], [\ion{Ni}{IV}], and [\ion{Ni}{V}] features that are several times stronger than observed, while the [\ion{Ni}{II}] feature is only marginally weaker than in SN~2022pul. Reconciling these differences would require a substantial shift towards lower ionisation states. However, strong micro-clumping alone is unlikely to resolve this discrepancy, as \cite{Blondin2023} found that such conditions lead to a strong [\ion{Ni}{I}] feature in the NIR. The [\ion{S}{IV}] 10.511\microns and [\ion{Co}{II}] 10.523\microns features are also significantly weaker than observed, further challenging this particular violent merger model as an explanation for SN~2022pul.

\subsubsection{Upper MIR Comparison}
\label{sec:average_spectra_UMIR}

In this section, we discuss the 14--30\microns region for several normal SNe~Ia, including SN~2021aefx, SN~2022aaiq, and SN~2024gy \citep{Kwok2023,Kwok2025}. We present the model predictions in Figure~\ref{fig:combined_spectra} and focus on comparisons with SN~2022aaiq and SN~2024gy in Figure~\ref{fig:Panchromatic_avg}. This region is dominated by IGEs, with prominent features from [\ion{Ni}{IV}] 17.283\microns, [\ion{Ni}{V}] 16.660\microns, [\ion{Co}{II}] 14.739\microns, [\ion{Co}{III}] 16.391\microns and 24.068\microns, [\ion{Co}{IV}] 15.647\microns and 22.710\microns, [\ion{Fe}{II}] 17.936\microns, 24.519\microns, and 25.988\microns, [\ion{Fe}{III}] 22.925\microns, and [\ion{Fe}{V}] 20.851\microns and 25.934\microns. The only prominent IME features are [\ion{Ne}{III}] 15.555\microns and [\ion{S}{III}] 18.713\microns.

Comparing the models to SN~2022aaiq and SN~2024gy, we find that most observed features are reproduced at comparable strengths. The main exceptions are [\ion{Ni}{IV}] 17.283\microns, which appears relatively broad and flat, and [\ion{Fe}{V}] 20.851\microns, which is significantly stronger than observed. Both the 1D and 3D treatments also predict strong [\ion{Fe}{V}] and [\ion{Fe}{II}] emission at 25.934\microns and 25.988\microns. These features may be present in the observed spectra, but they lie at the edge of the wavelength range and suffer from low signal-to-noise, making them difficult to confirm. The integrated fluxes of several IME and IGE features also differ between the 1D and 3D calculations, including [\ion{S}{III}] 18.713\microns, [\ion{Fe}{II}] 17.936\microns, and [\ion{Fe}{V}] 20.851\microns.

Overall, the upper MIR region is strongly affected by blending of IGEs and IMEs, complicating observational line identifications and feature fitting, particularly when combined with viewing-angle variations. Among the predicted features, [\ion{Fe}{III}] 22.925\microns is the strongest and least affected by blending. The main discrepancy is the strong [\ion{Fe}{V}] emission, which, together with the broader ionisation issues discussed above, suggests that the model remains too highly ionised. A lower ionisation state would likely shift some of this emission from [\ion{Fe}{V}] to lower ionisation stages, including [\ion{Fe}{IV}], potentially strengthening the [\ion{Fe}{IV}] feature around 2.9\microns. Compared to previous modelling efforts by \cite{Blondin2023}, our [\ion{Ni}{IV}] feature at 17\microns is stronger than in their violent merger calculation, and our [\ion{Fe}{V}] emission is not present at comparable strength in any of their explosion models. Despite these differences, there is generally good agreement with many of the predicted upper-MIR features presented by \cite{Blondin2022}.

\subsection{Violent Merger Orientation Effects} 
\label{sec:3DViolentMerger_Orientation_Effects}

We now examine the orientation dependence of the \threedvm model, focusing on comparisons with SN~2022pul and SN~2021aefx. Figures~\ref{fig:Equator_viewing_angle} and~\ref{fig:Pole_viewing_angle} show panchromatic spectra for representative orientations in the merger plane and around the azimuthal angle. We note that the viewing-angle spectra are shown at higher spectral resolution than the direction-averaged spectra. These spectra are shown at an intermediate epoch between the observations to allow both events to be compared without repetition of figures. As such direct luminosity comparisons should be treated with caution\footnote{Appendix~\ref{apen:pul_em} provides an direction-averaged comparison with SN~2022pul at its corresponding epoch.}. We emphasise that the viewing-angle variation is considerable. The orientations shown here are selected to capture the majority of this variation while maintaining clarity for the reader. The full set of 60 viewing-angle spectra are included with the released simulation outputs (see Section~\ref{sec:data_availability}).

Figures~\ref{fig:velocity_stacked_vm_aefx} and~\ref{fig:velocity_stacked_vm_pul} show the ten merger-plane orientations for key diagnostic species. These figures compare the model to SN~2021aefx and SN~2022pul at 270 and 338 days post-explosion, respectively. To quantify the variation of different IGE features, we fit Gaussian profiles and extract the velocity shifts and FWHM values, shown in Figure~\ref{fig:velocity_fits}.  Gaussian profiles are adopted as IGEs are expected to be centrally concentrated, although they remain an approximation to the true ejecta distribution and may therefore be inadequate in some cases.

The fitting procedure is as follows. Each spectrum is first normalised to its peak value, and a region of interest is defined around the target line. Continuum points on the red and blue sides of the feature are then selected interactively, as velocity offsets vary with observer orientation and make automated placement unreliable. For multiplets, the line separations are fixed to their known wavelength differences. Loose priors are imposed on the Gaussian width, requiring it to be larger than one grid cell ($\sim$300$\,\mathrm{km}\,\mathrm{s}^{-1}$) and smaller than the full ejecta width.
Compared to \cite{Pollin2025}, we use an improved fitting routine. Following \cite{Maguire2018}, each Gaussian component in a multiplet is weighted by the oscillator strength of the corresponding transition. Uncertainties are estimated using a Monte Carlo approach, motivated by the fact that continuum placement is the dominant source of uncertainty when fitting nebular spectra \citep{Maguire2018}. The interactively selected continuum points are randomly perturbed by up to $\sim$0.5\%, repeated 1000 times, and the velocity and FWHM are computed for each fit. We report the mean values and corresponding standard deviations across all 1000 fits. These uncertainties capture the effect of continuum placement, but not additional uncertainty from line blending and superposition in the complex ejecta. They should therefore be regarded as lower limits.

\subsubsection{Iron Group Element Orientation Effects} 
\label{sec:IGE_viewing_angl}

\begin{figure*} 
    \centering
    \includegraphics[width=0.99\textwidth,height=6.5cm]{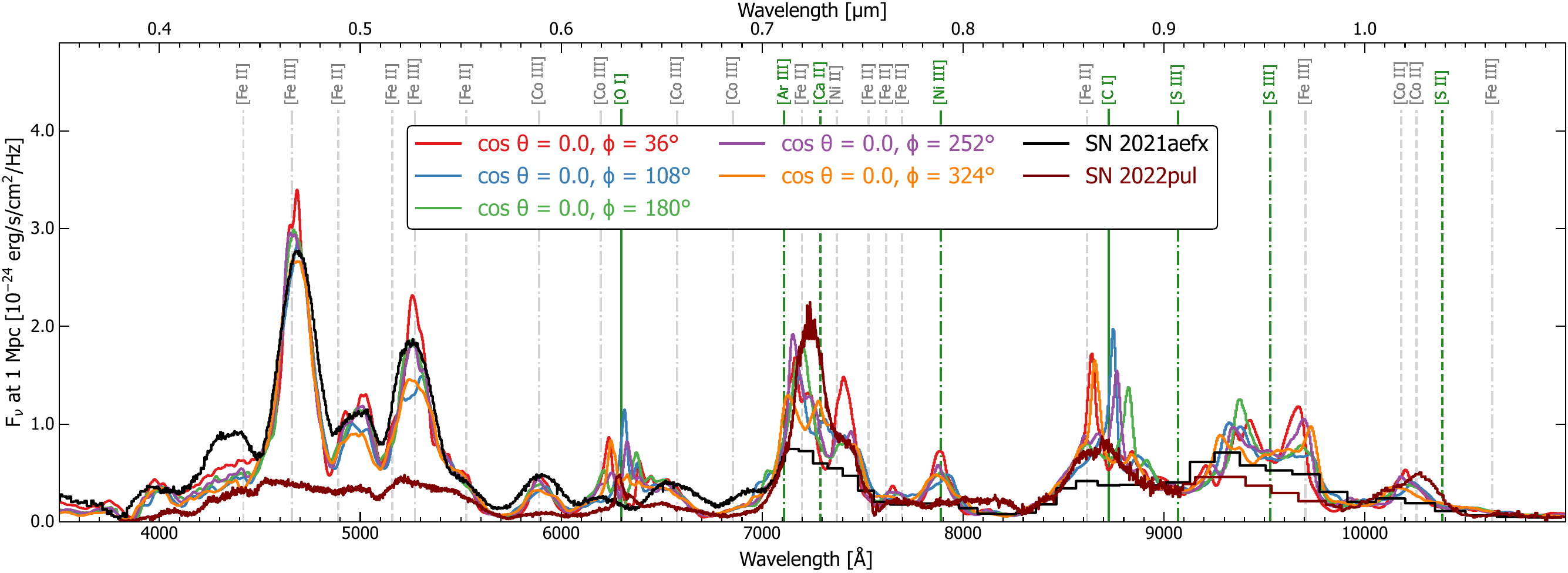}
    
    \includegraphics[width=0.99\textwidth,height=6.5cm]{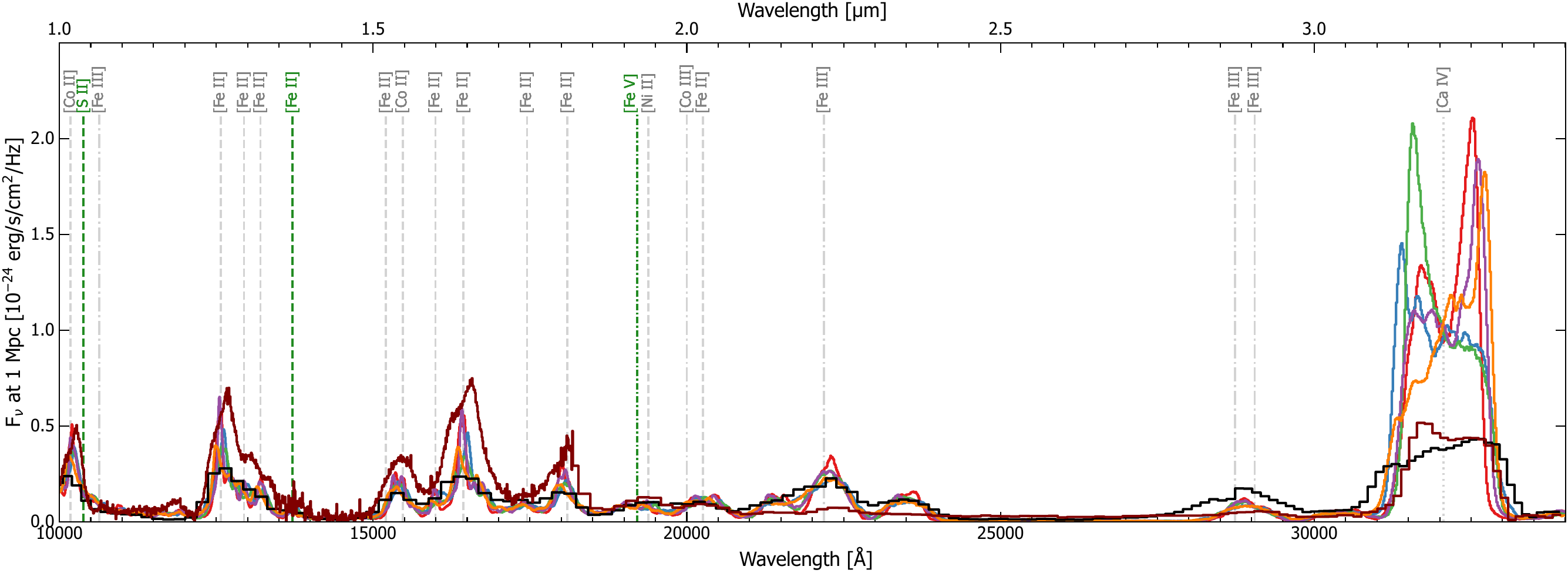}   
    
    \includegraphics[width=0.99\textwidth,height=6.5cm]{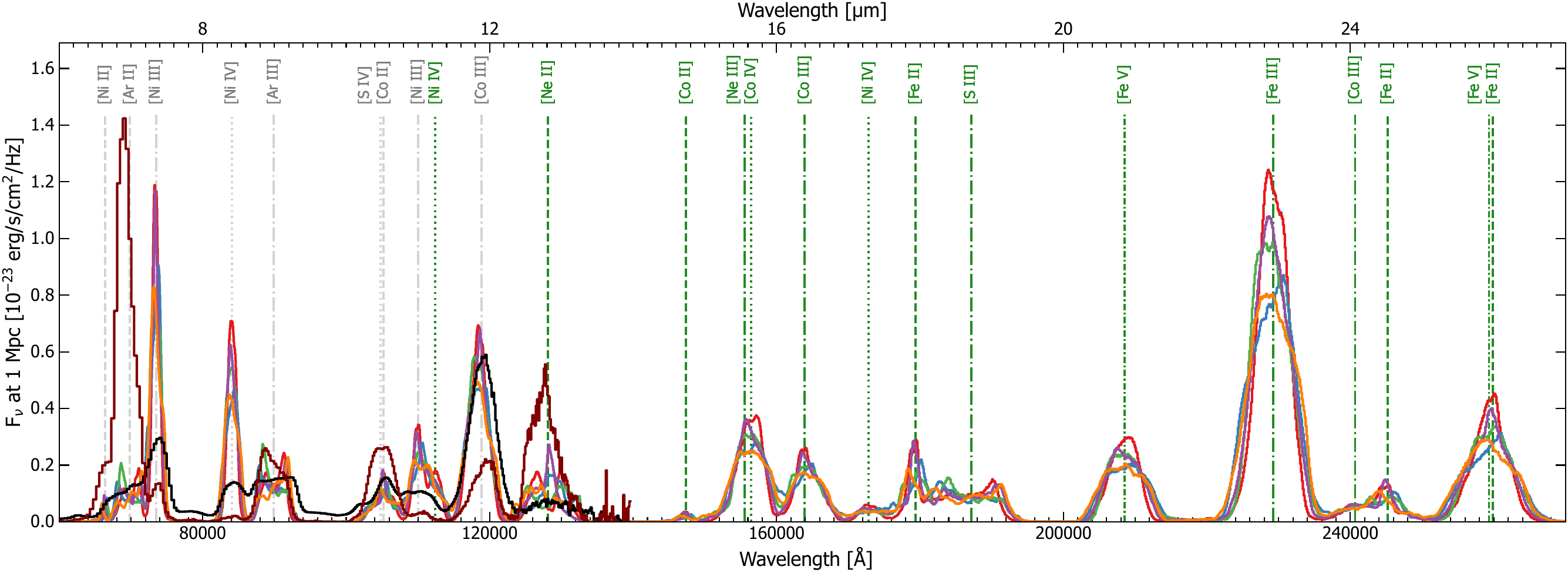}   
    
    \caption{Spectra of the \threedvm model for different viewing angles at $\sim$300 days post-explosion, shown for the optical (top), NIR (middle), and MIR (bottom). The lines of sight are oriented around the merger plane (i.e., $\cos(\theta)=0.0$), where significant variation in the synthetic observables occurs. Observed spectra of SN~2021aefx \citep[270 days;][]{Kwok2023} and SN~2022pul \citep[338 days;][]{Kwok2024} are overplotted and corrected for redshift and extinction \protect\citep{Hosseinzadeh2022,Siebert2024}, with all spectra scaled to a distance of 1 Mpc.  
    Vertical grey lines indicate the rest wavelengths of prominent features identified by \protect\cite{Flores2020} and \protect\cite{Kwok2023}, while green lines highlight significant model features that diverge from observations or lie outside the spectral range of SN~2021aefx. The line styles of the vertical markers denote ionisation stage: solid for neutral species, dashed for singly ionised, dash-dotted for doubly ionised, dotted for triply ionised, and a short dash-dot pattern for quadruply ionised species.  
    The observed and theoretical spectra are not shown at the exact same epochs; instead, an intermediate epoch is chosen to facilitate comparison between the two observations. A Savitzky--Golay filter has been applied to the synthetic spectra to reduce Monte Carlo noise. 
}   
    \label{fig:Equator_viewing_angle}
\end{figure*}

\begin{figure*} 
    \centering
    \includegraphics[width=0.99\textwidth,height=6.5cm]{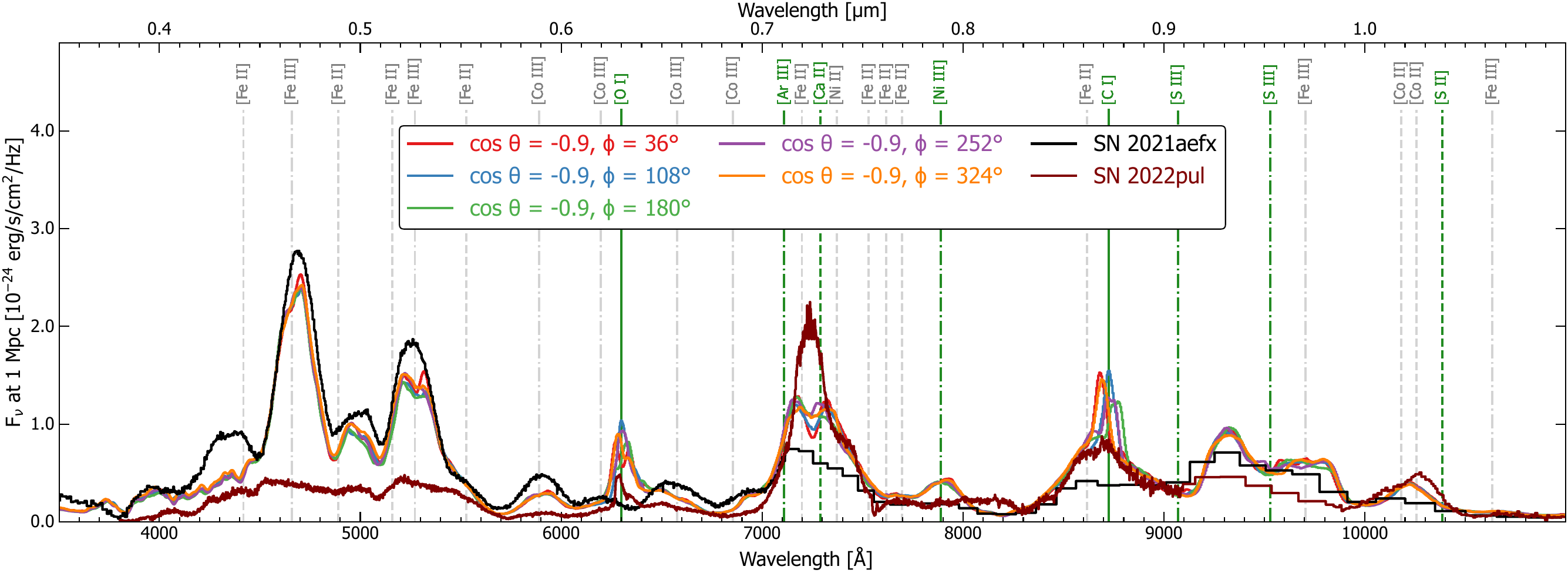}
    
    \includegraphics[width=0.99\textwidth,height=6.5cm]{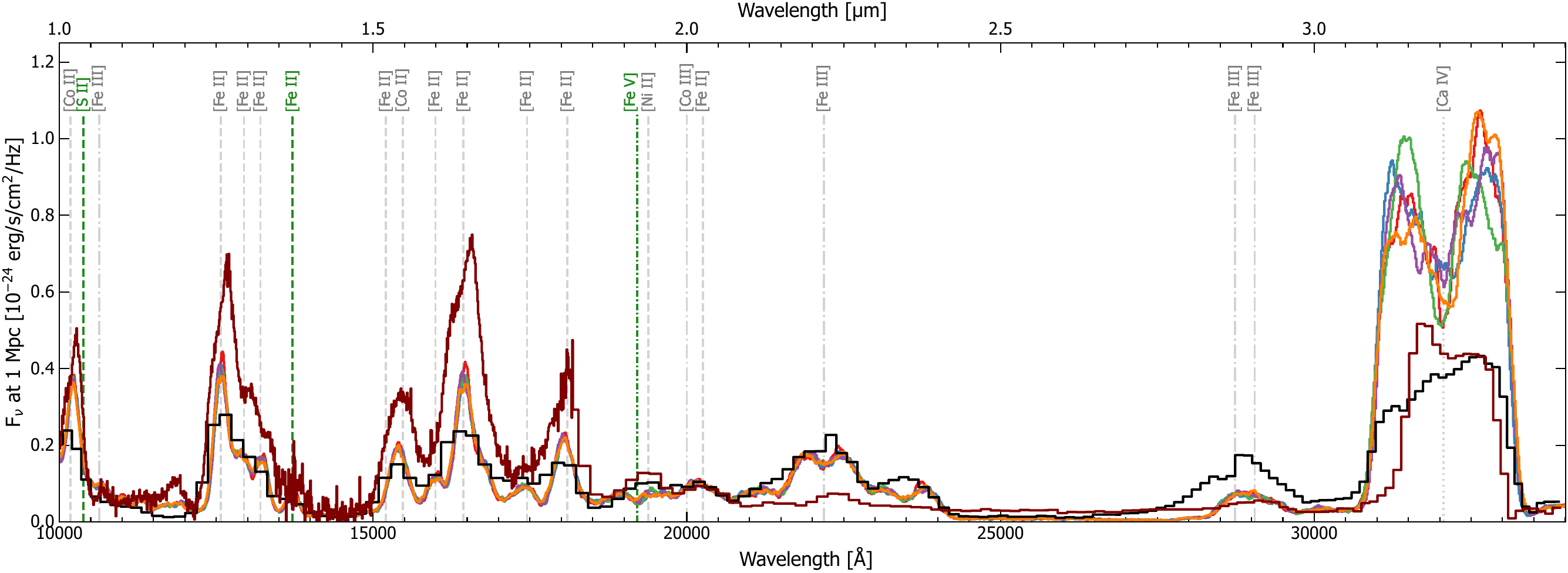}   
    
    \includegraphics[width=0.99\textwidth,height=6.5cm]{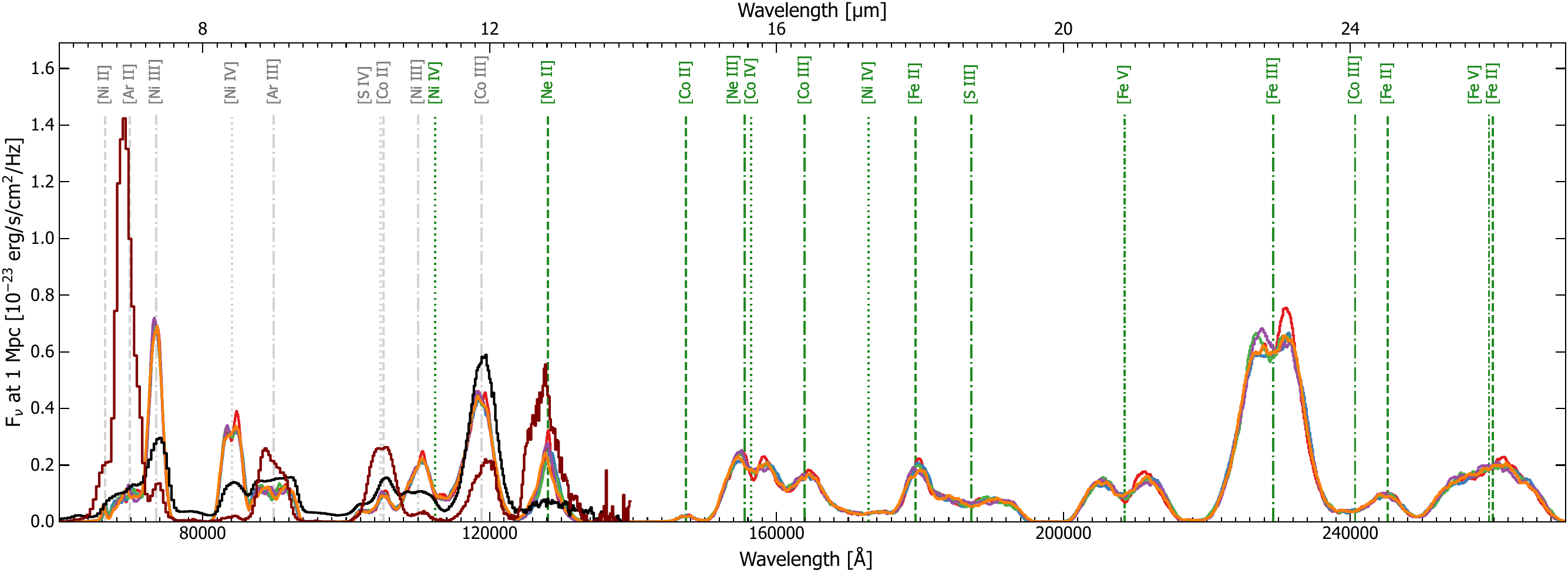}   
    
    \caption{    
    Same as Figure~\ref{fig:Equator_viewing_angle}, but for rotation about the azimuthal angle at $\phi=252\degree$.}    
    \label{fig:Pole_viewing_angle}
\end{figure*}

\begin{figure*} 
    \centering
    \includegraphics[height=8.5cm,width=3.5cm,trim=4 0 2 0, clip]{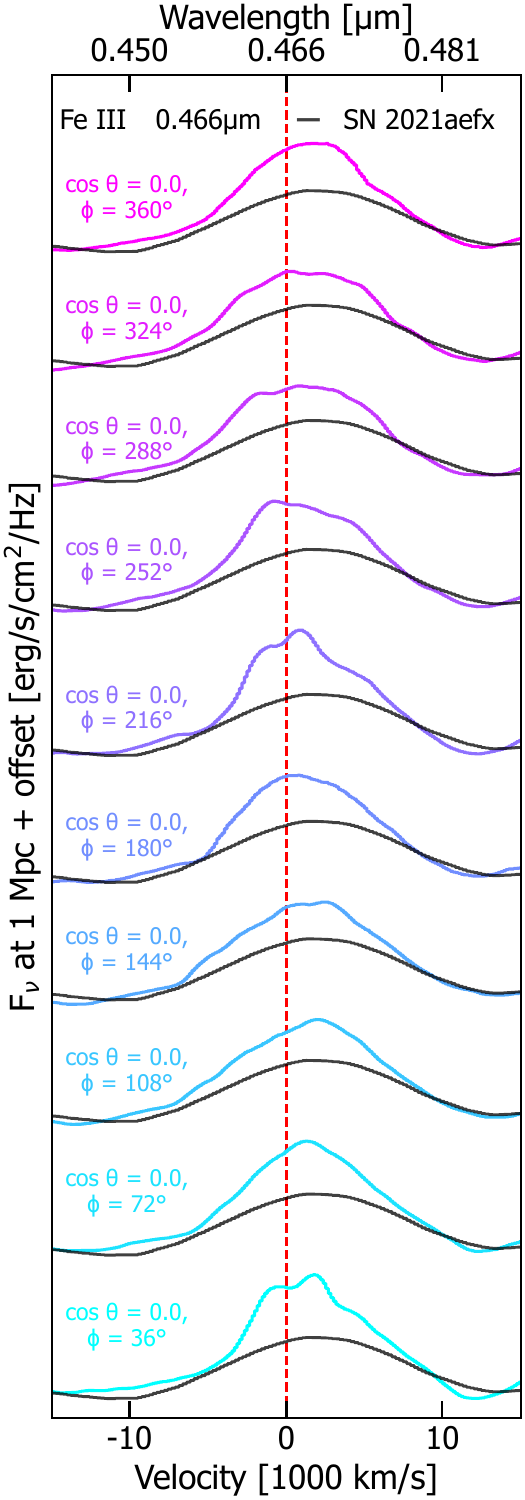}
    \includegraphics[height=8.5cm,width=3.5cm,trim=24 0 0 0, clip]{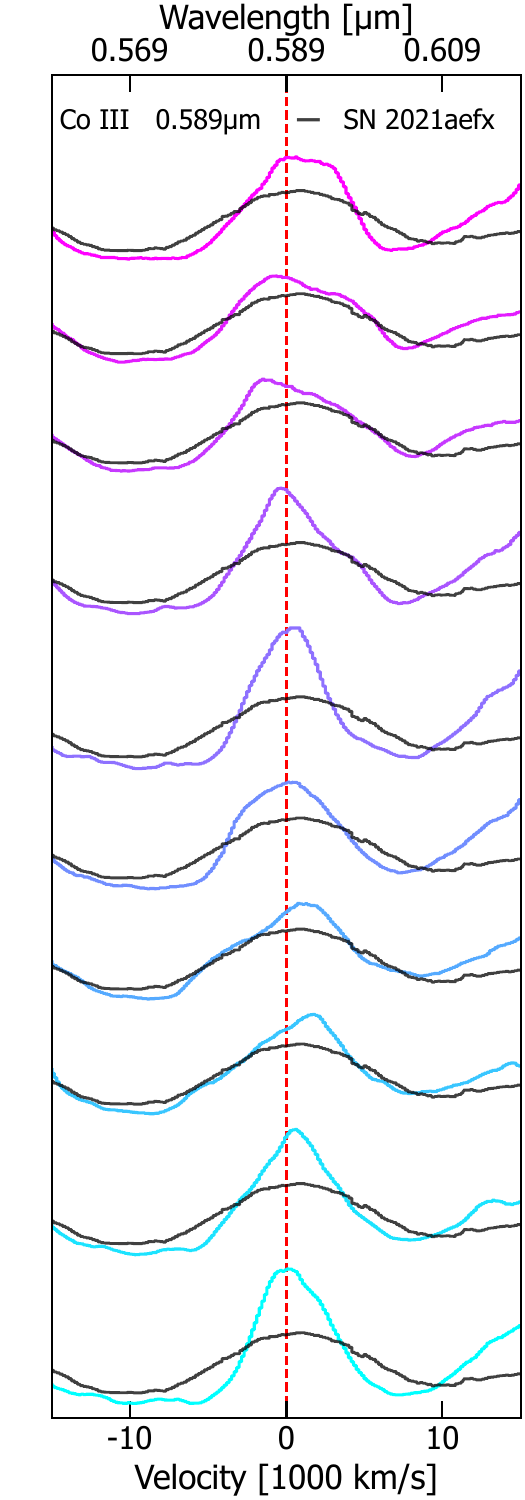}    
    \includegraphics[height=8.5cm,width=3.5cm,trim=24 0 0 0, clip]{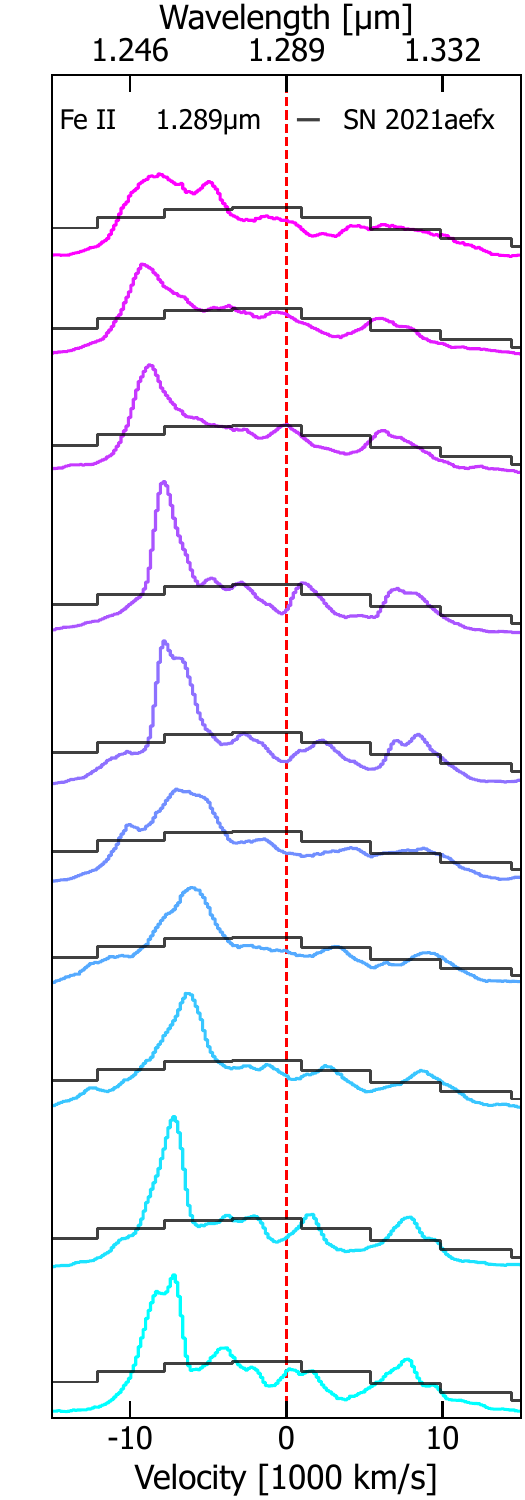}  
    \includegraphics[height=8.5cm,width=3.5cm,trim=24 0 0 0, clip]{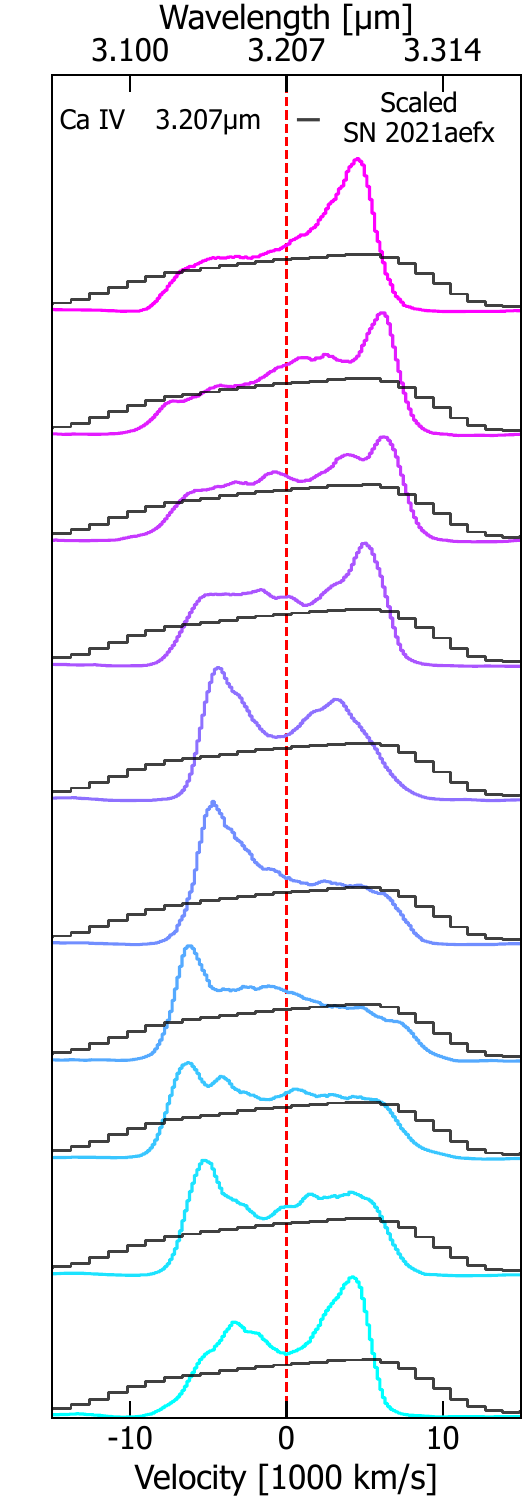}
    \includegraphics[height=8.5cm,width=3.5cm,trim=24 0 0 0, clip]{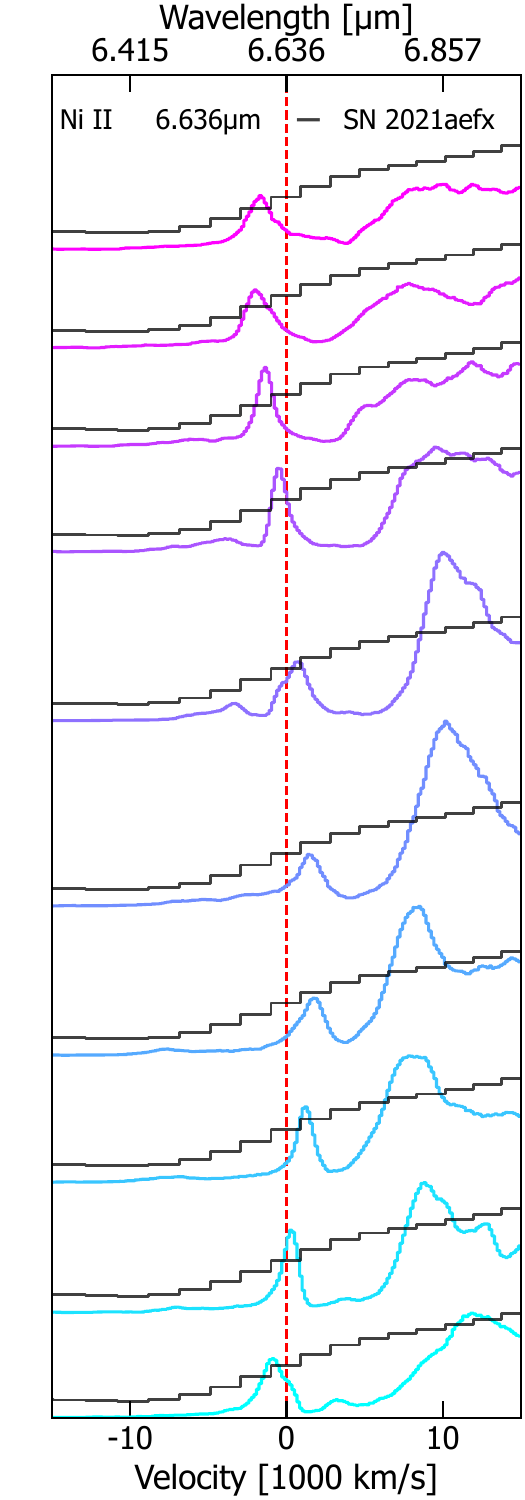}\\
    \includegraphics[height=8.5cm,width=3.5cm,trim=4 0 0 0, clip]{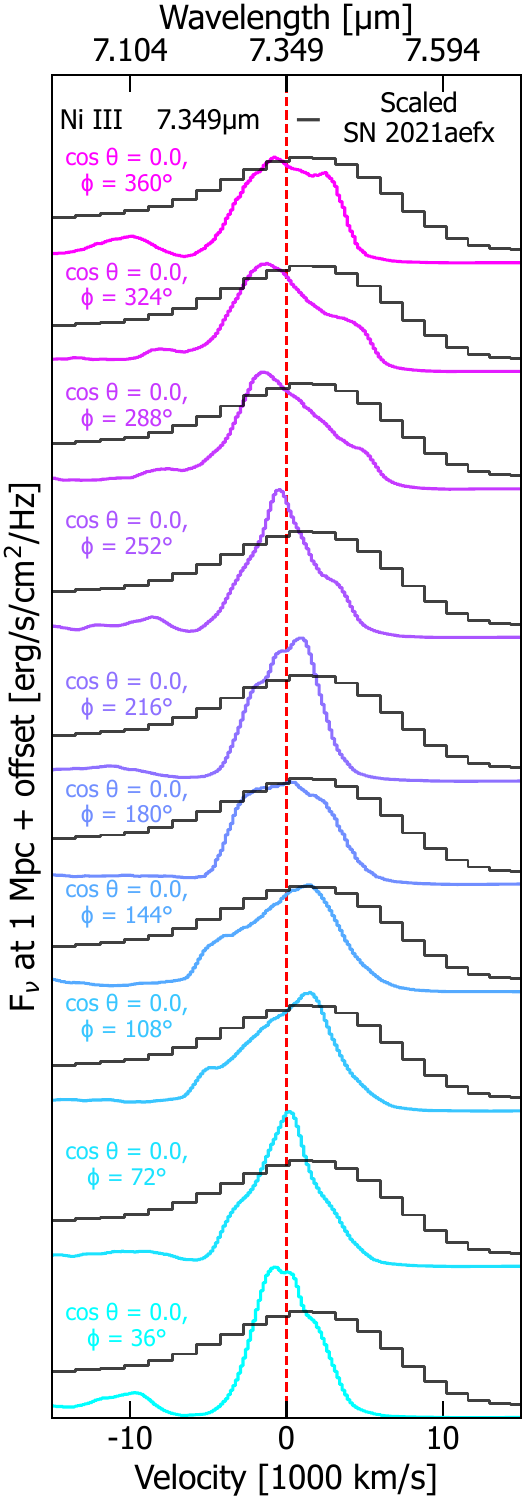}
    \includegraphics[height=8.5cm,width=3.5cm,trim=24 0 0 0, clip]{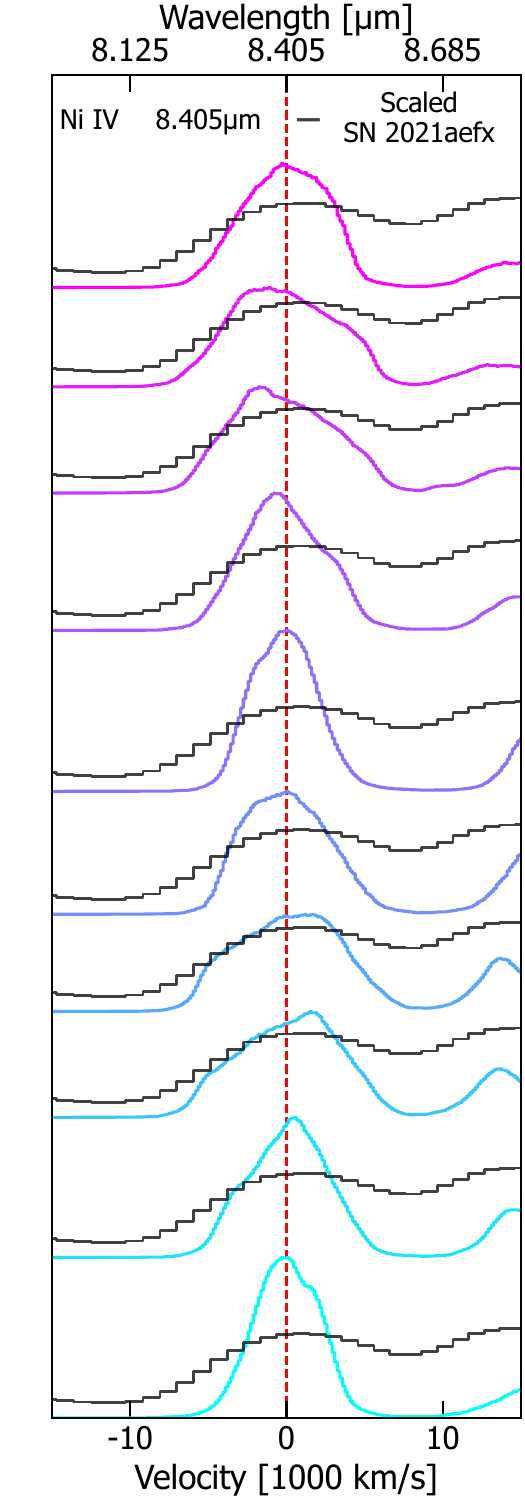}
    \includegraphics[height=8.5cm,width=3.5cm,trim=24 0 0 0, clip]{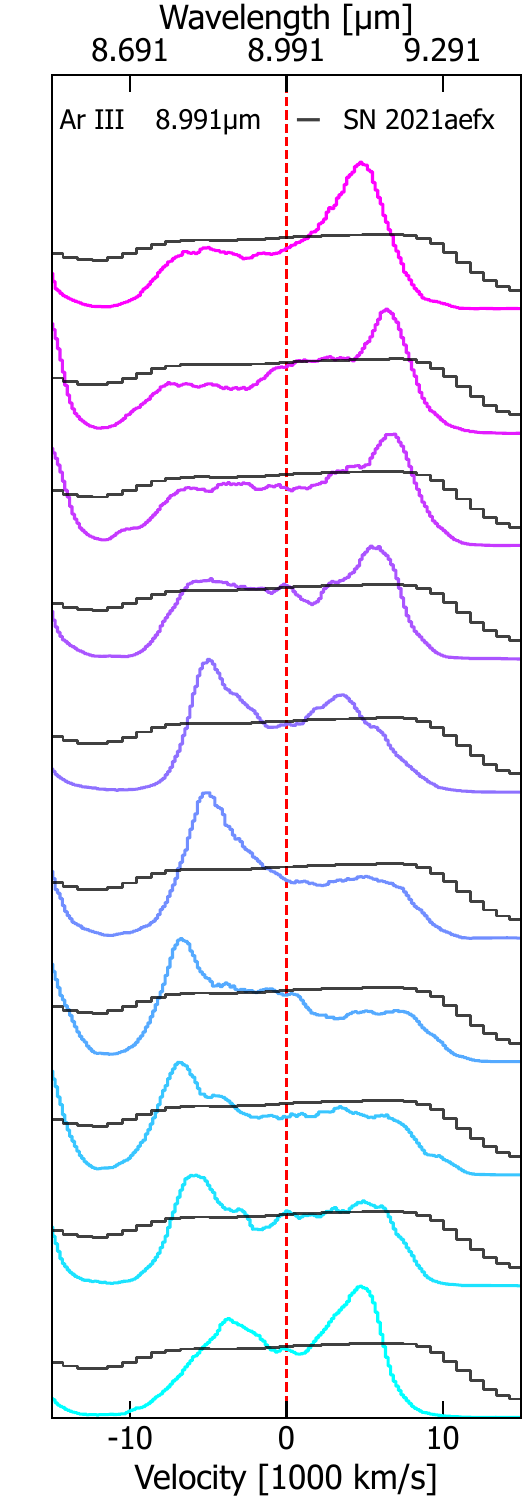}
    \includegraphics[height=8.5cm,width=3.5cm,trim=24 0 0 0, clip]{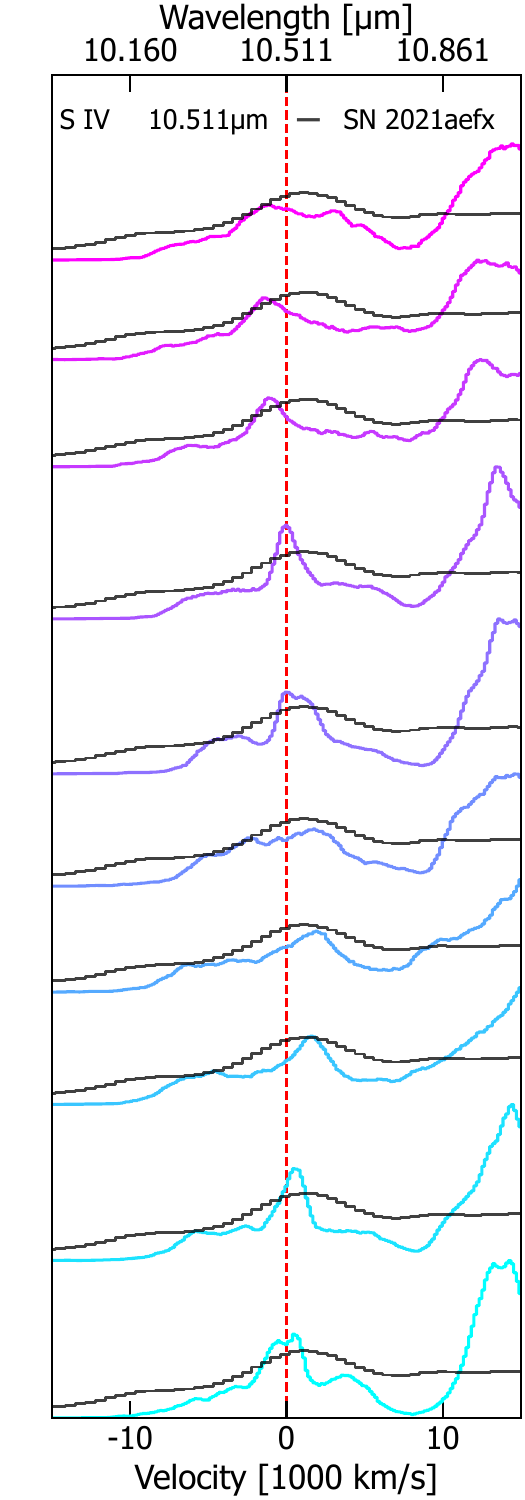}    
    \includegraphics[height=8.5cm,width=3.5cm,trim=24 0 0 0, clip]{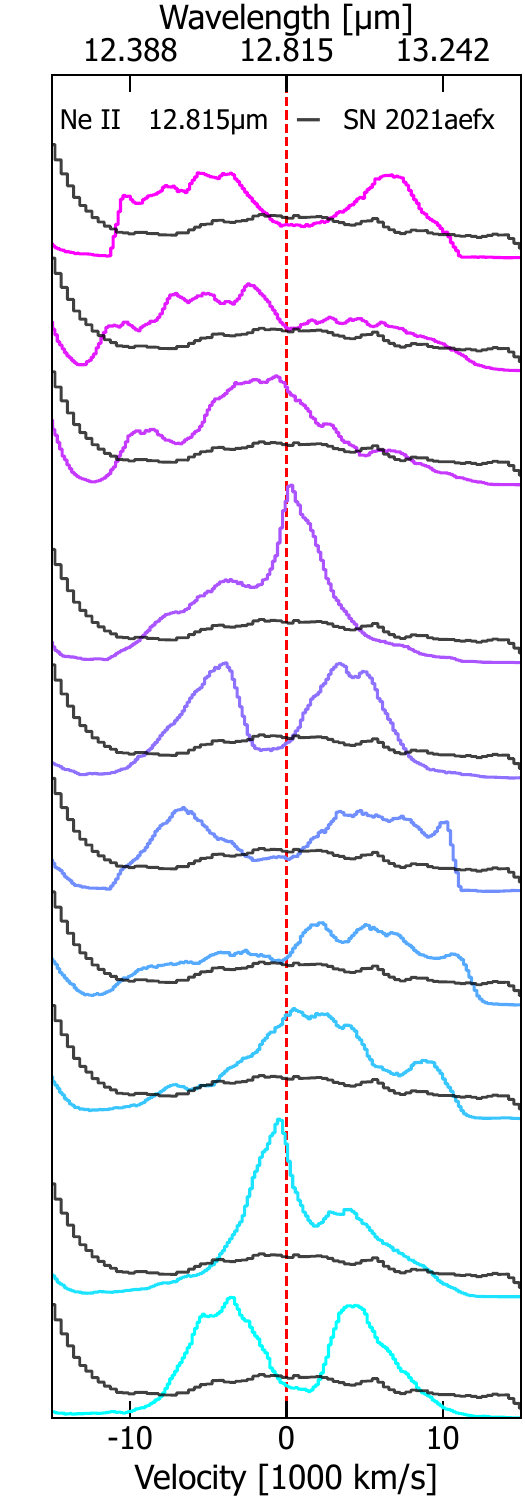}
    
    \caption{Viewing-angle spectra for the \threedvm model at 270 days post-explosion in the merger plane ($\cos(\theta)=0$), with the corresponding orientations ($\phi=0$--$360\degree$) indicated in the first panel of each row. The spectra are compared to the spectrum of SN~2021aefx \citep[black;][]{Kwok2023}, which has been corrected for redshift and reddening. Prominent IGE and IME features are shown: [\ion{Fe}{III}] 0.466\microns, [\ion{Co}{III}] 0.589\microns, [\ion{Fe}{II}] 1.289\microns, [\ion{Ca}{IV}] 3.207\microns, [\ion{Ni}{II}] 6.636\microns, [\ion{Ni}{III}] 7.349\microns and 8.405\microns, [\ion{Ar}{III}] 8.991\microns, [\ion{S}{IV}] 10.511\microns, and [\ion{Ne}{II}] 12.811\microns. Red dashed lines mark the rest wavelengths. Note that some panels contain blended contributions from neighbouring transitions, particularly [\ion{Ni}{II}] from nearby [\ion{Ar}{II}] emission and [\ion{S}{IV}] from nearby [\ion{Co}{II}] emission.}
    
    \label{fig:velocity_stacked_vm_aefx}
\end{figure*}

\begin{figure*} 
    \centering
    \includegraphics[height=8.5cm,width=3.5cm,trim=4 0 2 0, clip]{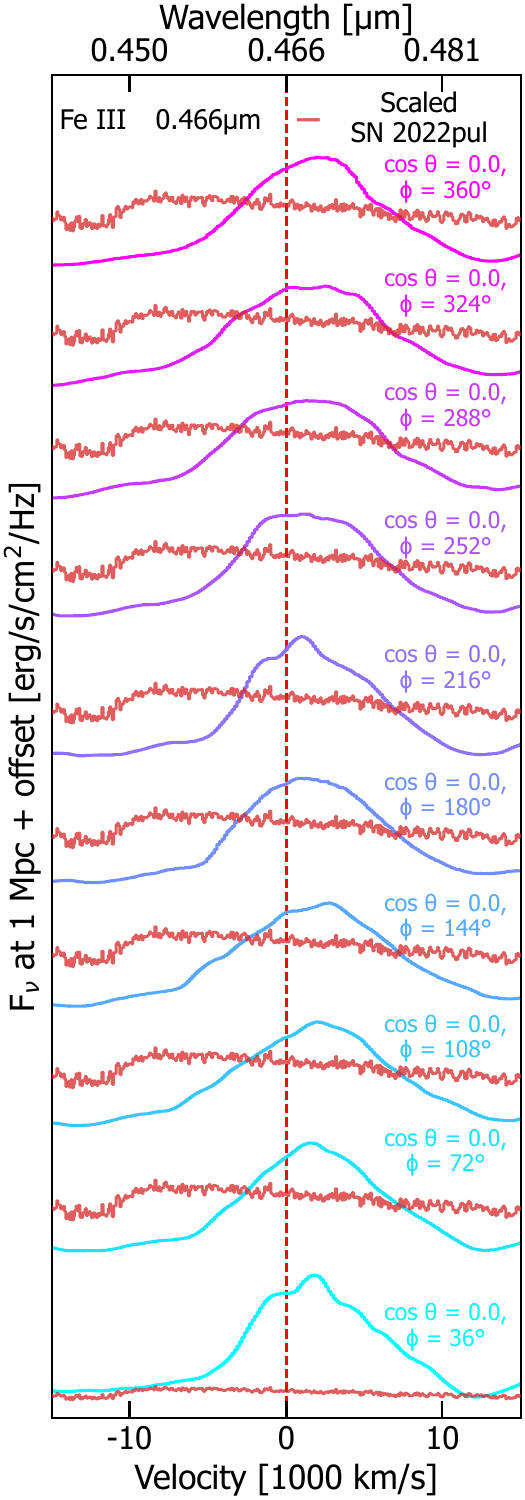}
    \includegraphics[height=8.5cm,width=3.5cm,trim=24 0 0 0, clip]{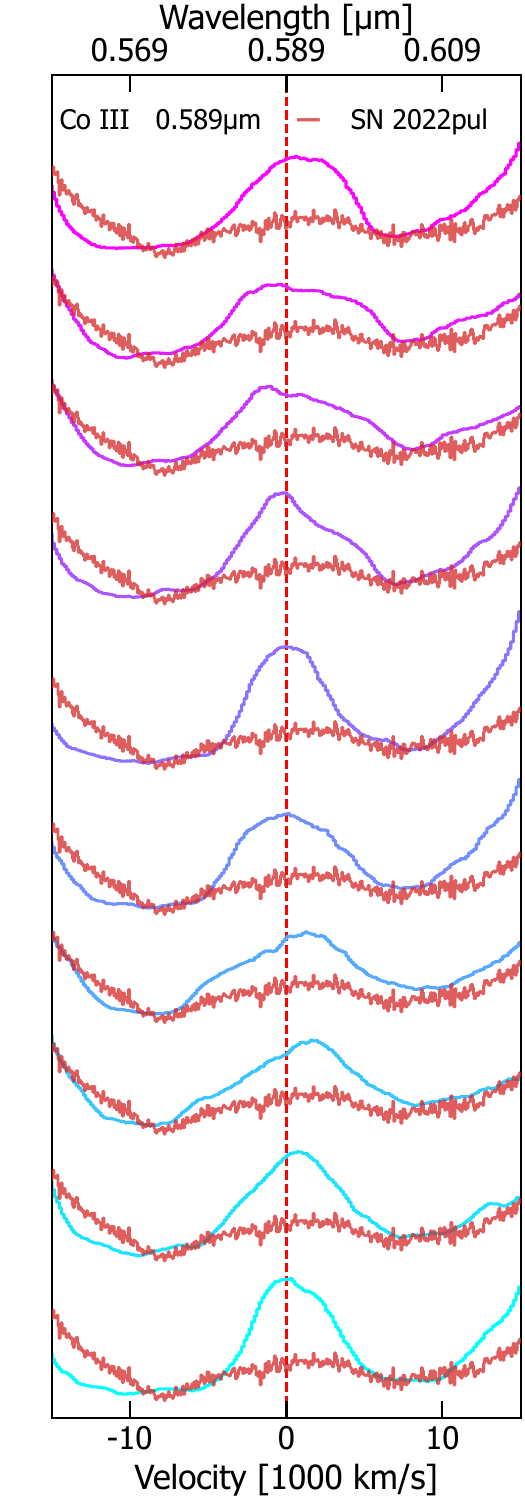}
    \includegraphics[height=8.5cm,width=3.5cm,trim=24 0 0 0, clip]{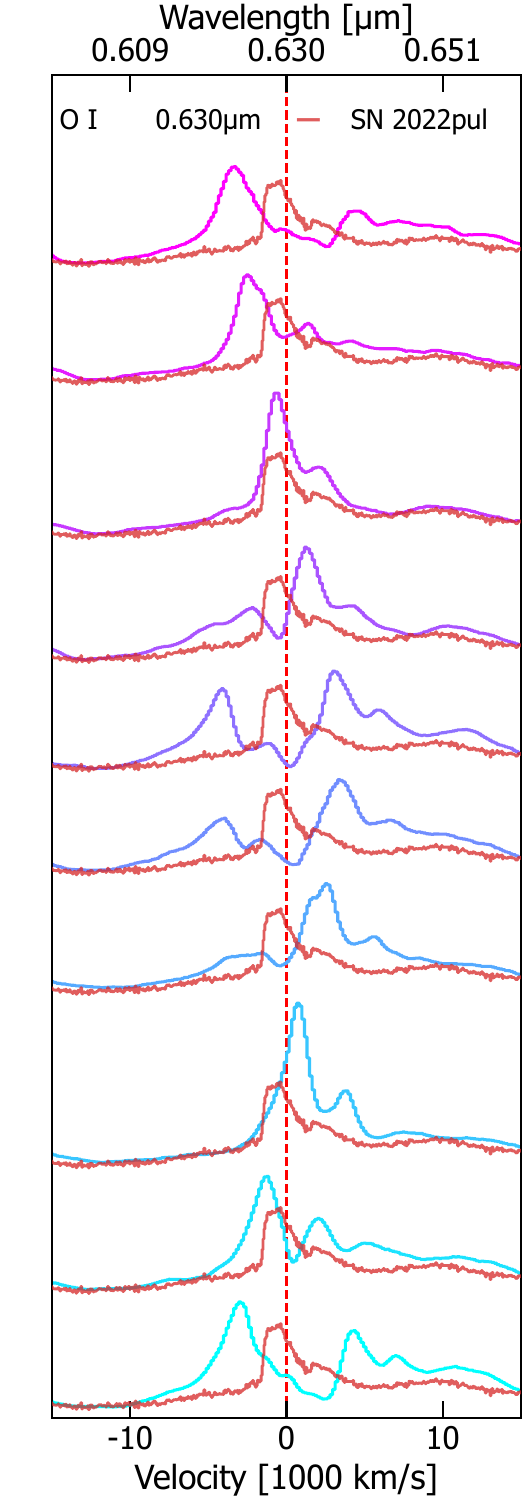}
    \includegraphics[height=8.5cm,width=3.5cm,trim=24 0 0 0, clip]{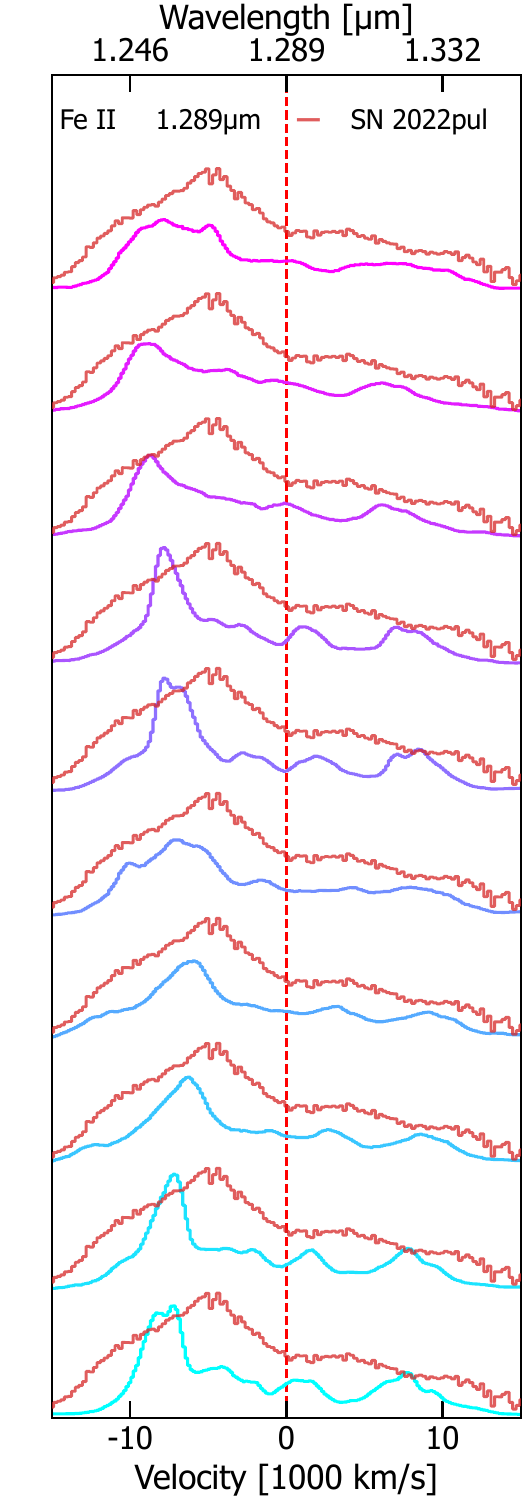}    
    \includegraphics[height=8.5cm,width=3.5cm,trim=24 0 0 0, clip]{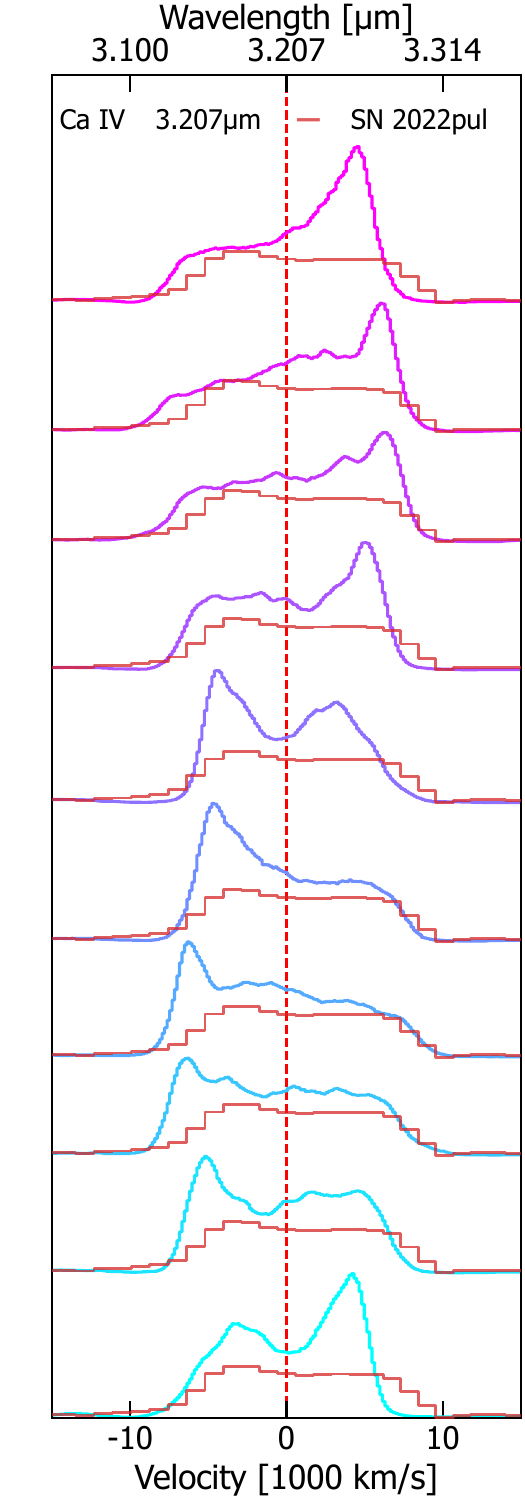}\\
    \includegraphics[height=8.5cm,width=3.5cm,trim=4 0 0 0, clip]{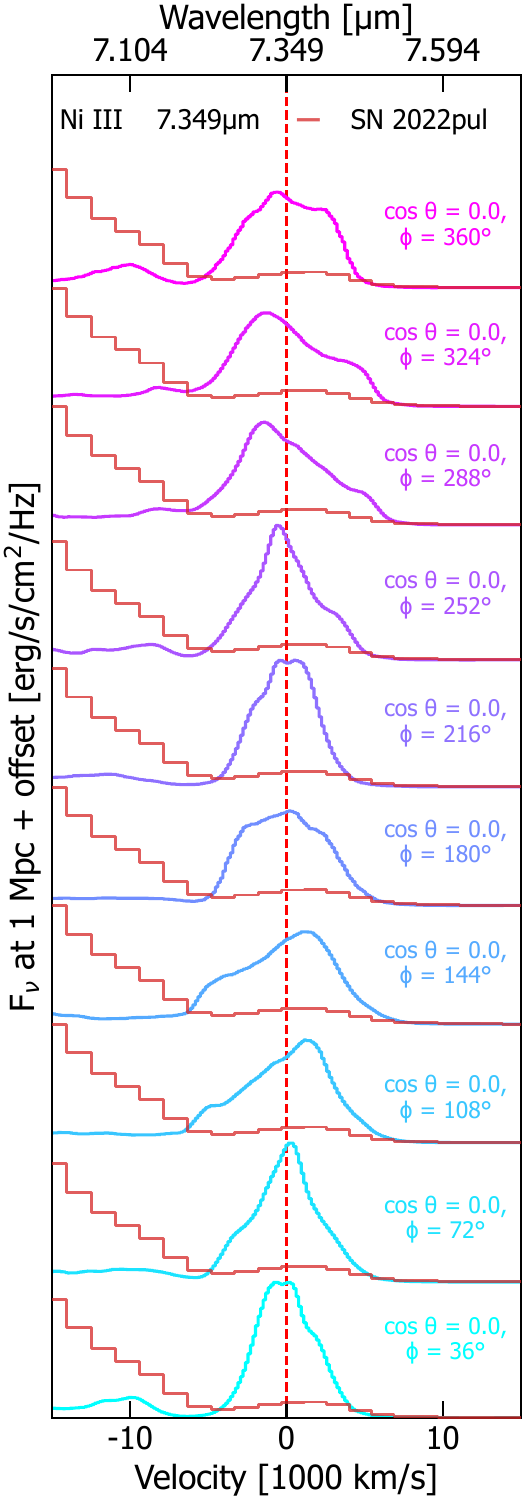}    
    \includegraphics[height=8.5cm,width=3.5cm,trim=24 0 0 0, clip]{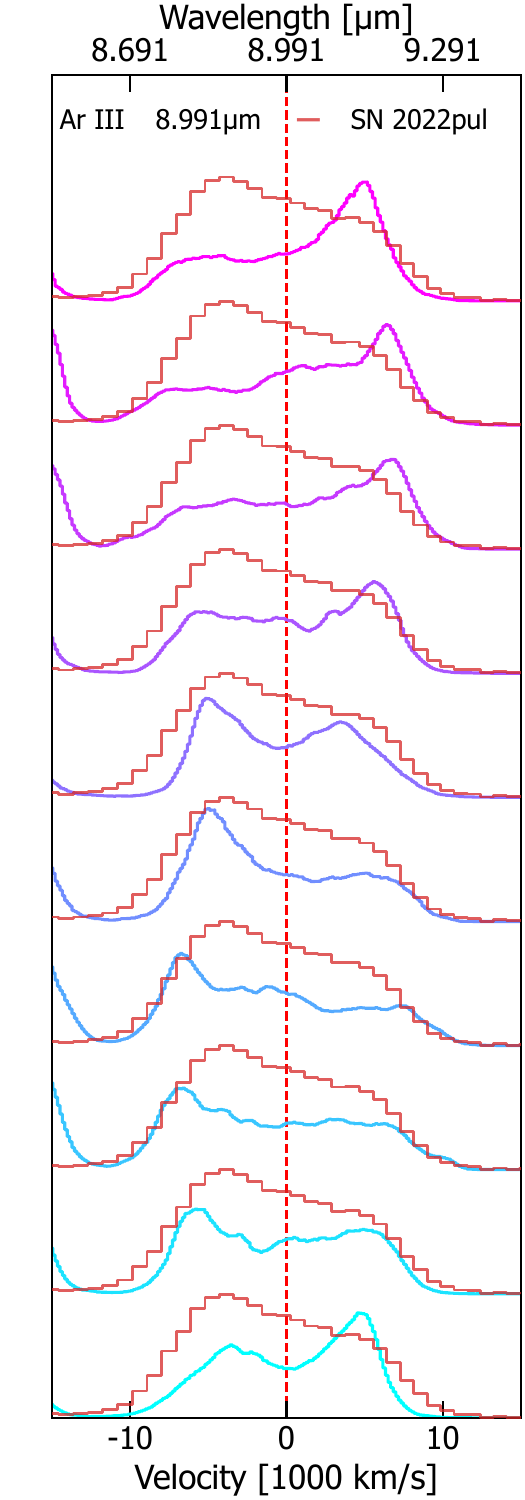}
    \includegraphics[height=8.5cm,width=3.5cm,trim=24 0 0 0, clip]{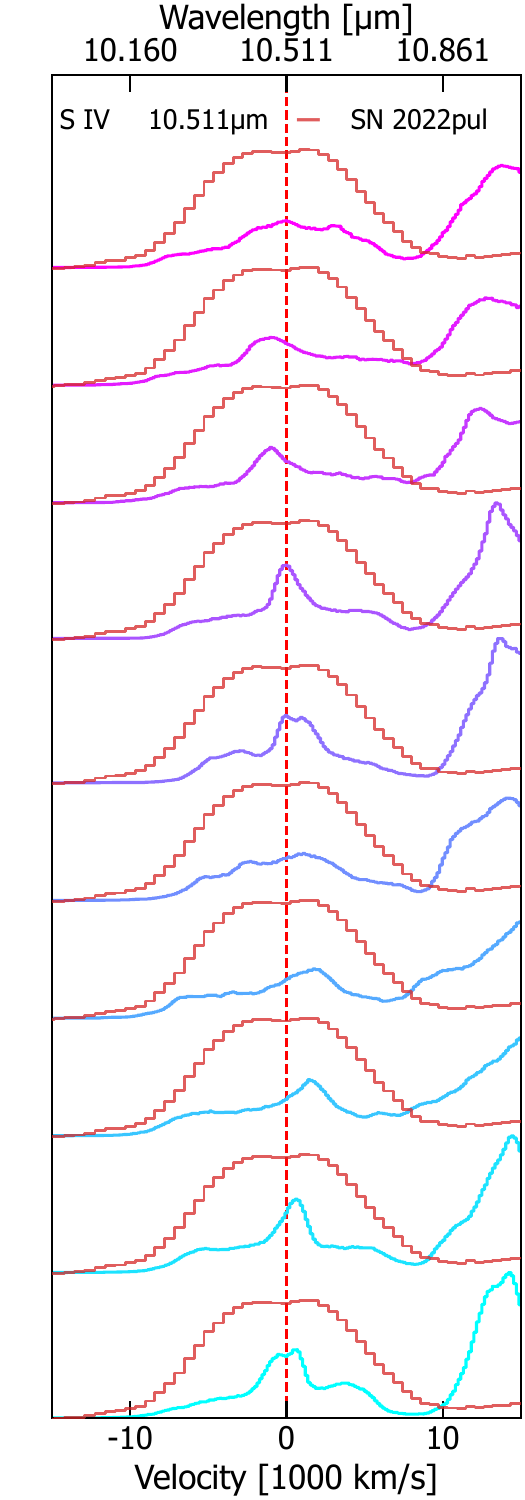}
    \includegraphics[height=8.5cm,width=3.5cm,trim=24 0 0 0, clip]{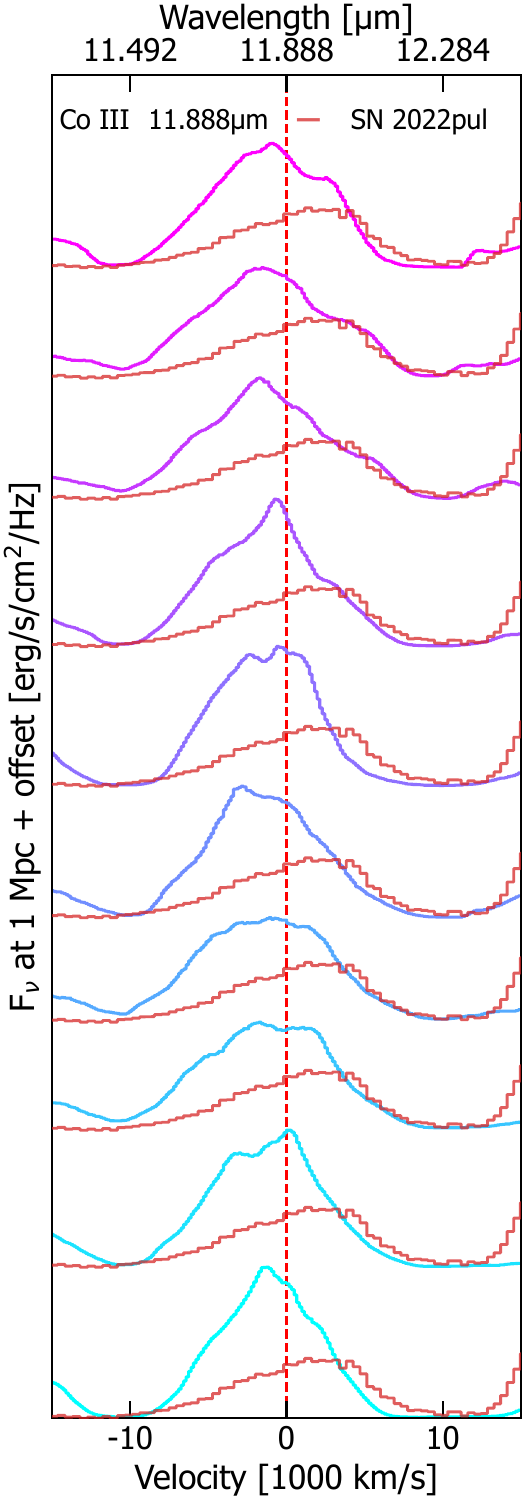}
    \includegraphics[height=8.5cm,width=3.5cm,trim=24 0 0 0, clip]{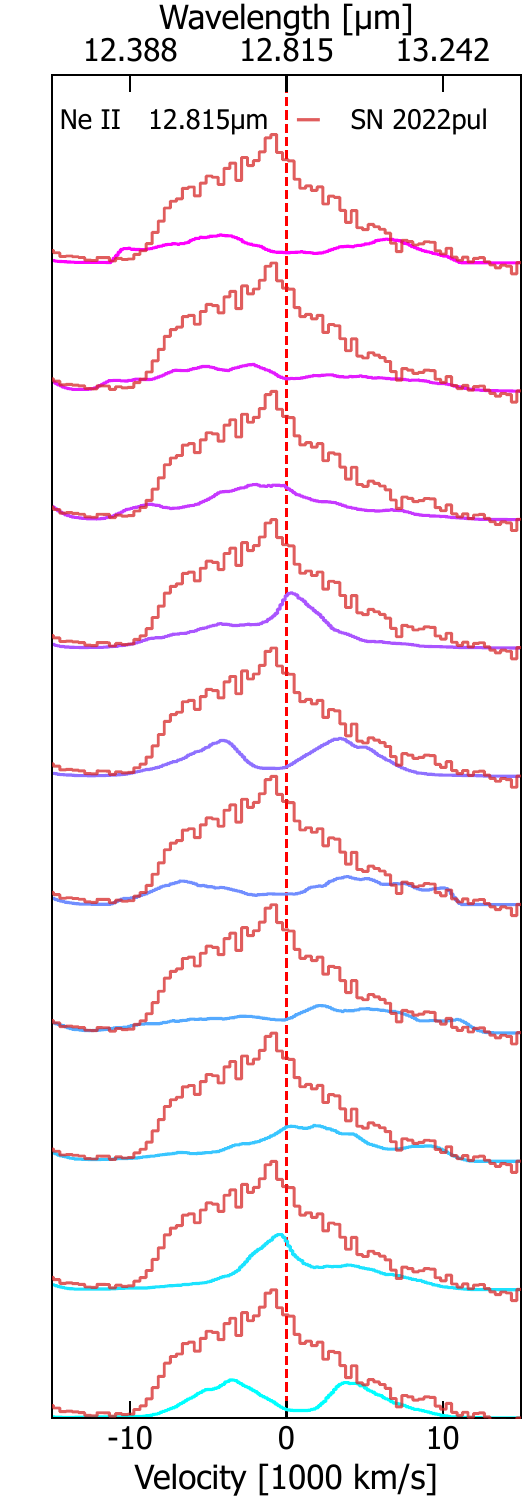}  \caption{    
    Same as Figure~\ref{fig:velocity_stacked_vm_aefx}, but compared to the spectrum of SN~2022pul \citep{Kwok2024} at 338 days post-explosion, which has also been corrected for redshift and reddening. The highlighted IGE and IME features are [\ion{Fe}{III}] 0.466\microns, [\ion{Co}{III}] 0.589\microns, [\ion{O}{I}] 0.630\microns, [\ion{Fe}{II}] 1.289\microns, [\ion{Ca}{IV}] 3.207\microns, [\ion{Ni}{III}] 7.349\microns, [\ion{Ar}{III}] 8.991\microns, [\ion{S}{IV}] 10.511\microns, [\ion{Co}{III}] 11.888\microns, and [\ion{Ne}{II}] 12.811\microns.}      
    \label{fig:velocity_stacked_vm_pul}
\end{figure*}

\begin{figure} 
    \centering
    \includegraphics[width=0.49\textwidth]{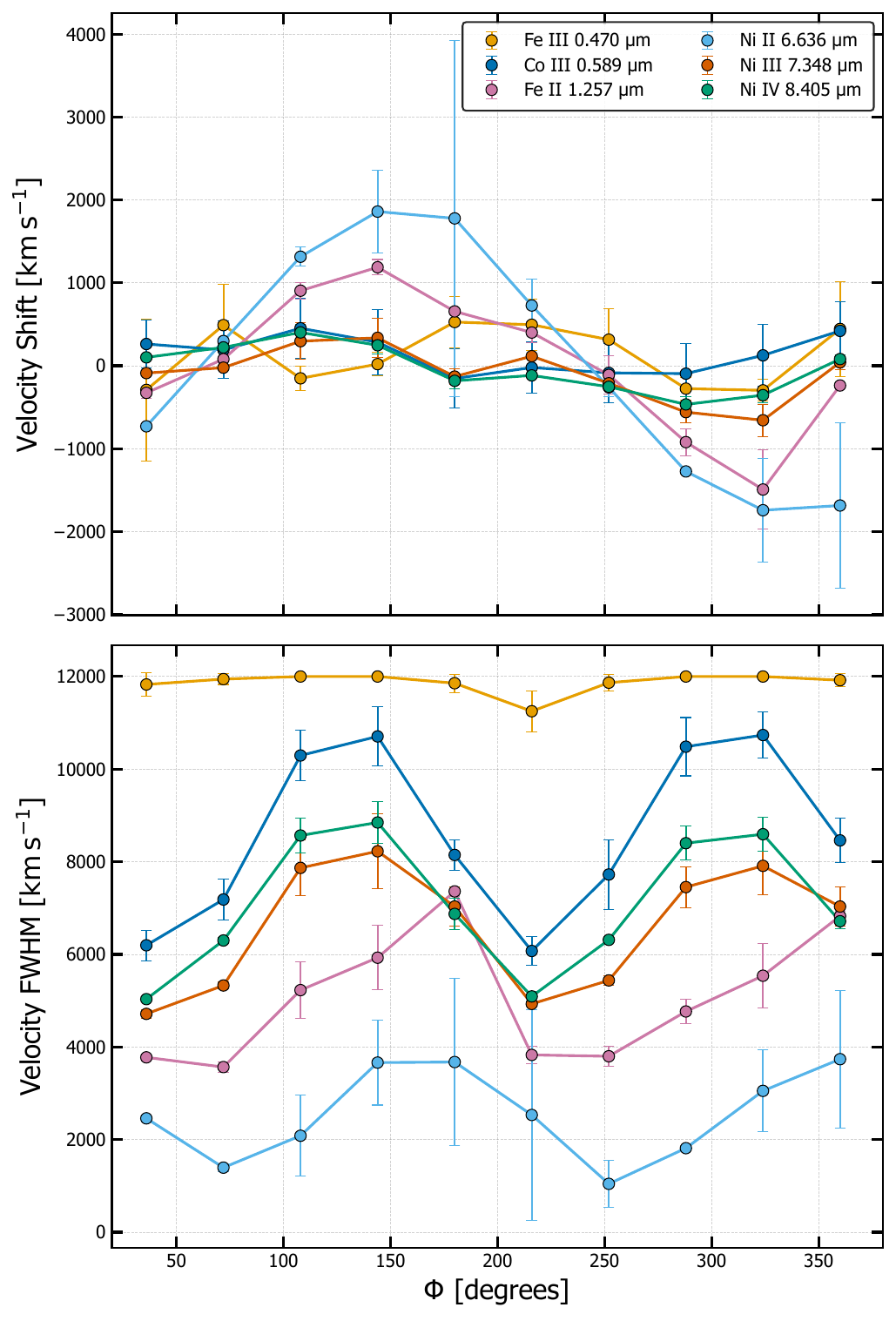}        
    \caption{Velocity shifts (top) of the [\ion{Fe}{II}] 1.257\microns, [\ion{Fe}{III}] 0.470\microns, [\ion{Co}{III}] 0.589\microns, [\ion{Ni}{III}] 7.348\microns and [\ion{Fe}{III}] 22.925\microns features for different observer orientations at $\cos(\theta) = 0$, $\phi = 0$--$360\degree$ (i.e., the merger plane), and the corresponding FWHM (bottom) for the \threedvm model.
    }
    \label{fig:velocity_fits}
\end{figure}

Figure~\ref{fig:Equator_viewing_angle} shows that the \threedvm model exhibits significant orientation dependence across several spectral features. At $\phi=36\degree$, the 0.470\microns [\ion{Fe}{III}] feature appears more sharply peaked, whereas at $\phi=324\degree$ it is noticeably broader. Examination of these orientations in Figure~\ref{fig:3d_ion_plot} shows that $\phi=36\degree$ probes a more compact emitting region, while $\phi=324\degree$ samples emission distributed over a more extended structure. This naturally explains the change in line-profile morphology, and is supported by the systematically larger FWHM of the [\ion{Co}{III}] feature at $\phi=324\degree$ compared to $\phi=36\degree$. However, Figure~\ref{fig:velocity_stacked_vm_aefx} shows that the [\ion{Co}{III}] and [\ion{Fe}{III}] morphologies do not track each other exactly. The optical [\ion{Fe}{III}] 0.470\microns feature is a complex blend of multiple ionisation stages and species, including contributions from IMEs, making this region difficult to interpret. Although fits are presented in Figure~\ref{fig:velocity_fits}, we do not consider this feature further, as it provides limited diagnostic value, particularly given that the optical depth in this region remains non-negligible.

Our calculation predicts strong NIR [\ion{Fe}{II}] emission, from which we extract velocity information using the [\ion{Fe}{II}] 1.289\microns feature. This feature shows sharply peaked substructure, similar to that seen in the \one model of \cite{Pollin2025}. The [\ion{Fe}{II}] emitting region is distinct from those of [\ion{Ni}{II}] and [\ion{Co}{III}], producing its own pattern of velocity shifts and widths with rotation. In the merger plane, the feature typically spans $\sim-1,000$ to $\sim1,000$\kms, with FWHM values of $4,000$--$7,000$\kms. Orientations such as $\phi=360\degree$ probe more extended emitting regions and produce broader profiles, while angles such as $\phi=252\degree$ yield narrower features. This behaviour can be understood from the \ion{Fe}{II} distribution in Figure~\ref{fig:3d_ion_plot}. Lines of sight aligned with the elongated IGE-rich structure, shaped by shielding from the secondary, produce broader emission, whereas perpendicular orientations yield narrower profiles. Compared to the \one model of \cite{Pollin2025}, which shows velocity shifts of $\sim-1,500$ to $\sim1,500$\kms and FWHM values of $2,000$--$3,000$\kms, the overall velocity range is similar. However, the [\ion{Fe}{II}] emitting region is more extended in the \threedvm model, bringing the predicted FWHM values closer to those observed in SN~2021aefx \citep{Kwok2023}.

The updated atomic data set also allows us to predict the key Ni features across the optical, NIR, and MIR. In particular, it includes [\ion{Ni}{IV}] forbidden transitions from NIST\footnote{https://www.nist.gov/pml/atomic-spectra-database} (\citealt{Kramida4NISTASD}; original data from \citealt{Hansen1984}), with collision strengths from \cite{Menchero2019}. Here, we examine three Ni features in detail: [\ion{Ni}{II}] 6.636\microns, [\ion{Ni}{III}] 7.349\microns, and [\ion{Ni}{IV}] 8.405\microns. This detailed analysis of [\ion{Ni}{II--IV}] is only possible with observations from JWST, as the optical and NIR regions contain features that are significantly contaminated by other species and do not display any [\ion{Ni}{IV}] emission.

As shown in Figure~\ref{fig:velocity_stacked_vm_aefx}, the predicted [\ion{Ni}{II}] feature is systematically narrower and more centrally peaked than observed. It spans velocities from $-2,000$\kms to $2,000$\kms, with FWHM values of $\sim1,500$--$\sim4,000$\kms, although some orientations have uncertainties of up to $\sim2,000$\kms (see Figure~\ref{fig:velocity_fits}). The broadest profiles occur at $\phi=144\degree$ and $\phi=180\degree$, and their antipodes, while orientations such as $\phi=72\degree$ produce narrower features. This behaviour reflects the \ion{Ni}{II} distribution in Figure~\ref{fig:3d_ion_plot}. \ion{Ni}{II} is concentrated near the outer edge of the IGE-rich core, adjacent to the inner unburned material from the secondary. Orientations tangential to this structure, such as $\phi=72\degree$, probe a more confined region and produce narrower emission, whereas orientations such as $\phi=144\degree$ align more closely with the elongated \ion{Ni}{II} distribution and give broader profiles. Compared to SN~2021aefx, which shows a [\ion{Ni}{II}] velocity of $4,000 \pm 1,500$\kms and a FWHM of $12,000 \pm 3,500$\kms, the model reproduces the characteristic velocity but underpredicts the width. This suppressed [\ion{Ni}{II}] emission was also seen in our previous work \citep{Pollin2025}, and likely reflects over-ionisation in the model.

In contrast to [\ion{Ni}{II}], the [\ion{Ni}{III}] and [\ion{Ni}{IV}] emission originates from a more extended, elongated structure produced by the detonation of the primary WD (see Figure~\ref{fig:3d_ion_plot}). This is reflected in the spectra: orientations such as $\phi=36\degree$ produce relatively narrow [\ion{Ni}{III}] features, while angles between $\phi=108\degree$ and $\phi=144\degree$ yield broader profiles by probing along the elongated distribution. The [\ion{Ni}{III}] and [\ion{Ni}{IV}] features closely track each other in both velocity and FWHM, typically agreeing within one standard deviation across orientations. Their velocity shifts remain small, generally within $\pm500$\kms. This contrasts sharply with the models of \cite{Pollin2025}, where the \one model shows [\ion{Ni}{III}] shifts of up to $\pm2,000$\kms and FWHM values of $3,000$--$7,000$\kms, while the \two model reaches shifts of up to $\pm4,500$\kms and FWHM values of $4,000$--$10,000$\kms. Although the \threedvm model produces a more favourable ionisation state, its [\ion{Ni}{III}] emitting region is less extended and shows reduced velocity variation. More broadly, this demonstrates that the geometry and extent of stable Ni-emitting regions vary significantly between explosion scenarios, producing distinct velocity offsets and line widths.

Although the strongest asymmetries occur in the merger plane, viewing-angle effects are also present for rotations around the azimuthal angle, as shown in Figure~\ref{fig:Pole_viewing_angle}. These variations further demonstrate that the ejecta structure is imprinted on several spectral features. A notable case is the MIR [\ion{Fe}{V}] feature at 20.851\microns, which shows a markedly different profile compared to equatorial orientations. When viewed along this orientation, the line exhibits a central dip in flux, giving the appearance of two separate components. This highlights the need for caution when interpreting such features observationally, as centrally peaked lines may instead arise from the geometric projection of a single transition and lead to misidentification. This profile results from projection effects in a vertically extended emitting region. More subtle effects are also present, for example, the MIR [\ion{Ni}{IV}] feature shows additional structure at its line centre, caused by two lobes of emitting material.

\subsubsection{Intermediate Mass Element Orientation Effects} 
\label{sec:IME_viewing_angl}

In this section, we discuss the viewing-angle variation of several IME features. While the optical and NIR spectra of SNe~Ia are dominated mainly by IGEs, the MIR contains many IME features that probe the stratification of the ejecta. Figures~\ref{fig:velocity_stacked_vm_aefx} and~\ref{fig:velocity_stacked_vm_pul} show several of these features in comparison with SN~2021aefx and SN~2022pul. We find considerable viewing-angle variation in [\ion{O}{I}] 0.630\microns, [\ion{Ca}{IV}] 3.207\microns, [\ion{Ar}{II}] 6.985\microns, [\ion{Ar}{III}] 8.991\microns, [\ion{S}{IV}] 10.511\microns, and [\ion{Ne}{II}] 12.815\microns, with the degree of variation set by the geometry of each emitting region.

One of the most striking features of our explosion model is the strong optical [\ion{O}{I}] emission from unburned material in the secondary WD. This feature is absent in normal SNe~Ia but has been detected in a small number of other subclasses, including 03fg-like SNe~Ia. It is clearly predicted in our 3D calculation but absent in 1D, demonstrating that multidimensional effects can determine not only the morphology of spectral features, but also whether they appear at all. Figure~\ref{fig:velocity_stacked_vm_pul} shows that the [\ion{O}{I}] feature exhibits substantial viewing-angle variation, with velocity shifts ranging from $\sim-3,000$ to $\sim3,000$\kms. For orientations such as $\phi=288\degree$, the profile is broadly similar to the observed spectra, showing a double-peaked structure with a dominant central component and a weaker shoulder. At $\phi=216\degree$, the feature instead separates into two distinct emission components. This behaviour can be understood from Figure~\ref{fig:3d_ion_plot}: this orientation views the ejecta along an axis where one \ion{O}{I}-rich region is moving towards the observer while the other is moving away, naturally producing two components, with the \ion{O}{I}-deficient central region producing the dip between them. This feature illustrates how the global IME distribution in this explosion model can produce complex, strongly separated line profiles. Although the merger-plane slices in Figure~\ref{fig:model_abundances} and Figure~\ref{fig:3d_ion_plot} show many IMEs concentrated in two distinct lobes, the full ejecta structure is more intricate, with the material distributed differently across other planes (see Appendix~\ref{apen:Additional Viewing Angles}).

In Figure~\ref{fig:velocity_stacked_vm_aefx}, we show the [\ion{Ca}{IV}] 3.207\microns feature, which exhibits the strongest viewing-angle dependence in the NIR. As the observer angle changes, the line profile evolves systematically. At $\phi=36\degree$, the feature is relatively narrow, with two horn-like components. As the angle increases, a stronger horn emerges on the blue side while the red side becomes flatter, as seen at $\phi=144\degree$. By $\phi=216\degree$, the profile again shows a distinct double-horned structure, similar to the $\phi=36\degree$ case, while by $\phi=324\degree$ the stronger horn has shifted to the red side.

This behaviour can be understood from the underlying ion distribution (see Figure~\ref{fig:3d_ion_plot}). In the \threedvm model, calcium is primarily synthesised in the primary and, like other IMEs, is located mainly in the outer layers surrounding the IGE-rich core. At $\phi=36\degree$, the line of sight intersects two relatively compact emitting regions, with stronger \ion{Ca}{IV} populations in the south-west quadrant than in the north-east, producing a profile with a stronger red-side peak. The most blueshifted profile occurs at $\phi=144\degree$, where the viewing angle aligns with the strongest \ion{Ca}{IV} emission, corresponding to the lobe of material between $5,000$ and $10,000$\kms in the north-west quadrant. This orientation is also aligned with the \ion{Ca}{IV} lobe extending between $5,000$ and $10,000$\kms in the south-east quadrant, making the profile more extended than at other orientations. At $\phi=324\degree$, offset by $180\degree$ from $\phi=144\degree$, the line of sight probes the reverse geometry, causing the horn to appear on the red side rather than the blue.

The [\ion{Ca}{IV}] feature therefore provides a clear example of how asymmetric ion structure maps directly to the emergent line profile. This complexity is not limited to the merger plane. As shown in Figure~\ref{fig:Pole_viewing_angle}, the feature can appear markedly different at other inclinations, producing a more double-peaked profile with a pronounced central dip reminiscent of a disc-like structure with an inner cavity, rather than simple horn-like asymmetry. The difference between the merger-plane and azimuthal-angle orientations highlight how the underlying morphology can produce an especially rich diversity of line-profile shapes, with a similar double-peaked structure also visible in the heavily blended [\ion{Ca}{II}] feature around 0.7\microns.

There are several key IME features in the MIR; here we focus on [\ion{Ar}{III}] 8.991\microns and [\ion{Ne}{II}] 12.815\microns, as they clearly illustrate the effects of viewing-angle variation. As shown in Figure~\ref{fig:velocity_stacked_vm_pul}, both features vary significantly with rotation, but in distinct ways that reflect their different ejecta distributions. During the primary explosion, the secondary is still being tidally disrupted, producing an extended distribution of carbon and oxygen that is only partially burned. This leads to neon in the inner ejecta, which appears at late times as [\ion{Ne}{II}] (see Figure~\ref{fig:3d_ion_plot}), with a total neon mass around one third lower than that of previous violent merger models \citep{Pakmor2012b}. In contrast, argon is synthesised in the primary WD explosion, so [\ion{Ar}{III}] emission traces material originating from the primary and thus has a different stratification.

At $\phi=36\degree$, both features appear broadly similar, although [\ion{Ne}{II}] separates more clearly into two components at velocities of approximately $\pm5,000$\kms. This double-peaked structure arises from two emitting regions separated by an IGE-rich core. By comparison, the [\ion{Ar}{III}] emission is more confined and includes a stronger contribution from lower-velocity material below $5,000$\kms, particularly in the north-east and south-west quadrants. This produces its double-peaked profile, while material between $5,000$ and $10,000$\kms contributes to the flat-topped base. As a result of these distinct distributions, the two features evolve differently with rotation.

At $\phi=72\degree$, the [\ion{Ne}{II}] feature becomes sharply peaked near $0$\kms, with an additional red-side component, while [\ion{Ar}{III}] develops a blue-side horn. This reflects the underlying geometry: more [\ion{Ar}{III}] material is moving towards the observer than at $\phi=36\degree$, similar to the NIR [\ion{Ca}{IV}] behaviour, whereas the [\ion{Ne}{II}] distribution includes an additional emitting region in the southern quadrants that produces the red-side component, similar to optical [\ion{O}{I}]. With further rotation, [\ion{Ne}{II}] evolves more strongly than [\ion{Ar}{III}]. Between $\phi=108\degree$ and $\phi=144\degree$, [\ion{Ne}{II}] becomes broader and flatter, consistent with a line of sight through the central region that is relatively deficient in IMEs. In contrast, [\ion{Ar}{III}] retains an overall flat-topped profile, reflecting its origin in the outer ejecta, but develops horn-like structures due to its lobe-like distribution. At azimuthal inclinations, such as $\cos(\theta)=-0.9,\phi=324\degree$ (see Figure~\ref{fig:Pole_viewing_angle}), [\ion{Ar}{III}] can show two central peaks superimposed on a flat-topped profile, while [\ion{Ne}{II}] remains centrally peaked. This contrasts with the merger-plane case, where double-peaked [\ion{Ar}{III}] can coincide with two distinct [\ion{Ne}{II}] components, highlighting the complex interplay between IME features and the 3D explosion structure.

Compared to our \dsix model investigation, the merger-plane behaviour differs markedly between the scenarios. In the \one model, [\ion{Ne}{II}] is extremely faint \citep[see figure~12 of][]{Pollin2025}, while [\ion{Ar}{III}] shows only modest viewing-angle variation, with a small bump from the wake left by the surviving secondary. This is much less pronounced than the horn-like structures in the \threedvm model. In the \two model, the secondary also detonates, producing a large amount of central [\ion{Ar}{III}] and therefore more centrally peaked profiles, although some orientations appear somewhat flat-topped \citep[see figure~15 of][]{Pollin2025}. By contrast, the \threedvm model shows more substructure, but does not produce a globally centrally peaked [\ion{Ar}{III}] profile. The \two model also produces strong [\ion{Ne}{II}], but with a sloping profile that shifts redward or blueward with orientation, rather than the distinct multi-component structure seen in the \threedvm model. These comparisons show that the MIR [\ion{Ar}{III}] 8.991\microns and [\ion{Ne}{II}] 12.815\microns features are particularly diagnostic of the explosion mechanism. The strength and morphology of [\ion{Ne}{II}] features trace the partially burned secondary material, while [\ion{Ar}{III}] features probes whether IMEs are confined to the primary ejecta or extend into the central regions through a secondary detonation. With sufficient MIR observations, these features therefore provide a direct way to distinguish violent mergers from double- and quadruple-detonation scenarios.

\subsection{Comparison with Observed SNe~Ia}
\label{sec:comparison}

\subsubsection{Comparison with the 03fg-like SN~2022pul}
In Figure~\ref{fig:velocity_stacked_vm_pul}, the $\cos(\theta)=0$, $\phi=288\degree$ orientation produces an [\ion{O}{I}] profile broadly consistent with SN~2022pul. However, this same orientation does not reproduce other key IME features, such as MIR [\ion{Ne}{II}]. Orientations approximately $180\degree$ offset from this line of sight still provide a reasonable match to the [\ion{O}{I}] feature and show better agreement with the other IME features. Even for these orientations, however, the model does not reproduce the redward-declining [\ion{Ar}{III}] profile observed in SN~2022pul, and the [\ion{Ne}{II}] feature can appear too red or too sharply peaked.

Even for the best-matching orientations (see Figure~\ref{fig:Panchromatic_avg}), several discrepancies remain. The IGE emission, while stronger than in our previous 3D investigation \citep{Pollin2025}, still does not reproduce the correct blended strength of the region around 0.7\microns. The IME features also show additional structure at all orientations that is not observed in SN~2022pul, most prominently in the NIR [\ion{Ca}{IV}] feature (see Figure~\ref{fig:velocity_stacked_vm_pul}). Even when [\ion{Ar}{III}] is better reproduced, the model predicts a horn-like structure on the edge of [\ion{Ca}{IV}] that is not evident in the observations. However, the strongest discrepancy is the absence of the strong MIR [\ion{Ar}{II}] emission that dominates the observed spectrum of SN~2022pul.

Taken together, these comparisons suggest that, while the model can reproduce the distinctive optical [\ion{O}{I}] feature associated with 03fg-like SNe~Ia, it does not simultaneously reproduce the optical, NIR, and MIR properties of SN~2022pul. The IME features, particularly NIR [\ion{Ca}{IV}], suggest that the IME distribution in this model is too asymmetric or too strongly separated in velocity space, since no orientation produces a profile as flat-topped as observed. A more promising merger realisation may therefore require a smoother IME distribution, while still retaining enough unburned material to produce optical [\ion{O}{I}]. At the same time, our model produces a substantial amount of $^{56}$Ni ($0.72\,\mathrm{M_\odot}$), potentially up to $32\%$ more than inferred for SN~2022pul ($0.66\pm0.17\,\mathrm{M_\odot}$), suggesting that a lower-mass primary WD may be required to reduce the overall IGE production. A more complete disruption of the secondary, similar to that found in quadruple-detonation models, may also be required to produce centrally peaked [\ion{Ar}{II}] emission. Future lower-mass violent merger realisations with more extensively disrupted secondaries should therefore be investigated.

\subsubsection{Comparison with normal events}

Violent mergers of massive WDs are unlikely to account for the majority of SNe~Ia, primarily as their expected rates are too low \citep[see e.g.,][]{Ruiter2013}, however, comparison with normal SNe~Ia remains useful to identify which nebular MIR features are generic across explosion scenarios and which are specific to particular progenitor channels. Moreover, the main difference between the scenario considered here and quadruple-detonation models is the detonation timing and the state of the secondary when the primary explodes, both of which depend on the initial helium shells. It is therefore plausible that this model may reproduce several key properties of normal SNe~Ia.

Overall, aside from [\ion{O}{I}] at 0.630\microns, [\ion{Fe}{V}] at $\sim$1.9\microns, and [\ion{Ni}{V}] at $\sim$11.2\microns, the model reproduces many general characteristics of normal SNe~Ia. In the optical, it captures much of the Fe-dominated structure between $\sim$0.4--0.55\microns, produces a pronounced [\ion{Co}{III}] feature, and yields a substantial [\ion{Fe}{II}] contribution around 0.7\microns. In the NIR, the strong [\ion{Fe}{II}] emission is broadly consistent in both strength and morphology with SN~2021aefx (see Figures~\ref{fig:Equator_viewing_angle} and~\ref{fig:Pole_viewing_angle}). The model also produces a weak [\ion{Ni}{II}] feature at 1.9\microns and a MIR [\ion{Ni}{II}]--[\ion{Ar}{II}]--[\ion{Ni}{III}] complex that more closely resembles SN~2021aefx than SN~2022pul. For certain viewing angles, [\ion{S}{IV}] 10.511\microns is particularly well reproduced. The model further predicts strong MIR [\ion{Co}{III}] emission and a prominent [\ion{Ni}{III}] feature, neither of which was present at comparable strength in the \dsix explosion models explored by \cite{Pollin2025}. The predicted [\ion{Ne}{II}] profile shows notable viewing-angle variation and can become significantly broadened, for example in the $\cos(\theta)=0$, $\phi=324\degree$ orientation compared with SN~2021aefx (see Figure~\ref{fig:Panchromatic_avg}). In several orientations, the feature separates into a clear double-peaked profile, although this is not present for all lines of sight. This behaviour is also absent in the 1D calculation, where the feature appears centrally peaked and therefore more similar to normal SNe~Ia. Interestingly, none of the JWST spectra shown in Figure~\ref{fig:Panchromatic_avg} display a clearly double-peaked [\ion{Ne}{II}] profile. However, this absence does not by itself rule out the scenario, since the separated profile appears only for a subset of orientations. Instead, it may represent a distinctive prediction of this violent merger model, particularly in contrast to the \dsix scenario, where [\ion{Ne}{II}] does not clearly separate into two components. Further calculations will be required to determine whether this morphology is a robust indicator of this specific model or of violent merger scenarios more generally.

A broader comparison with the JWST sample highlights the diversity of the MIR [\ion{Ni}{II}]--[\ion{Ar}{II}]--[\ion{Ni}{III}] complex. Some objects show a clearly separated [\ion{Ni}{II}] feature distinct from neighbouring [\ion{Ar}{II}] emission, while others show a more continuous blend between these components. Both behaviours are reproduced in our model depending on viewing angle, for example in the $\cos(\theta)=0$, $\phi=36\degree$ and $\cos(\theta)=-0.3$, $\phi=288\degree$ orientations compared with SN~2022aaiq (see Figure~\ref{fig:Panchromatic_avg}). This agreement extends to longer MIR wavelengths ($\gtrsim14$\microns), where all viewing angles provide a reasonable match to SN~2024gy. In particular, the double-peaked structure between 15--17\microns in the $\cos(\theta)=0.9$, $\phi=324\degree$ orientation agrees well with the blended emission from [\ion{Co}{III}], [\ion{Co}{IV}], and other trace species in this region. Comparisons with SN~2022aaiq beyond 19\microns are limited by the increasing noise in the observational data, although some orientations remain broadly consistent. Despite this overall agreement, notable discrepancies remain. The model predicts a [\ion{Ni}{IV}] feature that is not observed at these longer wavelengths. While changes in the ionisation state could suppress this feature, it may also strengthen the already prominent [\ion{Ni}{III}] emission. Additionally, all viewing angles produce [\ion{Ca}{IV}] features that are stronger and more structured than those observed in SN~2021aefx, SN~2022aaiq, and SN~2024gy, suggesting that these events may arise from a more symmetric or less strongly structured IME distribution.

\subsubsection{Kinematic trends of IGE features}

In Figure~\ref{fig:velocity_fits}, we present the velocity shifts and FWHM measurements for several key IGE features. The [\ion{Ni}{II}] 6.636\microns feature shows the largest viewing-angle variation, spanning approximately $-2,000$ to $+2,000$\kms, followed by [\ion{Fe}{II}] 1.257\microns, which spans roughly $-1,000$ to $+1,000$\kms. These features are also among the narrowest predicted by the model, with FWHM values of $1,500$--$4,000$\kms for [\ion{Ni}{II}] and $\sim4,000$--$7,000$\kms for [\ion{Fe}{II}].
Compared to SN~2021aefx, for which \cite{Kwok2023} report a [\ion{Ni}{II}] velocity of $4,000 \pm 1,500$\kms and a FWHM of $12,000 \pm 3,500$\kms, the model reproduces comparable velocity shifts only for the most extreme orientations, but predicts substantially narrower profiles. The same behaviour is seen for [\ion{Fe}{II}]: the predicted velocity shifts are broadly consistent with the observed value of $1,300 \pm 1,500$\kms, but the FWHM values remain below the observed $9,400 \pm 1,500$\kms range, with only the $\phi=180\degree$ orientation approaching the lower bound.
Although the [\ion{Ni}{II}] feature is too narrow to provide a good match to SN~2021aefx, the presence of a sharp [\ion{Ni}{II}] peak across several orientations is nevertheless interesting. This resembles the narrow [\ion{Ni}{II}] spikes seen in SN~2024gy and SN~2022aaiq, although our model lacks the broader underlying base present in those observations. A lower ionisation state in the outer ejecta may shift more emission into [\ion{Ni}{II}], potentially producing broader emission beneath the narrow component and improving agreement with these sharp observed [\ion{Ni}{II}] features.

We also measure the velocities and FWHM of the [\ion{Ni}{III}] 7.348\microns and [\ion{Ni}{IV}] 8.405\microns features. These features show similar behaviour, with velocities and widths typically consistent within one standard deviation, although [\ion{Ni}{IV}] is slightly broader on average. For SN~2021aefx, \cite{Kwok2023} report [\ion{Ni}{III}] and [\ion{Ni}{IV}] velocities of $3,000 \pm 1,400$\kms and $1,300 \pm 1,200$\kms, respectively, with corresponding FWHM values of $11,200 \pm 1,300$\kms and $13,600 \pm 600$\kms. The predicted [\ion{Ni}{III}] velocities fall more than one standard deviation below the observed range, while the [\ion{Ni}{IV}] velocities show better agreement, with approximately half of the viewing angles lying within the observational constraints. However, the FWHM values of both features remain systematically smaller than observed, typically by around two standard deviations. Finally, we consider the [\ion{Co}{III}] 0.589\microns feature. The model predicts velocity shifts between $-500$ and $+500$\kms, with FWHM values of $\sim6,000$--$10,500$\kms. These are broadly consistent with SN~2021aefx, for which \cite{Kwok2023} report a velocity of $1,000 \pm 500$\kms and a FWHM of $10,800 \pm 2,500$\kms. In summary, the model produces velocity shifts for [\ion{Fe}{II}], [\ion{Co}{III}], [\ion{Ni}{II}], and [\ion{Ni}{IV}] that are generally consistent with SN~2021aefx. However, the [\ion{Ni}{III}] velocities are lower than observed, and the stable Ni features are systematically too narrow.

We conclude this section by comparing our results to the broader population of SNe~Ia \citep[e.g.,][]{Maeda2010,Maeda2011,Maguire2018,Flores2018}, rather than focusing solely on one observed object. Such comparisons are subject to additional uncertainty from line blending in the optical and NIR, which can affect inferred velocities and FWHM values, as well as differences in fitting methodology. Observationally, singly ionised species such as [\ion{Fe}{II}] and [\ion{Ni}{II}] typically show maximum velocity shifts of $\sim2,000$--$2,500$\kms and FWHM values of $\sim6,000$--$9,000$\kms \citep[see figure 7 of][]{Maguire2018}, with the [\ion{Fe}{II}] and [\ion{Ni}{II}] features often offset from one another. The velocity shifts of both [\ion{Fe}{II}] and [\ion{Ni}{II}] in our model are consistent with these ranges. The [\ion{Fe}{II}] FWHM values are also in good agreement, and a small number of observations presented by \cite{Maguire2018} show [\ion{Ni}{II}] FWHM values comparable to those predicted here. For doubly ionised species such as [\ion{Co}{III}], observed velocities are typically no greater than $\sim1,000$\kms, with FWHM values spanning $\sim8,500$--$11,500$\kms. Our model therefore produces [\ion{Co}{III}] features consistent with observational samples of normal SNe~Ia.

\section{Discussion and Conclusions}
\label{sec:Discussion_and_Conclusions}

In this work, we present the first 3D NLTE nebular radiative-transfer calculations for a violent merger explosion model, with a $1.1\mass + 0.7\mass$ WD binary. We compute spectra from $\sim190$ to $\sim410$ days after explosion, across both the merger-plane and azimuthal angles. We compare these predictions to our previous 3D double- and quadruple-detonation calculations \citep{Pollin2025}, and to panchromatic observations of the normal SNe~2021aefx, 2022aaiq, and 2024gy, and the 03fg-like SN~2022pul \citep{Kwok2023,Kwok2024,Kwok2025}. Our main conclusions are as follows:

\begin{enumerate}  

\item \textbf{Ionisation state and stable-Ni features}: 
The \threedvm model achieves a substantially more favourable ionisation state for normal SNe~Ia than the \one and \two models \citep{pakmor_2021,Pollin2025}. This is most evident in the NIR and around 0.7\microns, where it avoids strong [\ion{Ar}{III}] emission. It also reproduces a MIR [\ion{Co}{III}] feature with a strength comparable to that observed in SN~2021aefx.
However, some features remain too highly ionised, particularly the stable-Ni emission, with strong MIR [\ion{Ni}{III--V}] emission relative to both observations and the [\ion{Co}{III}] feature. 
This illustrates that prominent stable-Ni signatures can be produced by \submch\xspace explosion models, and that weak [\ion{Ni}{II}] emission does not necessarily imply insufficient stable-Ni production, but may instead reflect ionisation-state effects.
This highlights the need for panchromatic comparisons that probe multiple ionisation stages when testing explosion-model predictions. 

\item \textbf{Multidimensional signatures}: 
Many important spectral features emerge only in our 3D calculation. The clearest examples are [\ion{O}{I}] 0.630\microns and [\ion{C}{I}] 0.873\microns, both of which originate from unburned material associated with the tidally disrupted secondary. 
These features are absent from the 1D calculation, where the ejecta are artificially redistributed into compositions that do not correspond to any physical region of the 3D structure. This demonstrates that differences between 1D and 3D treatments are substantial in the nebular phase. We therefore caution that spherically averaged ejecta can suppress true signatures of a model and may not faithfully represent the predictions of the underlying explosion model.

\item \textbf{Viewing-angle dependence and progenitor geometry}:
The \threedvm model exhibits strong viewing-angle dependence across all wavelength regions, with signatures that differ from the double- and quadruple-detonation models of \citet{Pollin2025}. The MIR stable-Ni features show relatively small [\ion{Ni}{III}] and [\ion{Ni}{IV}] velocity shifts ($\lesssim500$~\kms), compared to those in the quadruple ($\sim4,000$~\kms) and double-detonation ($\sim2,000$~\kms) models. IME features, including [\ion{O}{I}], [\ion{Ca}{IV}], [\ion{Ar}{III}], and [\ion{Ne}{II}], show strong orientation dependence and frequently develop complex double-peaked or horn-like profiles. For the MIR [\ion{Ne}{II}] feature, some orientations produce profiles that are sufficiently separated to resemble two distinct features, reflecting the widely separated distribution of the emitting material. Since [\ion{O}{I}] traces material associated with the secondary WD, while [\ion{Ar}{III}] primarily traces material burned in the primary, these profiles provide a way to probe the initial progenitor configuration. Viewing-angle effects are therefore not only a source of spectral diversity, but also a diagnostic of the underlying explosion geometry, with violent mergers, double detonations, and quadruple detonations producing distinct nebular signatures.

\item \textbf{Remarkable agreement with normal SNe~Ia}:
The model reproduces characteristics of normal SNe~Ia particularly well, despite expectations that it would more closely resemble the peculiar 03fg-like subclass. The model captures the central optical [\ion{Fe}{II}]/[\ion{Fe}{III}] complex, the strong [\ion{Co}{III}] 0.589\microns feature, NIR [\ion{Fe}{II}], and several key MIR features, including [\ion{Ni}{II--IV}], [\ion{Co}{III}], and the [\ion{S}{IV}]/[\ion{Co}{II}] complex for specific viewing angles. Apart from optical [\ion{O}{I}], this scenario is therefore, in many respects, a better match to normal SNe~Ia than the \dsix models. However, clear discrepancies remain in the IME features: [\ion{Ca}{IV}] is systematically too strong, while [\ion{Ar}{III}] develops horn-like profiles that are absent from the JWST spectra. This suggests that the IME-emitting material is too asymmetric, too strongly separated in velocity space, has an ionisation structure that is not fully consistent with normal SNe~Ia, or some combination of these effects. As violent merger and \dsix-like scenarios are closely related, small changes to the progenitor configuration may preserve the favourable ionisation state found here while removing optical [\ion{O}{I}] and producing a more uniform IME distribution. This offers a possible pathway towards unifying these explosion scenarios within a common framework.

\item \textbf{SN~2022pul and the need for a different merger configuration}:
Although the \threedvm model predicts an [\ion{O}{I}] feature consistent with SN~2022pul, it does not reproduce the strong, centrally peaked [\ion{Ar}{II}] 6.985\microns feature. The MIR [\ion{Ne}{II}] feature, identified as a key diagnostic supporting a violent-merger interpretation \citep[][see also \citealt{Blondin2023}]{Kwok2024}, is also considerably weaker than in the violent-merger model of \citet{Pakmor2012b}. This reflects the smaller neon mass in our ejecta, only about one third of that in the \citet{Pakmor2012b} model, where the secondary is fully disrupted. Since [\ion{Ne}{II}] is already the dominant neon ion in our calculation, the weak emission is not primarily an ionisation effect. Instead, the main discrepancy is abundance and ejecta structure: our model lacks both sufficient neon-rich material and the central IME concentration needed to reproduce the strong [\ion{Ne}{II}] and centrally peaked [\ion{Ar}{II}] features observed in SN~2022pul. This suggests that SN~2022pul requires a merger in which the secondary is fully disrupted, or burns more completely, naturally producing stronger [\ion{Ne}{II}] emission and a more central argon distribution. A broader range of violent-merger configurations, particularly lower-mass systems and systems that lead to full disruption or detonation of the secondary, should therefore be explored. We also note that, with the updated argon atomic data \citep{Pelan1995,Tayal1996} adopted here, quadruple-detonation models may also produce a more centrally concentrated [\ion{Ar}{II}] feature.

The \threedvm model may instead be more promising for other bright 03fg-like events. For example, SN~2021zny \citep{Dimitriadis2023} was notably more luminous and exhibited an optical ionisation state more similar to normal SNe~Ia, while also showing a sharp [\ion{O}{I}] feature and a double-peaked [\ion{Fe}{II}]--[\ion{Ni}{II}]--[\ion{Ca}{II}] structure around 0.7\microns.

\item \textbf{Distinguishing explosion models with MIR IME features}: 
JWST enables detailed measurements of MIR IME features that were largely inaccessible in previous optical and NIR studies, and only sparsely sampled by low signal-to-noise MIR spectra from facilities such as \textit{Spitzer} \citep{Gerardy2007,Telesco2015}. These IME features provide some of the clearest diagnostics of the explosion channel, as they trace how burning proceeds through both the primary and secondary WDs. Our previous investigation of double- and quadruple-detonation models showed that the [\ion{Ar}{III}] profile is particularly sensitive to the explosion mechanism, while the \threedvm model further demonstrates the diagnostic value of combining [\ion{Ar}{III}] with [\ion{Ne}{II}]: [\ion{Ar}{III}] primarily traces material associated with the primary WD explosion, whereas [\ion{Ne}{II}] also traces material from the tidally disrupted secondary. This produces an IME stratification that differs markedly from double- and quadruple-detonation models, but is difficult to infer from ground-based observations alone, where key features are heavily blended. With a sufficiently large number of JWST observations, these orientation-dependent MIR signatures should become directly testable, providing a clear route to distinguish between competing explosion channels across different progenitor masses and SNe~Ia subtypes.

\end{enumerate}

This work adds to a growing set of 3D NLTE predictions from modern explosion models, intended to support the interpretation of current and future nebular observations. Even within the still limited JWST sample, substantial diversity is already apparent, much of which reflects viewing-angle effects. 
We encourage the community to make use of the synthetic spectra presented here for detailed comparisons with observations, including studies of diversity among brighter 03fg-like events. At the same time, these calculations represent only one merger configuration, and 3D NLTE nebular modelling of modern explosion models is still maturing. Larger suites of calculations will therefore be needed to identify which spectral features are robust signatures of specific explosion scenarios. Future work should carefully consider the choice of computational domain, atomic data set, and epochs of interest. Only through this combination of expanded theoretical grids and high-quality JWST observations will it be possible to establish firm connections between explosion models and the progenitor systems of both peculiar and normal SNe~Ia.

\section*{Acknowledgements}

JMP thanks Aysha Aamer for discussions related to the observational spectra and flux conversions, and Kate Maguire for helpful comments on the manuscript. JMP also thanks Ravi Seth and Andreas Fl{\"o}rs for helpful discussions at the XIX Würzburg Winter Workshop. FPC and SAS acknowledge funding from STFC grant ST/X00094X/1. This work used the DiRAC Memory Intensive service Cosma8 at Durham University, managed by the Institute for Computational Cosmology on behalf of the STFC DiRAC HPC Facility (www.dirac.ac.uk). The DiRAC service at Durham was funded by BEIS, UKRI and STFC capital funding, Durham University and STFC operations grants. DiRAC is part of the UKRI Digital Research Infrastructure. 
LJS and SAS acknowledge funding by the European Union (ERC, HEAVYMETAL, 101071865).
CEC is funded by the European Union’s Horizon Europe
research and innovation programme under the Marie Skłodowska-Curie grant
agreement No.~101152610. 
The work of FKR  is supported by the Klaus Tschira Foundation, by the Deutsche Forschungsgemeinschaft (DFG, German Research Foundation) -- RO 3676/7-1, project number 537700965, and by the European Union (ERC, ExCEED, project number 101096243). Views and opinions expressed are, however, those of the authors only and do not necessarily reflect those of the European Union or the European Research Council Executive Agency. Neither the European Union nor the granting authority can be held responsible for them.
The authors gratefully acknowledge the Gauss Centre for Supercomputing e.V. (www.gauss-centre.eu) for funding this project by providing computing time on the GCS Supercomputer SuperMUC-NG at Leibniz Supercomputing Centre (www.lrz.de) via the project pn76fu. JMP acknowledges the use of Grammarly for proofreading and grammar checking.
We acknowledge Numpy \citep{harris2020array}, SciPy \citep{2020SciPy-NMeth}, Matplotlib \citep{Hunter:2007} and
\href{https://zenodo.org/records/8302355} {\textsc{artistools}}\footnote{\href{https://github.com/artis-mcrt/artistools/}{https://github.com/artis-mcrt/artistools/}} \citep{artistools2024} for data processing and plotting.
\section*{Data Availability}
\label{sec:data_availability}

The spectra presented here will be made publicly available via the Heidelberg Supernova Model Archive (HESMA; \citealt{Kromer2017}; https://hesma.hits.org). They may also be obtained directly from the
corresponding author upon request.
 
\bibliographystyle{mnras}
\bibliography{bibliography} 

\appendix
\section{Computational Setup and Requirements}
\label{apen:comp_cost}

We use \textsc{ARTIS} v2025.08.01 \citep{artis_collaboration_2025}. This version includes developments to the NLTE solver in \textsc{artis} that enables ion stages with negligible populations to be removed from the NLTE solution dynamically in order to avoid numerical problems in the solution. These developments are described in detail by Callan et al. (2026, in prep.) Our primary 3D nebular simulations employ a $36^3$ grid resolution. This choice reflects the substantial computational cost associated with increasing the number of grid cells, compounded by the significantly larger number of NLTE levels employed in this work compared to \cite{Pollin2025} (5121 levels versus 3316, corresponding to a $\sim$55\% increase).
Both the 1D and 3D calculations were performed on 3200 CPU cores across 25 nodes (128 cores per node; 2 × 64-core AMD EPYC 7H12 processors; 1 TB RAM per node). The 1D nebular simulation required $\sim$36,600 CPU core hours, while the $36^3$ simulation required $\sim$990,000 CPU core hours. The total computational cost of the production-level 1D and 3D calculations is therefore $\sim$1,027,000 CPU core hours.

\section{Additional Viewing Angles}
\label{apen:Additional Viewing Angles}

\begin{figure*}
\begin{subfigure}{\textwidth}
    \centering
    \includegraphics[width=0.99\textwidth]{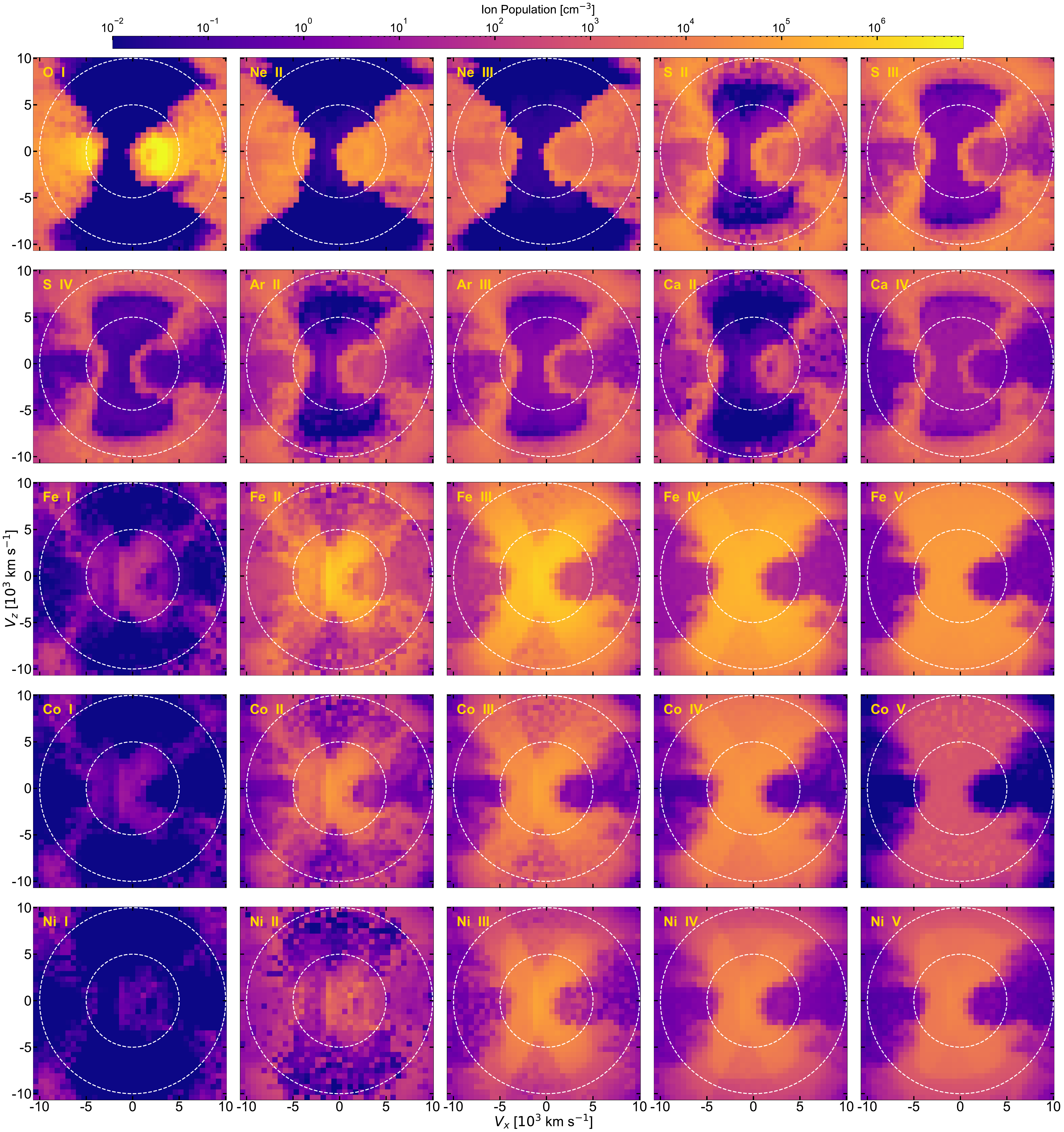}    
\end{subfigure}
        \caption{Same as Figure~\ref{fig:3d_ion_plot} but for the $x$-$z$ plane}
\label{fig:3d_ion_plot_x_z}
\end{figure*}

\begin{figure*}
\begin{subfigure}{\textwidth}
    \centering
    \includegraphics[width=0.99\textwidth]{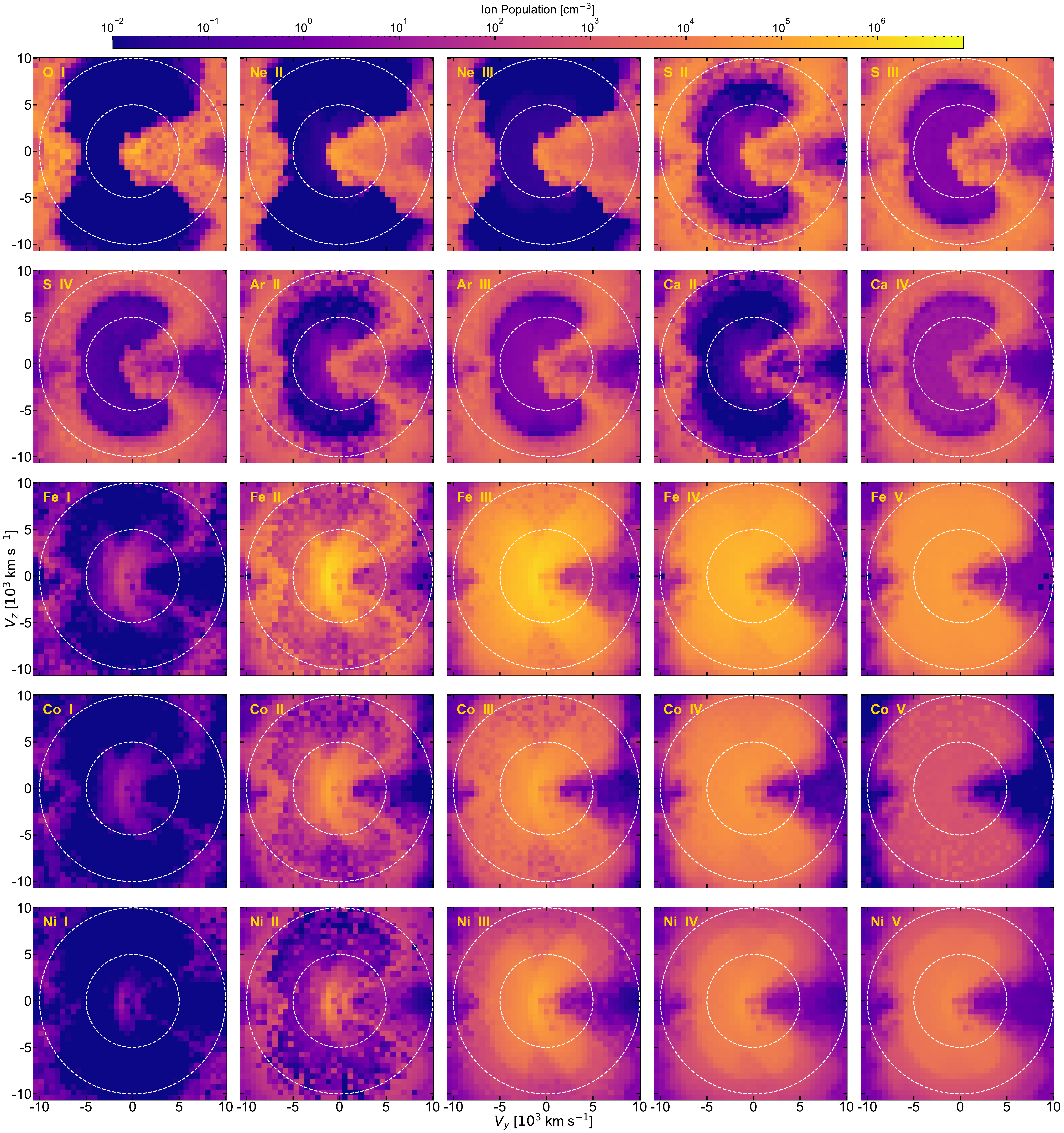}    
\end{subfigure}
        \caption{Same as Figure~\ref{fig:3d_ion_plot} but for the $y$-$z$ plane}
\label{fig:3d_ion_plot_y_Z}
\end{figure*}

We primarily focus on the $x$-$y$ plane when interpreting spectral features, as the merger plane synthetic observables exhibit a broad range of characteristics that probe much of the spectral diversity. However, explosion models are inherently multidimensional, and 1D temperatures, densities, and ion distributions do not capture the full dynamic range of the 3D structure.
For completeness, we also present the $x$-$z$ and $y$-$z$ slices of the \threedvm model in Figures~\ref{fig:3d_ion_plot_x_z} and~\ref{fig:3d_ion_plot_y_Z}.

\section{Emission and Absorption Decomposition at 338 Days}
\label{apen:pul_em}

\begin{figure*} 
    \centering
    \includegraphics[width=0.99\textwidth,height=6.5cm]{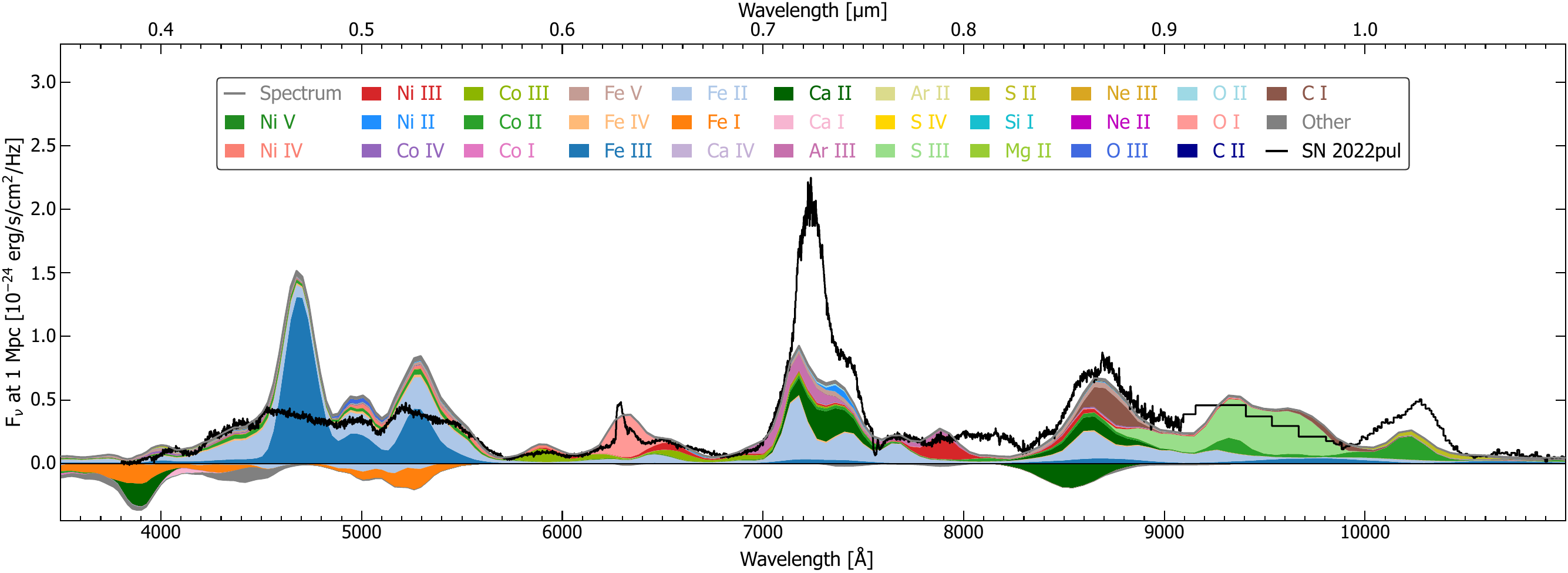}
    
    \includegraphics[width=0.99\textwidth,height=6.5cm]{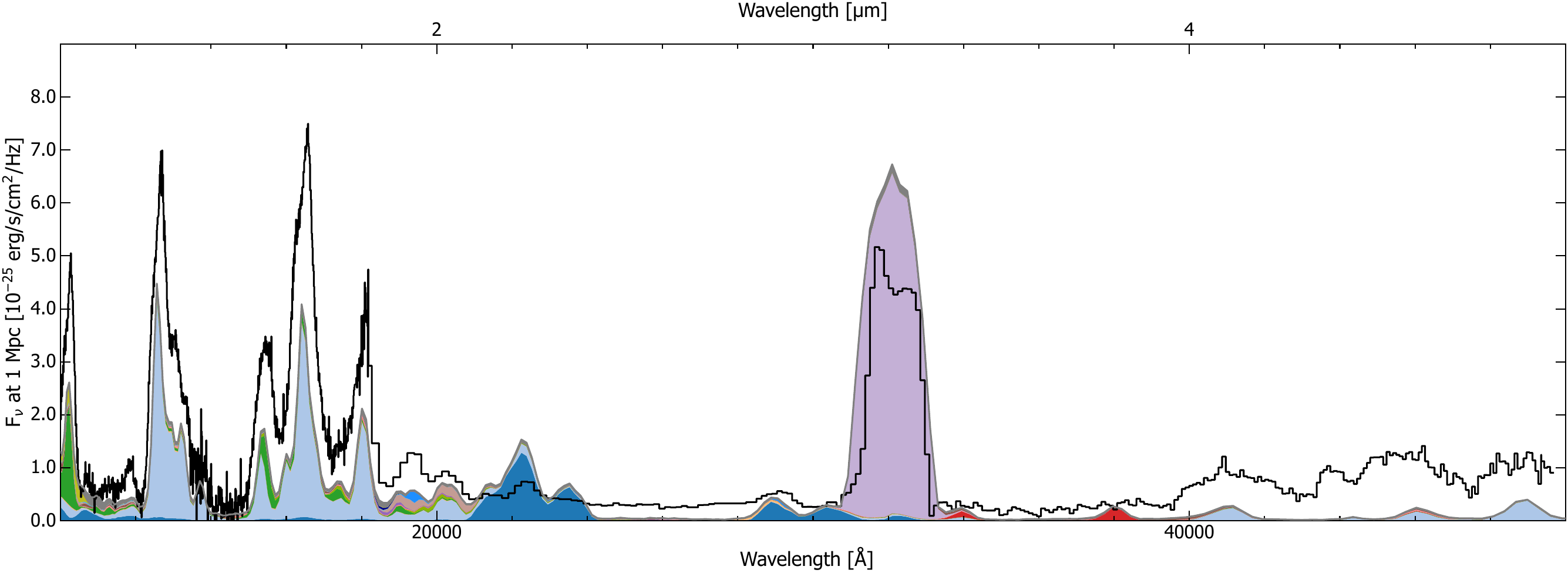}   
    
    \includegraphics[width=0.99\textwidth,height=6.5cm]{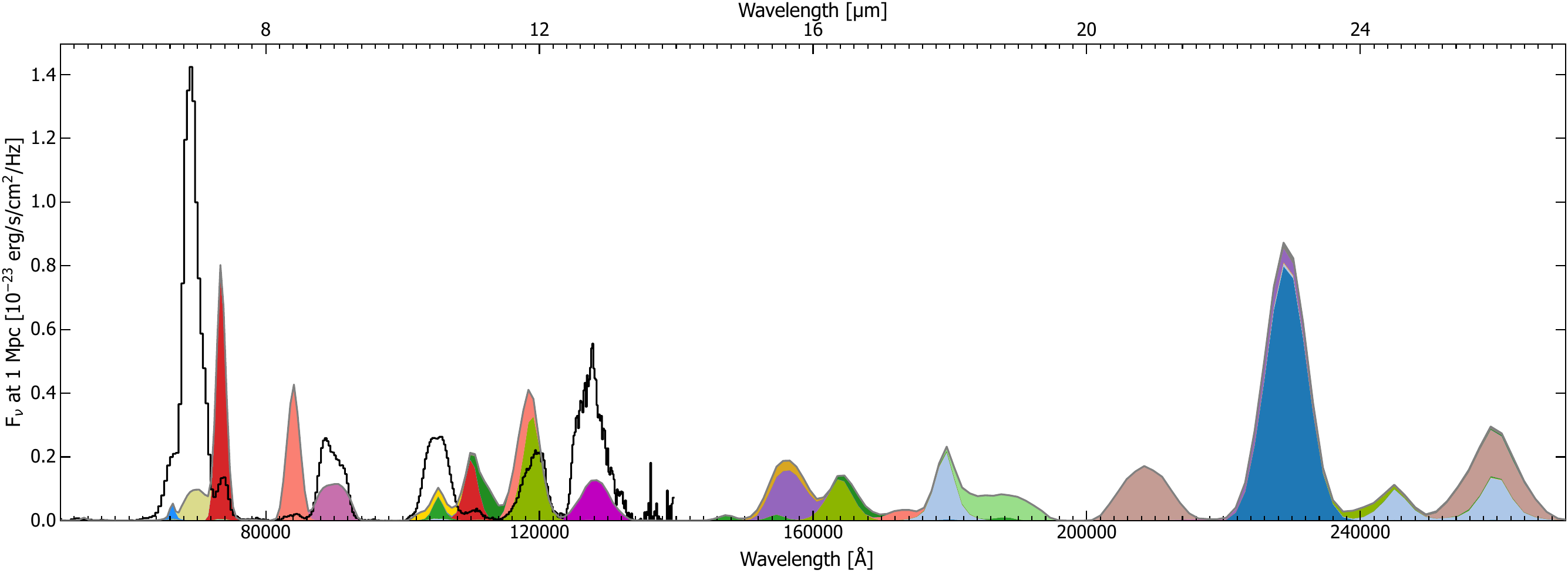}   
    
    \caption{Nebular emission and absorption spectra for the \threedvm model at 338 days across all wavelength ranges, from top to bottom: optical, NIR, and MIR ($\sim$6–29\microns).
    The positive axis is colour-coded to indicate the emitting ions, based on each Monte Carlo packet’s thermal emission type. The negative axis shows the corresponding absorption contributions from each ion, which is only present in the optical region. The total spectrum is overlaid as a thick grey curve, with the shaded regions indicating the contribution of individual ions. Observations of SN~2022pul \protect\citep{Kwok2024} are included for comparison and have been corrected for redshift and reddening.}
    \label{fig:Kromer_3DViolentMerger_pul}
\end{figure*}

\begin{figure*} 
    \centering
    \includegraphics[width=0.99\textwidth,height=6.5cm]{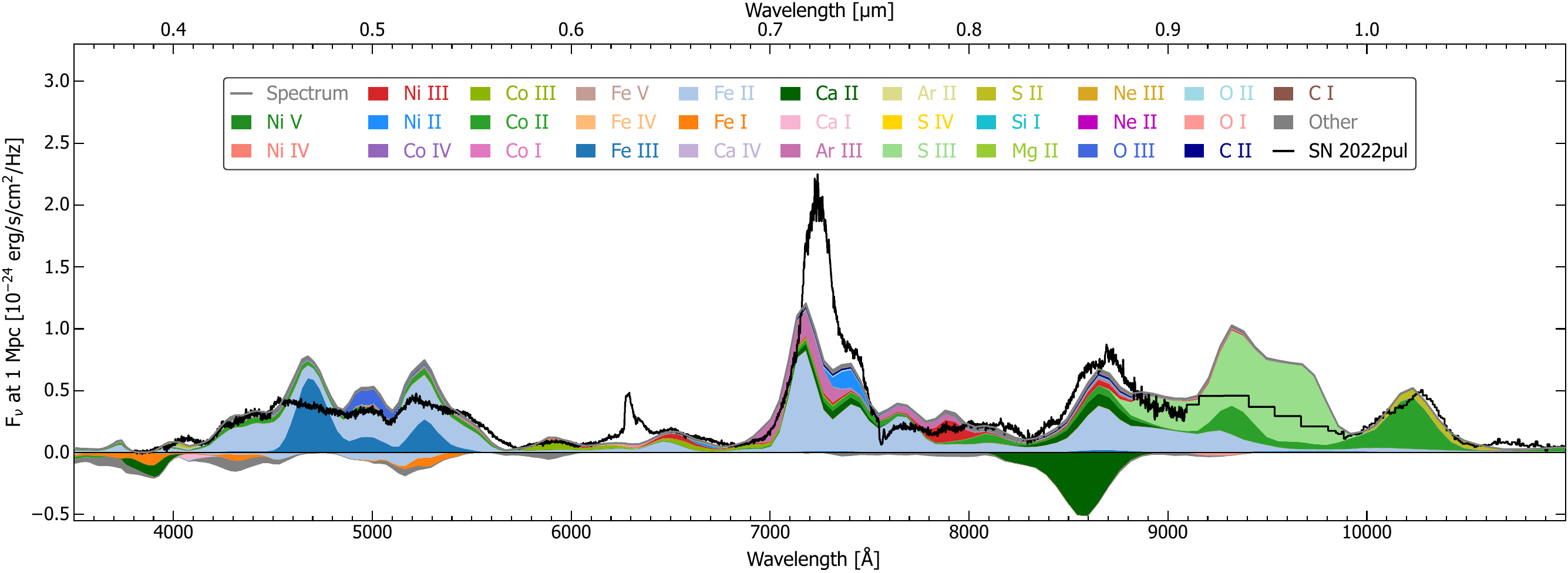}
    
    \includegraphics[width=0.99\textwidth,height=6.5cm]{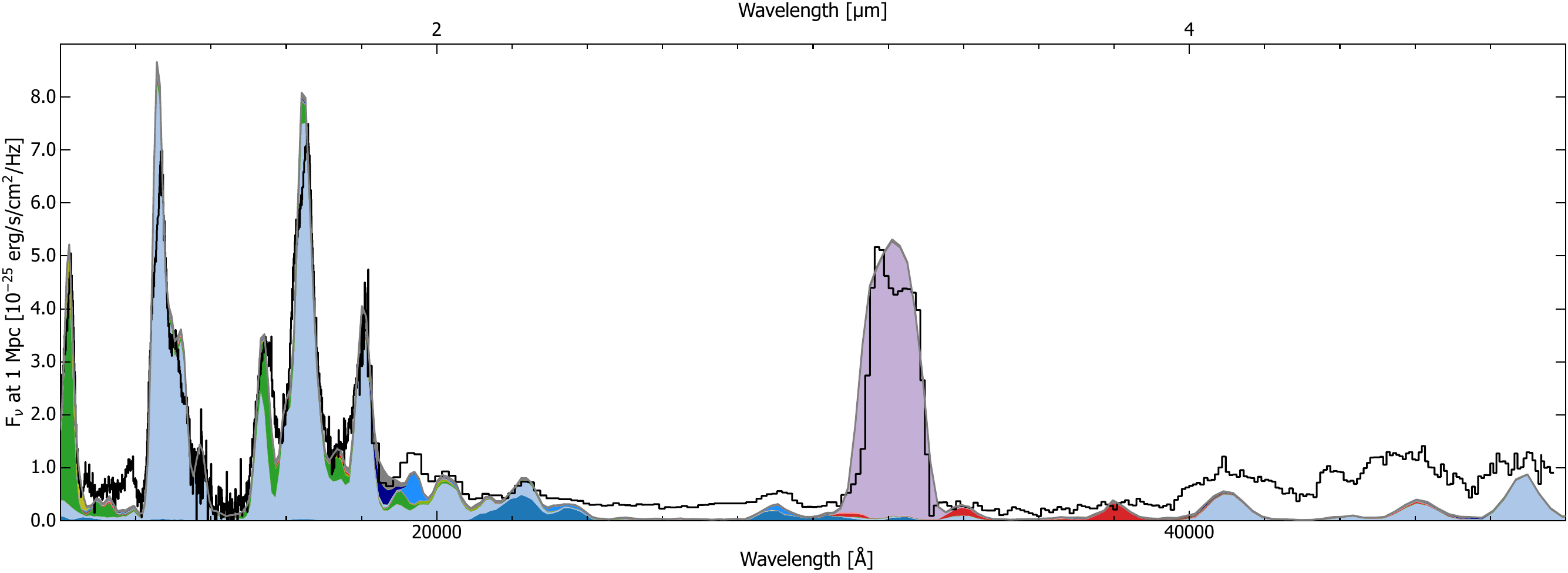}   
    
    \includegraphics[width=0.99\textwidth,height=6.5cm]{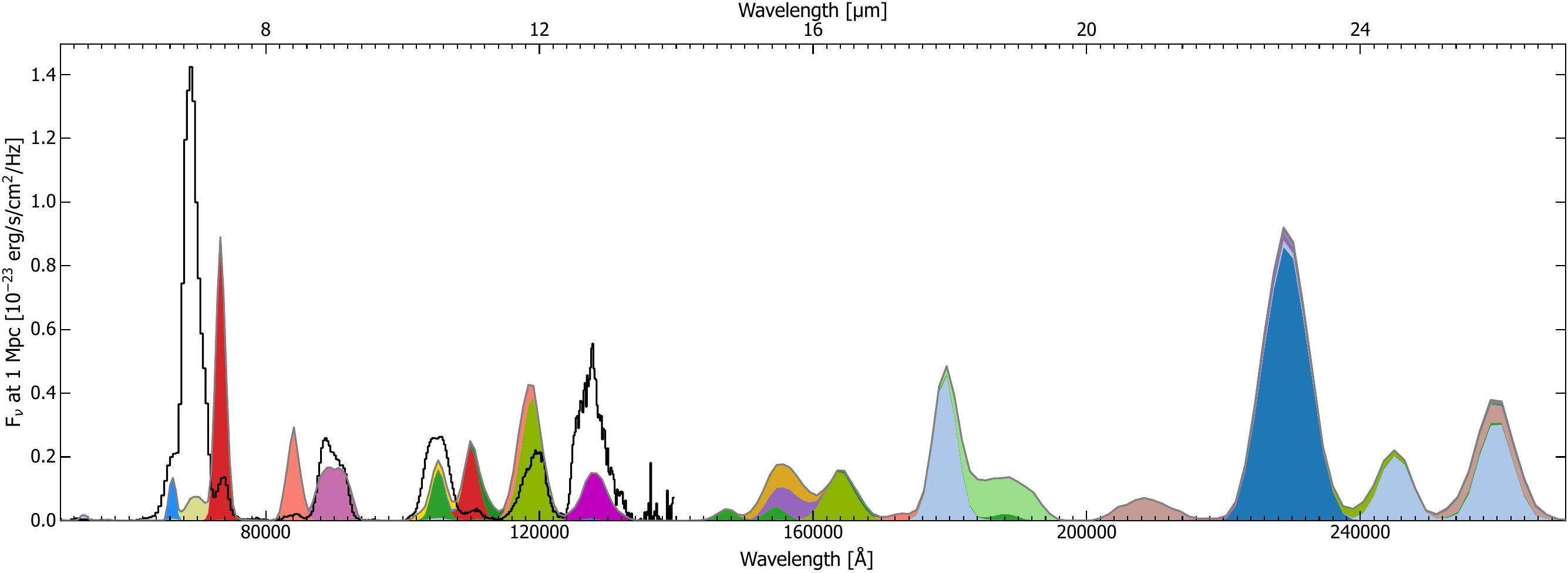} 
    
    \caption{Same as Figure~\ref{fig:Kromer_3DViolentMerger_pul} but for the \onedvm model.}
    \label{fig:Kromer_1DViolentMerger_pul}
\end{figure*}

Throughout the discussion, comparisons to SN~2022pul are made using model spectra at a slightly earlier epoch than the observations. Although nebular features evolve slowly, we present the spectral element decomposition for both the \onedvm and \threedvm models at the exact reference epoch of SN~2022pul here (Figure~\ref{fig:Kromer_3DViolentMerger_pul} and~~\ref{fig:Kromer_1DViolentMerger_pul}) for completeness.

\bsp	
\label{lastpage}
\end{document}